\documentclass[11pt]{article}
\usepackage{jheppub}
\usepackage{epsfig}
\usepackage{amssymb}
\usepackage{amsmath}
\usepackage{graphicx, subfigure, array, placeins, float}

\usepackage{tcolorbox}
\tcbuselibrary{fitting}
\usepackage{makecell}
\usepackage{multirow}
\usepackage{epstopdf}
\usepackage{extarrows}

\usepackage{dsfont,bbm}
\usepackage{shuffle}
\usepackage{blkarray}

\graphicspath{{./figures/}}

\newcommand{\im}{{\rm i}}
\newcommand{\WL}{{\rm WL}}

\newcommand{\ap}{{\rm a}}
\newcommand{\am}{\textrm{-a}}
\newcommand{\bp}{{\rm b}}
\newcommand{\bm}{\textrm{-b}}
\newcommand{\cp}{{\rm c}}
\newcommand{\cm}{\textrm{-c}}
\newcommand{\dpp}{{\rm d}}
\newcommand{\dmm}{\textrm{-d}}

\newcommand{\Bmu}{\boldsymbol{\mu}}

\usepackage{tikz}
\usetikzlibrary{decorations.markings}
\usetikzlibrary{arrows.meta}
\usepackage{adjustbox}
\tikzset{
    midarrow/.style={
        decoration={markings,
            mark=at position 0.58 with {\arrow{stealth}}
        },
        postaction={decorate}
    }
}
\tikzset{
    midreversearrow/.style={
        decoration={markings,
            mark=at position 0.42 with {\arrowreversed{stealth}}
        },
        postaction={decorate}
    }
}
\newcommand{\WLa}[1]{
\adjustbox{valign=c}{\begin{tikzpicture}[scale=0.5]
	\draw [#1] (0, 0)--(1, 0);
	\draw (1, 0)--(1, 1)--(0, 1)--(0, 0);
\end{tikzpicture}}}
\newcommand{\WLar}{
\adjustbox{valign=c}{\begin{tikzpicture}[scale=0.5]
	\draw (0, 0)--(1, 0);
	\draw [red,midarrow] (1, 0)--(1, 1);
	\node[scale=0.8] at (1.3,0.5) {$\epsilon$};
	\draw (1, 1)--(0, 1)--(0, 0);
\end{tikzpicture}}}
\newcommand{\WLbr}{
\adjustbox{valign=c}{\begin{tikzpicture}[scale=0.5]
	\draw [red,midarrow] (1, 0)--(2, 0);
	\node[scale=0.8] at (1.5, -0.3) {$\epsilon$};
	\draw (2, 0)--(2, 0.2)--(1, 0.2)--(1, 1)--(0, 1)--(0, 0)--(1,0);
\end{tikzpicture}}}
\newcommand{\WLb}[1]{
\adjustbox{valign=c}{\begin{tikzpicture}[scale=0.5]
	\draw [#1] (0, 0)--(1, 0);
	\draw (1, 0)--(2, 0)--(2, 1)--(1, 1)--(0, 1)--(0, 0);
\end{tikzpicture}}}
\newcommand{\WLc}[1]{
\adjustbox{valign=c}{\begin{tikzpicture}[scale=0.5]
	\draw [#1] (0, 0)--(1, 0);
	\draw (1, 0)--(1.2, 1);
	\draw [fill=white, draw=white] (1.1,0.5) circle (0.2);
	\draw (1.2, 1)--(2.2, 1)--(2.2, 0)--(1.2, 0)--(1, 1)--(0, 1)--(0, 0);
\end{tikzpicture}}}
\newcommand{\WLcr}{
\adjustbox{valign=c}{\begin{tikzpicture}[scale=0.5]
	\draw (1.2, 1)--(2.2, 1)--(2.2, 0)--(1.2, 0)--(1, 1)--(0, 1)--(0, 0)--(1, 0);
	\draw [fill=white, draw=white] (1.1,0.5) circle (0.2);
	\node[scale=0.8] at (1.5,0.5) {$\epsilon$};
	\draw [red, midarrow] (1, 0)--(1.2, 1);
\end{tikzpicture}}}
\newcommand{\WLd}[1]{
\adjustbox{valign=c}{\begin{tikzpicture}[scale=0.5]
	\draw [#1] (0.2, 0)--(1, 0);
	\draw (1, 0)--(1, 1);
	\draw [fill=white, draw=white] (1, 0.8) circle (0.1);
	\draw (1, 1)--(0, 1)--(0, -0.2)--(1.2, -0.2)--(1.2, 0.8)--(0.2, 0.8)--(0.2, 0);
\end{tikzpicture}}}
\newcommand{\WLe}[1]{
\adjustbox{valign=c}{\begin{tikzpicture}[scale=0.5]
	\draw [#1] (0, 0)--(1, 0);
	\draw (1, 0)--(1.2, 1.2);
	\draw [fill=white, draw=white] (1.1,0.5) circle (0.2);
	\draw (1.2, 1.2)--(2.5, 1.2)--(2.5, 0)--(1.3, 0);
	\draw [fill=white, draw=white] (2.3,0) circle (0.1);
	\draw (1.3, 0)--(1.3, 1)--(2.3, 1)--(2.3, -0.2)--(1.2, -0.2)--(1, 1)--(0, 1)--(0, 0);
\end{tikzpicture}}}
\newcommand{\WLf}[1]{
\adjustbox{valign=c}{\begin{tikzpicture}[scale=0.5]
	\draw (1.1, -0.2)--(1.3, 1);
	\draw [fill=white, draw=white] (1.2,0.5) circle (0.2);
	\draw (1.3, 1)--(2.3, 1)--(2.3, 0)--(1.3, 0)--(1.1, 1.2)--(-0.2, 1.2)--(-0.2, 0);
	\draw [#1] (-0.2, 0)--(1, 0);
	\draw [fill=white, draw=white] (0, 0) circle (0.1);
	\draw (1, 0)--(1, 1)--(0, 1)--(0, -0.2)--(1.1, -0.2);
\end{tikzpicture}}}
\newcommand{\WLg}[1]{
\adjustbox{valign=c}{\begin{tikzpicture}[scale=0.5]
	\draw [#1] (0, 0)--(1.4, 0);
	\draw (1.4, 0)--(1.4, 1.4)--(0.4, 1.4)--(0.4, 0.4);
	\draw [fill=white, draw=white] (0.4, 1.1) circle (0.2);
	\draw (0.4, 0.4)--(1.2, 0.4);
	\draw [fill=white, draw=white] (1, 0.4) circle (0.1);
	\draw (1.2, 0.4)--(1.2, 1.2)--(0.2, 1.2)--(0.2, 0.2);
	\draw [fill=white, draw=white] (0.2, 1) circle (0.1);
	\draw (0.2, 0.2)--(1, 0.2)--(1, 1)--(0, 1)--(0, 0);
\end{tikzpicture}}}
\newcommand{\WLh}[1]{
\adjustbox{valign=c}{\begin{tikzpicture}[scale=0.5]
	\draw [#1] (0, 0)--(1, 0);
	\draw (1, 0)--(2, 0)--(2, 1.2)--(1, 1.2)--(1, 0.2)--(1.8, 0.2)--(1.8, 1);
	\draw [fill=white, draw=white] (1, 1) circle (0.1);
	\draw (1.8, 1)--(1, 1)--(0, 1)--(0, 0);
\end{tikzpicture}}}
\newcommand{\WLi}[1]{
\adjustbox{valign=c}{\begin{tikzpicture}[scale=0.5]
	\draw (-0.2, -0.2)--(1.1, -0.2);
	\draw (1.1, -0.2)--(1.3, 1.2);
	\draw [fill=white, draw=white] (1.2,0.5) circle (0.2);
	\draw (1.3, 1.2)--(2.6, 1.2)--(2.6, -0.2)--(1.3, -0.2)--(1.1, 1.2)--(0, 1.2)--(0, 0);
	\draw [fill=white, draw=white] (0, 1) circle (0.1);
	\draw [#1] (0, 0)--(1, 0);
	\draw (1, 0)--(1.4, 1);
	\draw [fill=white, draw=white] (1.2,0.5) circle (0.12);
	\draw (1.4, 1)--(2.4, 1)--(2.4, 0)--(1.4, 0)--(1, 1)--(-0.2, 1)--(-0.2, -0.2);
\end{tikzpicture}}}
\newcommand{\WLj}[1]{
\adjustbox{valign=c}{\begin{tikzpicture}[scale=0.5]
	\draw [#1] (0, 0)--(1, 0);
	\draw (1, 0)--(2, 0)--(2, -1)--(1, -1);
	\draw [fill=white, draw=white] (1, 0) circle (0.1);
	\draw (1, -1)--(1, 0)--(1, 1)--(0, 1)--(0, 0);
\end{tikzpicture}}}
\newcommand{\WLk}[1]{
\adjustbox{valign=c}{\begin{tikzpicture}[scale=0.5]
	\draw [#1] (0, 0)--(0.8, 0);
	\draw (0.8, 0)--(0.8, -1)--(1.8, -1)--(1.8, 0)--(1, 0)--(1, 1)--(0, 1)--(0, 0);
\end{tikzpicture}}}
\newcommand{\WLl}[1]{
\adjustbox{valign=c}{\begin{tikzpicture}[scale=0.5]
	\draw (-1, 0)--(0, 0);
	\draw [#1] (0, 0)--(1, 0);
	\draw [fill=white, draw=white] (0, 0) circle (0.1);
	\draw (1, 0)--(1, 1)--(0, 1)--(0, 0)--(0, -1)--(-1, -1)--(-1, 0);
\end{tikzpicture}}}
\newcommand{\WLm}[1]{
\adjustbox{valign=c}{\begin{tikzpicture}[scale=0.5]
	\draw [#1] (0, 0)--(1, 0);
	\draw (1, 0)--(1, 1)--(0, 1)--(0, 0.2)--(-1, 0.2)--(-1, -0.8)--(0, -0.8)--(0, 0);
\end{tikzpicture}}}
\newcommand{\WLn}{
\adjustbox{valign=c}{\begin{tikzpicture}[scale=0.5]
	\draw [midarrow] (0, 0)--(1, 0);
	\draw (1, 0)--(2, 0)--(2, 1)--(2, 2)--(3, 2)--(3, 1);
	\draw [fill=white, draw=white] (2, 1) circle (0.1);
	\draw (3, 1)--(2, 1)--(1, 1)--(0, 1)--(0, 0);
\end{tikzpicture}}}
\newcommand{\WLdaa}[2]{
\adjustbox{valign=c}{\begin{tikzpicture}[scale=0.5]
	\draw [#1] (0, 0)--(1, 0);
	\draw (1, 0)--(1, 1)--(0, 1)--(0, 0);
	\draw [#2] (1.2, 0)--(2.2, 0);
	\draw (2.2, 0)--(2.2, 1)--(1.2, 1)--(1.2, 0);
\end{tikzpicture}}}
\newcommand{\WLdaar}{
\adjustbox{valign=c}{\begin{tikzpicture}[scale=0.5]
	\draw [midarrow] (0, 0)--(1, 0);
	\draw (1, 0)--(1, 1)--(0, 1)--(0, 0);
	\draw (1.2, 0)--(2.2, 0);
	\draw [red,midarrow] (2.2, 0)--(2.2, 1);
	\node[scale=0.8] at (2.5,0.5) {$\epsilon$};
	\draw (2.2, 1)--(1.2, 1)--(1.2, 0);
\end{tikzpicture}}}
\newcommand{\WLdab}[2]{
\adjustbox{valign=c}{\begin{tikzpicture}[scale=0.5]
	\draw [#1] (0, 0)--(1, 0);
	\draw (1, 0)--(1, 1)--(0, 1)--(0, 0);
	\draw [#2] (-0.2, -0.2)--(1.2, -0.2);
	\draw (1.2, -0.2)--(1.2, 1.2)--(-0.2, 1.2)--(-0.2, -0.2);
\end{tikzpicture}}}
\newcommand{\WLdabr}{
\adjustbox{valign=c}{\begin{tikzpicture}[scale=0.5]
	\draw [midarrow] (0, 0)--(1, 0);
	\draw (1, 0)--(1, 1)--(0, 1)--(0, 0);
	\draw [red,midarrow] (1.2, -0.2)--(1.2, 1.2);
	\node[scale=0.8] at (1.5,0.5) {$\epsilon$};
	\draw (1.2, 1.2)--(-0.2, 1.2)--(-0.2, -0.2)--(1.2, -0.2);
\end{tikzpicture}}}
\newcommand{\WLdac}[2]{
\adjustbox{valign=c}{\begin{tikzpicture}[scale=0.5]
	\draw [#1] (0, 0)--(1, 0);
	\draw (1, 0)--(1, 1)--(0, 1)--(0, 0);
	\draw [#2] (1.2, 0)--(2.2, 0);
	\draw (2.2, 0)--(2.2, -1)--(1.2, -1)--(1.2, 0);
\end{tikzpicture}}}
\newcommand{\WLdba}[2]{
\adjustbox{valign=c}{\begin{tikzpicture}[scale=0.5]
	\draw [#1] (0, 0)--(1, 0);
	\draw (1, 0)--(2, 0)--(2, 1)--(1, 1)--(0, 1)--(0, 0);
	\draw [#2] (-1.2, 0)--(-0.2, 0);	
	\draw (-0.2, 0)--(-0.2, 1)--(-1.2, 1)--(-1.2, 0);
\end{tikzpicture}}}
\newcommand{\WLdbb}[2]{
\adjustbox{valign=c}{\begin{tikzpicture}[scale=0.5]
	\draw [#1] (0, 0)--(1, 0);
	\draw (1, 0)--(2, 0)--(2, 1)--(1, 1)--(0, 1)--(0, 0);
	\draw [fill=white, draw=white] (0.2, 1) circle (0.1);
	\draw [fill=white, draw=white] (1.2, 1) circle (0.1);
	\draw [#2] (0.2, 0.2)--(1.2, 0.2);
	\draw (1.2, 0.2)--(1.2, 1.2)--(0.2, 1.2)--(0.2, 0.2);
\end{tikzpicture}}}
\newcommand{\WLdbc}[2]{
\adjustbox{valign=c}{\begin{tikzpicture}[scale=0.5]
	\draw [#1] (0, 0)--(1, 0);
	\draw (1, 0)--(2, 0)--(2, 1)--(1, 1)--(0, 1)--(0, 0);
	\draw [fill=white, draw=white] (0.8, 1) circle (0.1);
	\draw [fill=white, draw=white] (1.8, 1) circle (0.1);
	\draw [#2] (0.8, 0.2)--(1.8, 0.2);
	\draw (1.8, 0.2)--(1.8, 1.2)--(0.8, 1.2)--(0.8, 0.2);
\end{tikzpicture}}}
\newcommand{\WLdca}[2]{
\adjustbox{valign=c}{\begin{tikzpicture}[scale=0.5]
	\draw [#1] (0, 0)--(1, 0);
	\draw (1, 0)--(1.2, 1);
	\draw [fill=white, draw=white] (1.1,0.5) circle (0.2);
	\draw (1.2, 1)--(2.2, 1)--(2.2, 0)--(1.2, 0)--(1, 1)--(0, 1)--(0, 0);
	\draw [#2] (-1.2, 0)--(-0.2, 0);	
	\draw (-0.2, 0)--(-0.2, 1)--(-1.2, 1)--(-1.2, 0);
\end{tikzpicture}}}
\newcommand{\WLdda}[2]{
\adjustbox{valign=c}{\begin{tikzpicture}[scale=0.5]
	\draw [#1] (0.2, 0)--(1, 0);
	\draw (1, 0)--(1, 1);
	\draw [fill=white, draw=white] (1, 0.8) circle (0.1);
	\draw (1, 1)--(0, 1)--(0, -0.2)--(1.2, -0.2)--(1.2, 0.8)--(0.2, 0.8)--(0.2, 0);
	\draw [#2] (-1.2, 0)--(-0.2, 0);	
	\draw (-0.2, 0)--(-0.2, 1)--(-1.2, 1)--(-1.2, 0);
\end{tikzpicture}}}
\newcommand{\WLddb}[2]{
\adjustbox{valign=c}{\begin{tikzpicture}[scale=0.5]
	\draw [#1] (0.2, 0)--(1, 0);
	\draw (1, 0)--(1, 1);
	\draw [fill=white, draw=white] (1, 0.8) circle (0.1);
	\draw (1, 1)--(0, 1)--(0, -0.2)--(1.2, -0.2)--(1.2, 0.8)--(0.2, 0.8)--(0.2, 0);
	\draw [#2] (-0.2, -0.4)--(1.4, -0.4);	
	\draw (1.4, -0.4)--(1.4, 1.2)--(-0.2, 1.2)--(-0.2, -0.4);
\end{tikzpicture}}}
\newcommand{\WLdde}[2]{
\adjustbox{valign=c}{\begin{tikzpicture}[scale=0.5]
	\draw [#1] (0, 0)--(1, 0);
	\draw (1, 0)--(1.2, 1.2);
	\draw [fill=white, draw=white] (1.1,0.5) circle (0.2);
	\draw (1.2, 1.2)--(2.5, 1.2)--(2.5, -0.2)--(1.2, -0.2)--(1, 1)--(0, 1)--(0, 0);
	\draw [#2] (2.3, 0)--(1.3, 0);
	\draw (1.3, 0)--(1.3, 1)--(2.3, 1)--(2.3, 0);
\end{tikzpicture}}}
\newcommand{\WLddf}[2]{
\adjustbox{valign=c}{\begin{tikzpicture}[scale=0.5]
	\draw [#1] (-0.2, -0.2)--(1.1, -0.2);
	\draw (1.1, -0.2)--(1.3, 1);
	\draw [fill=white, draw=white] (1.2,0.5) circle (0.2);
	\draw (1.3, 1)--(2.3, 1)--(2.3, 0)--(1.3, 0)--(1.1, 1.2)--(-0.2, 1.2)--(-0.2, -0.2);
	\draw [#2] (0, 0)--(1, 0);
	\draw (1, 0)--(1, 1)--(0, 1)--(0, 0);
\end{tikzpicture}}}
\newcommand{\WLddg}[2]{
\adjustbox{valign=c}{\begin{tikzpicture}[scale=0.5]
	\draw [#1] (-0.2, -0.2)--(1.1, -0.2);
	\draw (1.1, -0.2)--(1.3, 1.2);
	\draw [fill=white, draw=white] (1.2,0.5) circle (0.2);
	\draw (1.3, 1.2)--(2.6, 1.2)--(2.6, -0.2)--(1.3, -0.2)--(1.1, 1.2)--(-0.2, 1.2)--(-0.2, -0.2);
	\draw [#2] (0, 0)--(1, 0);
	\draw (1, 0)--(1.4, 1);
	\draw [fill=white, draw=white] (1.2,0.5) circle (0.12);
	\draw (1.4, 1)--(2.4, 1)--(2.4, 0)--(1.4, 0)--(1, 1)--(0, 1)--(0, 0);
\end{tikzpicture}}}
\newcommand{\WLta}[3]{
\adjustbox{valign=c}{\begin{tikzpicture}[scale=0.5]
	\draw [#1] (0, 0)--(1, 0);
	\draw (1, 0)--(1, 1)--(0, 1)--(0, 0);
	\draw [#2] (1.2, 0)--(2.2, 0);
	\draw (2.2, 0)--(2.2, 1)--(1.2, 1)--(1.2, 0);
	\draw [#3] (-1.2, 0)--(-0.2, 0);
	\draw (-0.2, 0)--(-0.2, 1)--(-1.2, 1)--(-1.2, 0);
\end{tikzpicture}}}
\newcommand{\WLtb}[3]{
\adjustbox{valign=c}{\begin{tikzpicture}[scale=0.5]
	\draw [#1] (0, 0)--(1, 0);
	\draw (1, 0)--(1, 1)--(0, 1)--(0, 0);
	\draw [#2] (-0.2, -0.2)--(1.2, -0.2);
	\draw (1.2, -0.2)--(1.2, 1.2)--(-0.2, 1.2)--(-0.2, -0.2);
	\draw [#3] (-1.4, 0)--(-0.4, 0);
	\draw (-0.4, 0)--(-0.4, 1)--(-1.4, 1)--(-1.4, 0);
\end{tikzpicture}}}
\newcommand{\WLtc}[3]{
\adjustbox{valign=c}{\begin{tikzpicture}[scale=0.5]
	\draw [#1] (0, 0)--(1, 0);
	\draw (1, 0)--(1, 1)--(0, 1)--(0, 0);
	\draw [#2] (-0.2, -0.2)--(1.2, -0.2);
	\draw (1.2, -0.2)--(1.2, 1.2)--(-0.2, 1.2)--(-0.2, -0.2);
	\draw [#3] (1.4, 0)--(2.4, 0);
	\draw (2.4, 0)--(2.4, 1)--(1.4, 1)--(1.4, 0);
\end{tikzpicture}}}
\newcommand{\WLtd}[3]{
\adjustbox{valign=c}{\begin{tikzpicture}[scale=0.5]
	\draw [#1] (0, 0)--(1, 0);
	\draw (1, 0)--(1, 1)--(0, 1)--(0, 0);
	\draw [#2] (-0.2, -0.2)--(1.2, -0.2);
	\draw (1.2, -0.2)--(1.2, 1.2)--(-0.2, 1.2)--(-0.2, -0.2);
	\draw [#3] (-0.4, -0.4)--(1.4, -0.4);
	\draw (1.4, -0.4)--(1.4, 1.4)--(-0.4, 1.4)--(-0.4, -0.4);
\end{tikzpicture}}}
\newcommand{\WPa}{
\adjustbox{valign=c}{\begin{tikzpicture}[scale=0.5]
	\draw [fill=black] (0, 0) circle (0.05);
	\draw [fill=black] (0, 0.2) circle (0.05);
\end{tikzpicture}}}
\newcommand{\WPb}[1]{
\adjustbox{valign=c}{\begin{tikzpicture}[scale=0.5]
	\draw [fill=black] (0, 0) circle (0.05);
	\draw [fill=black] (0, 0.2) circle (0.05);
	\draw [#1] (0, 0)--(1, 0);
	\draw (1, 0)--(1, 1)--(0, 1)--(0, 0.2);
\end{tikzpicture}}}
\newcommand{\WPc}[1]{
\adjustbox{valign=c}{\begin{tikzpicture}[scale=0.5]
	\draw [fill=black] (1, 0) circle (0.05);
	\draw [fill=black] (1, 0.2) circle (0.05);
	\draw [#1] (0, 0)--(1, 0);
	\draw (1, 0.2)--(1, 1)--(0, 1)--(0, 0);
\end{tikzpicture}}}
\newcommand{\WPd}[1]{
\adjustbox{valign=c}{\begin{tikzpicture}[scale=0.5]
	\draw [fill=black] (0, 0) circle (0.05);
	\draw [fill=black] (0, 0.2) circle (0.05);
	\draw [#1] (0, 0)--(1.2, 0);
	\draw (1.2, 0)--(1.2, 1.2)--(0.2, 1.2)--(0.2, 0.2);
	\draw [fill=white, draw=white] (0.2, 1) circle (0.1);
	\draw (0.2, 0.2)--(1, 0.2)--(1, 1)--(0, 1)--(0, 0.2);
\end{tikzpicture}}}
\newcommand{\WPe}[1]{
\adjustbox{valign=c}{\begin{tikzpicture}[scale=0.5]
	\draw [fill=black] (0, 0) circle (0.05);
	\draw [fill=black] (0, 0.2) circle (0.05);
	\draw [#1] (0, 0)--(1.4, 0);
	\draw (1.4, 0)--(1.4, 1.4)--(0.4, 1.4)--(0.4, 0.4);
	\draw [fill=white, draw=white] (0.4, 1.1) circle (0.2);
	\draw (0.4, 0.4)--(1.2, 0.4);
	\draw [fill=white, draw=white] (1, 0.4) circle (0.1);
	\draw (1.2, 0.4)--(1.2, 1.2)--(0.2, 1.2)--(0.2, 0.2);
	\draw [fill=white, draw=white] (0.2, 1) circle (0.1);
	\draw (0.2, 0.2)--(1, 0.2)--(1, 1)--(0, 1)--(0, 0.2);
\end{tikzpicture}}}
\newcommand{\WPf}[1]{
\adjustbox{valign=c}{\begin{tikzpicture}[scale=0.5]
	\draw [fill=black] (0.9, 0) circle (0.05);
	\draw [fill=black] (1.1, 0) circle (0.05);
	\draw [#1] (0, 0)--(0.9, 0);
	\draw (1.1, 0)--(2, 0)--(2, 1)--(1, 1)--(0, 1)--(0, 0);
\end{tikzpicture}}}
\newcommand{\WPg}[1]{
\adjustbox{valign=c}{\begin{tikzpicture}[scale=0.5]
	\draw [fill=black] (0.8, 0) circle (0.05);
	\draw [fill=black] (1, 0) circle (0.05);
	\draw [#1] (0, 0)--(0.8, 0);
	\draw (1, 0)--(1.2, 1);
	\draw [fill=white, draw=white] (1.1, 0.5) circle (0.2);
	\draw (1.2, 1)--(2.2, 1)--(2.2, 0)--(1.2, 0)--(1, 1)--(0, 1)--(0, 0);
\end{tikzpicture}}}
\newcommand{\WPh}[1]{
\adjustbox{valign=c}{\begin{tikzpicture}[scale=0.5]
	\draw [fill=black] (0, 0) circle (0.05);
	\draw [fill=black] (0, 0.2) circle (0.05);
	\draw [#1] (0, 0)--(1, 0);
	\draw (1, 0)--(2, 0)--(2, 1)--(1, 1)--(0, 1)--(0, 0.2);
\end{tikzpicture}}}
\newcommand{\WPi}[1]{
\adjustbox{valign=c}{\begin{tikzpicture}[scale=0.5]
	\draw [fill=black] (0, 0) circle (0.05);
	\draw [fill=black] (0, 0.2) circle (0.05);
	\draw [#1] (0, 0)--(1, 0);
	\draw (1, 0)--(2, 0)--(3, 0)--(3, 1)--(2, 1)--(1, 1)--(0, 1)--(0, 0.2);
\end{tikzpicture}}}
\newcommand{\WPj}[1]{
\adjustbox{valign=c}{\begin{tikzpicture}[scale=0.5]
	\draw [fill=black] (1, 1) circle (0.05);
	\draw [fill=black] (1, 0.8) circle (0.05);
	\draw [#1] (0, 1)--(1, 1);
	\draw (0, 1)--(0, 0)--(1, 0)--(1, 0.8);
\end{tikzpicture}}}

\newcommand{\WPdaa}[1]{
\adjustbox{valign=c}{\begin{tikzpicture}[scale=0.5]
	\draw [#1] (0, 0)--(1, 0);
	\draw (1, 0)--(1, 1);
	\draw (1, 1)--(0, 1);
	\draw (0, 1)--(0, 0);
	\draw [fill=black] (-0.2, 0) circle (0.05);
	\draw [fill=black] (-0.2, 0.2) circle (0.05);
\end{tikzpicture}}}
\newcommand{\WPdab}[1]{
\adjustbox{valign=c}{\begin{tikzpicture}[scale=0.5]
	\draw [#1] (0, 0)--(1, 0);
	\draw (1, 0)--(1, 1);
	\draw (1, 1)--(0, 1);
	\draw (0, 1)--(0, 0);
	\draw [fill=black] (1.2, 0) circle (0.05);
	\draw [fill=black] (1.2, 0.2) circle (0.05);
\end{tikzpicture}}}
\newcommand{\WPdac}[2]{
\adjustbox{valign=c}{\begin{tikzpicture}[scale=0.5]
	\draw [#1] (0, 0)--(1, 0);
	\draw (1, 0)--(1, 1)--(0, 1)--(0, 0);
	\draw [#2] (-0.2, -0.2)--(1.2, -0.2);
	\draw (1.2, -0.2)--(1.2, 1.2)--(-0.2, 1.2)--(-0.2, 0);
	\draw [fill=black] (-0.2, -0.2) circle (0.05);
	\draw [fill=black] (-0.2, 0) circle (0.05);
\end{tikzpicture}}}
\newcommand{\WLdad}[2]{
\adjustbox{valign=c}{\begin{tikzpicture}[scale=0.5]
	\draw [#1] (0, 0)--(1, 0);
	\draw (1, 0)--(1, 1)--(0, 1)--(0, 0);
	\draw [#2] (1.2, 0)--(2.2, 0);
	\draw (2.2, 0)--(2.2, 1)--(1.2, 1)--(1.2, 0.2);
	\draw [fill=black] (1.2, 0) circle (0.05);
	\draw [fill=black] (1.2, 0.2) circle (0.05);
\end{tikzpicture}}}
\newcommand{\WPdba}[2]{
\adjustbox{valign=c}{\begin{tikzpicture}[scale=0.5]
	\draw [#1] (0, 0)--(1, 0);
	\draw (1, 0)--(2, 0)--(2, 1)--(1, 1)--(0, 1)--(0, 0.2);
	\draw [fill=white, draw=white] (0.2, 1) circle (0.1);
	\draw [fill=white, draw=white] (1.2, 1) circle (0.1);
	\draw [#2] (0.2, 0.2)--(1.2, 0.2);
	\draw (1.2, 0.2)--(1.2, 1.2)--(0.2, 1.2)--(0.2, 0.2);
	\draw [fill=black] (0, 0) circle (0.05);
	\draw [fill=black] (0, 0.2) circle (0.05);
\end{tikzpicture}}}
\newcommand{\WPdbb}[2]{
\adjustbox{valign=c}{\begin{tikzpicture}[scale=0.5]
	\draw [#1] (0, 0)--(1, 0);
	\draw (1, 0)--(2, 0)--(2, 1)--(1, 1)--(0, 1)--(0, 0.2);
	\draw [#1] (-1.2, 0)--(-0.2, 0);
	\draw (-0.2, 0)--(-0.2, 1)--(-1.2, 1)--(-1.2, 0);
	\draw [fill=black] (0, 0) circle (0.05);
	\draw [fill=black] (0, 0.2) circle (0.05);
\end{tikzpicture}}}
\newcommand{\WPdbc}[1]{
\adjustbox{valign=c}{\begin{tikzpicture}[scale=0.5]
	\draw [#1] (0, 0)--(1, 0);
	\draw (1, 0)--(2, 0)--(2, 1)--(1, 1)--(0, 1)--(0, 0);
	\draw [fill=black] (0.9, -0.2) circle (0.05);
	\draw [fill=black] (1.1, -0.2) circle (0.05);
\end{tikzpicture}}}
\newcommand{\WPdca}[2]{
\adjustbox{valign=c}{\begin{tikzpicture}[scale=0.5]
	\draw [#1] (0, 0)--(1, 0);
	\draw (1, 0)--(2, 0)--(3, 0)--(3, 1)--(2, 1)--(1, 1)--(0, 1)--(0, 0.2);
	\draw [fill=white, draw=white] (0.2, 1) circle (0.1);
	\draw [fill=white, draw=white] (1.2, 1) circle (0.1);
	\draw [#2] (0.2, 0.2)--(1.2, 0.2);
	\draw (1.2, 0.2)--(1.2, 1.2)--(0.2, 1.2)--(0.2, 0.2);
	\draw [fill=black] (0, 0) circle (0.05);
	\draw [fill=black] (0, 0.2) circle (0.05);
\end{tikzpicture}}}
\newcommand{\WPdcb}[2]{
\adjustbox{valign=c}{\begin{tikzpicture}[scale=0.5]
	\draw [#1] (0, 0)--(1, 0);
	\draw (1, 0)--(2, 0)--(3, 0)--(3, 1)--(2, 1)--(1, 1)--(0, 1)--(0, 0.2);
	\draw [#1] (-1.2, 0)--(-0.2, 0);
	\draw (-0.2, 0)--(-0.2, 1)--(-1.2, 1)--(-1.2, 0);
	\draw [fill=black] (0, 0) circle (0.05);
	\draw [fill=black] (0, 0.2) circle (0.05);
\end{tikzpicture}}}
\newcommand{\WPdcc}[1]{
\adjustbox{valign=c}{\begin{tikzpicture}[scale=0.5]
	\draw [#1] (0, 0)--(1, 0);
	\draw (1, 0)--(1.2, 1);
	\draw [fill=white, draw=white] (1.1,0.5) circle (0.2);
	\draw (1.2, 1)--(2.2, 1)--(2.2, 0)--(1.2, 0)--(1, 1)--(0, 1)--(0, 0);
	\draw [fill=black] (1, -0.2) circle (0.05);
	\draw [fill=black] (1.2, -0.2) circle (0.05);
\end{tikzpicture}}}
\newcommand{\WPdda}[2]{
\adjustbox{valign=c}{\begin{tikzpicture}[scale=0.5]
	\draw [#1] (0.2, 0)--(1, 0);
	\draw (1, 0)--(1, 1);
	\draw [fill=white, draw=white] (1, 0.8) circle (0.1);
	\draw (1, 1)--(0, 1)--(0, -0.2)--(1.2, -0.2)--(1.2, 0.8)--(0.2, 0.8)--(0.2, 0);
	\draw [#2] (-0.2, -0.4)--(1.4, -0.4);	
	\draw (1.4, -0.4)--(1.4, 1.2)--(-0.2, 1.2)--(-0.2, -0.2);
	\draw [fill=black] (-0.2, -0.2) circle (0.05);
	\draw [fill=black] (-0.2, -0.4) circle (0.05);
\end{tikzpicture}}}
\newcommand{\WPddb}[2]{
\adjustbox{valign=c}{\begin{tikzpicture}[scale=0.5]
	\draw [#1] (0.2, 0)--(1, 0);
	\draw (1, 0)--(1, 1);
	\draw [fill=white, draw=white] (1, 0.8) circle (0.1);
	\draw (1, 1)--(0, 1)--(0, -0.2)--(1.2, -0.2)--(1.2, 0.8)--(0.2, 0.8)--(0.2, 0);
	\draw [#2] (1.4, 0)--(2.4, 0);	
	\draw (2.4, 0)--(2.4, 1)--(1.4, 1)--(1.4, 0.2);
	\draw [fill=black] (1.4, 0) circle (0.05);
	\draw [fill=black] (1.4, 0.2) circle (0.05);
\end{tikzpicture}}}
\newcommand{\WPddc}[1]{
\adjustbox{valign=c}{\begin{tikzpicture}[scale=0.5]
	\draw [#1] (0.2, 0)--(1, 0);
	\draw (1, 0)--(1, 1);
	\draw [fill=white, draw=white] (1, 0.8) circle (0.1);
	\draw (1, 1)--(0, 1)--(0, -0.2)--(1.2, -0.2)--(1.2, 0.8)--(0.2, 0.8)--(0.2, 0);
	\draw [fill=black] (-0.2, 0) circle (0.05);
	\draw [fill=black] (-0.2, -0.2) circle (0.05);
\end{tikzpicture}}}
\newcommand{\WPdga}[1]{
\adjustbox{valign=c}{\begin{tikzpicture}[scale=0.5]
	\draw [#1] (0, 0)--(1.4, 0);
	\draw (1.4, 0)--(1.4, 1.4)--(0.4, 1.4)--(0.4, 0.4);
	\draw [fill=white, draw=white] (0.4, 1.1) circle (0.2);
	\draw (0.4, 0.4)--(1.2, 0.4);
	\draw [fill=white, draw=white] (1, 0.4) circle (0.1);
	\draw (1.2, 0.4)--(1.2, 1.2)--(0.2, 1.2)--(0.2, 0.2);
	\draw [fill=white, draw=white] (0.2, 1) circle (0.1);
	\draw (0.2, 0.2)--(1, 0.2)--(1, 1)--(0, 1)--(0, 0);
	\draw [fill=black] (-0.2, 0.2) circle (0.05);
	\draw [fill=black] (-0.2, 0) circle (0.05);
\end{tikzpicture}}}

\title{Bootstrapping SU(3) Lattice Yang-Mills Theory}

\author[a,b]{Yuanhong Guo,}
\emailAdd{guoyuanhong@itp.ac.cn}
\author[a,b]{Zeyu Li,}
\emailAdd{lizeyu@itp.ac.cn}
\author[a,b,c]{Gang Yang,}
\emailAdd{yangg@itp.ac.cn}
\author[a,b]{and Guorui Zhu}
\emailAdd{zhuguorui@itp.ac.cn}
\affiliation[a]{CAS Key Laboratory of Theoretical Physics, Institute of Theoretical Physics, \\Chinese Academy of Sciences, Beijing 100190, China}
\affiliation[b]{School of Physical Sciences, University of Chinese Academy of Sciences, Beijing 100049, China}
\affiliation[c]{School of Fundamental Physics and Mathematical Sciences, Hangzhou Institute for Advanced Study, UCAS, Hangzhou 310024, China}

\abstract{
We apply the positivity bootstrap approach to SU(3) lattice Yang-Mills (YM) theory, extending previous studies of large N and SU(2) theories by incorporating multiple-trace Wilson loop operators. By utilizing Hermitian and reflection positivity conditions, alongside Schwinger-Dyson (SD) loop equations, we compute rigorous bounds for the expectation values of plaquette Wilson loops in 2D, 3D, and 4D YM theories. Our results exhibit clear convergence and are consistent with known analytic or numerical results. To enhance the approach, we introduce a novel twist-reflection positivity condition, which we prove to be exact in 2D YM theory. Additionally, we propose a dimensional-reduction truncation, where Wilson loop operators are effectively restricted to a lower-dimensional subplane, significantly simplifying computations. SD equations for double-trace Wilson loops are also derived in detail. Our findings suggest that the positivity bootstrap method is broadly applicable to higher-rank gauge theories beyond single-trace cases, providing a solid foundation for further non-perturbative investigations of gauge theories using positivity-based methods.
}

\begin{document}

\maketitle

\setcounter{footnote}{0}

\section{Introduction}

Understanding the strong coupling dynamics of non-Abelian gauge theories remains one of the most challenging problems in theoretical physics. Among these theories, Quantum Chromodynamics (QCD) is particularly significant, as it describes the strong interaction, a fundamental pillar of the Standard Model. 
While the lattice gauge theory approach based on numerical simulation has proven to be the most effective non-perturbative framework for studying QCD, it remains a fundamental problem to understand it by other means \cite{JaffeWitten}.
Notable achievements have been made for solving exactly certain four-dimensional gauge theories, such as the Seiberg-Witten solution for ${\cal N}=2$ supersymmetric gauge theories \cite{Seiberg:1994rs}, the integrability of planar ${\cal N}=4$ super-Yang-Mills theory \cite{Beisert:2010jr}, and the bootstrap method for conformal field theories \cite{Rattazzi:2008pe}.

A novel bootstrap method incorporating positivity constraints and loop equations for lattice gauge theories was introduced by Anderson and Kruzenski in 2016 \cite{Anderson:2016rcw}.
In contrast to other powerful non-perturbative methods, this approach does not require special symmetries (such as supersymmetry, conformal symmetry, or integrability). Instead, it relies solely on the universal unitarity principles alongside Schwinger-Dyson equations which can be derived from fundamental actions. Moreover, it applies to general spacetime dimensions,
making it a promising framework for exploring realistic gauge theories.

Important improvements to \cite{Anderson:2016rcw} were made in the recent work of Kazakov and Zheng in \cite{Kazakov:2022xuh, Kazakov:2024ool}, 
where reflection positivity constraints and symmetry reductions were incorporated. 
A further key observation in \cite{Kazakov:2024ool} is the reduction of operators in finite rank gauge theories: in the SU($N$) Yang-Mills (YM) theory, all multiple-trace operators can be reduced to those with at most $(N-1)$ traces. For SU(2), this reduction simplifies analyses to single-trace operators, facilitating an efficient bootstrap approach as demonstrated in \cite{Kazakov:2024ool}.
The abelian U(1) lattice gauge theory was also considered in \cite{Li:2024wrd}.\footnote{
The similar bootstrap strategy has been applied in other contexts, including random matrix models \cite{Lin:2020mme, Kazakov:2021lel}, 
lattice Ising models \cite{Cho:2022lcj, Cho:2023ulr},
quantum mechanics systems \cite{Han:2020bkb, Berenstein:2021dyf, Berenstein:2022unr, Nancarrow:2022wdr, Lin:2023owt, Lin:2024vvg, Guo:2023gfi, Mathaba:2023non, Hessam:2021byc, Berenstein:2024ebf, Khalkhali:2024kiv, Cho:2024kxn, Scheer:2024eyu, Cho:2024owx, Lawrence:2024mnj}, 
as well as evaluating Feynman integral \cite{Zeng:2023jek}.
Similar ideas appeared also in studying quantum many-body systems, see \emph{e.g.}~\cite{MazziottiPRA2001, NakataEtal2001, MazziottiPRL2004, Barthel:2012Solving, Baumgratz:2012Lower, Lin:2020Variational, Haim:2020VariationalCorrelations, Wang:2023hss, Gao:2024etm}.
}

Extending these methods to higher-rank gauge theories presents significant challenges. In particular, SU(3) Yang-Mills theory plays a crucial role for two reasons: first, it is the smallest case requiring multi-trace operators beyond the single-trace sector; second, it directly connects to QCD, the theory governing the strong interaction in nature. In this work, we present the first concrete application of the positivity bootstrap method to SU(3) Yang-Mills theory.

This study makes several advancements, summarized as follows:

\begin{itemize}

\item 
Considering multiple-trace operators is significantly more complicated than focusing solely on single-trace operators. We address the complexities of incorporating both double- and triple-trace operators, tackling technical challenges related to loop representations, loop equations, and the construction of positivity matrices. 

\item
We introduce a new positivity condition that incorporates a twisted symmetry mapping, which we refer to as twist-reflection positivity.
We prove that this positivity condition is strictly valid in two-dimensional lattice Yang-Mills theory. We also demonstrate it imposes very stringent new constraints.

\item
We propose a new dimensional-truncation strategy for Wilson loop operators, demonstrating that in a $d$-dimensional lattice theory, it is beneficial to consider operators within an effectively $(d-1)$-dimensional subspace. This strategy significantly simplifies computations while preserving essential physical characteristics.

\end{itemize}

Despite the increased complexity introduced by multiple-trace operators, our SU(3) results yield bounds similar to those obtained for SU(2). This suggests that the positivity bootstrap method is broadly applicable to more general higher-rank gauge theories.
Compared to traditional lattice gauge theory methods, which rely purely on numerical simulations, 
the bootstrap method offers a fundamentally different perspective:  it directly incorporates positivity (\emph{i.e.}, unitarity) constraints and uses detailed knowledge of loop variables—their shape, structure, and dynamics. This new approach is expected to provide new insights into non-perturbative gauge theories and offer promising avenues for exploring hadronic physics and strong interactions. 

The remainder of this paper is organized as follows. 
In Section~\ref{sec:setup}, we review the basic setup of the positivity bootstrap method.
Section~\ref{sec:positivity} explains the positivity conditions, with an emphasis on the novel twist-reflection positivity.
In Section~\ref{sec:loopequation}, we consider loop equations including the double-trace loop equations and the SU(3) reduction relations.
Section~\ref{sec:pathselection} explain the strategy of the selection of path operators in constructing the positivity matrices. 
Section~\ref{sec:results2D} and \ref{sec:results3D4D} present the details of bootstrap constructions and results in 2D, 3D, and 4D YM respectively.
Section~\ref{sec:outlook} provides a summary and outlook.
We also provide several appendices discussing loop representation and symmetry decomposition, SU(3) identities, as well as perturbative expansions in both strong- and weak-coupling regimes.

\section{Setup and strategy}
\label{sec:setup}

In this section, we first review the setup of the lattice gauge theory and the Wilson loop operators. 
Then we describe the bootstrap strategy based on the positivity conditions and loop equations, which will be elaborated on in later sections.

\subsection{Lattice YM theory and Wilson loops}

Lattice gauge theory provides a non-perturbative framework to study quantum chromodynamics (QCD) by discretizing spacetime into a grid \cite{Wilson:1974sk}. 
We will work with the standard cubic Euclidean lattice in $d$ dimensions
\begin{equation}
\Gamma_d = \{ x:  \ x = a \sum_{\mu=1}^d n_\mu  \, \hat\mu , \quad n_\mu \in \mathbb{Z} \} \,.
\end{equation}
Here, $n_\mu$ are integers representing the coordinates along each axis, and $\hat{\mu}$ are unit vectors in the respective directions. The lattice spacing $a$ sets the scale of discretization.
We will consider $d=2,3,4$ cases.
To each edge connecting neighboring points $x$ and $x+\hat\mu$, we assign an SU(N) matrix $U_\mu(x)$. These matrices represent gauge fields in the discrete framework and are called link variables.
The Hermitian conjugate of a link variable is to reverse the direction of the link, thus the link variable from the site $x+\hat\mu$ to $x$ is denoted as $U_\mu^\dagger(x)$:
\begin{equation}
U_\mu^\dagger(x) = U_{-\mu}(x+\hat\mu) \,.
\end{equation}

\begin{figure}[t]
\centering
\includegraphics[width=0.35\linewidth]{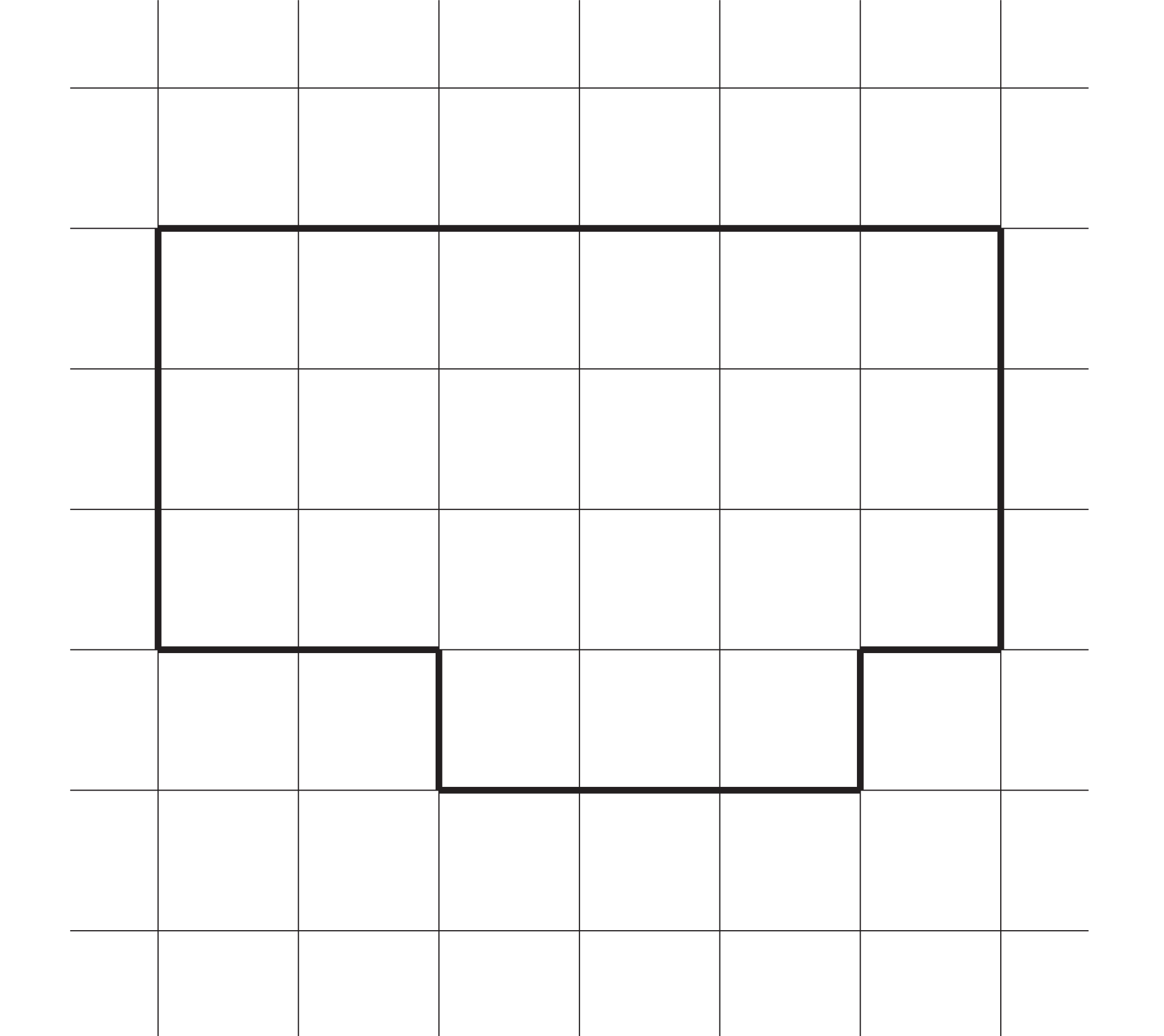}
\caption{A Wilson loop on a 2D square lattice.}
\label{fig:latticeloop}
\end{figure}

The link variables transform under gauge symmetry as
\begin{equation}
U_\mu(x) \rightarrow \Omega(x) U_\mu(x) \Omega^\dagger(x+\hat\mu) \,,
\end{equation}
where $\Omega(x)$ is also an element of SU(N).
A gauge invariant operator can be formed by taking the trace of products of link variables that form a closed path on the lattice. Such operators are called \emph{Wilson loops}. 
The simplest example is a \emph{plaquette}:
\begin{equation}
	{\rm tr}(U_P) = {\rm tr} \big[ U_\mu(x) U_\nu(x+\hat\mu) U_\mu^\dagger(x+\hat\nu) U_\nu^\dagger(x) \big] =
	\adjustbox{valign=c}{\begin{tikzpicture}[scale=1]
		\node at (0.5,-0.3) {$U_\mu$};
		\draw [midarrow] (0, 0)--(1, 0);
		\node at (1.3,0.5) {$U_\nu$};
		\draw [midarrow] (1, 0)--(1, 1);
		\node at (0.5,1.3) {$U_\mu^\dagger$};
		\draw [midarrow] (1, 1)--(0, 1);
		\node at (-0.3,0.5) {$U_\nu^\dagger$};
		\draw [midarrow] (0, 1)--(0, 0);
	\end{tikzpicture} 
	} \,,
\end{equation}
which is the most basic loops one can form on the cubic lattice.
A more complicated loop that spans multiple plaquettes is given in Figure~\ref{fig:latticeloop}.
We define the \emph{length} of a Wilson loop as the number of its link variables, so the plaquette has length 4 and the loop in Figure~\ref{fig:latticeloop} is of length 20.

The dynamics of the lattice YM theory are governed by the Wilson action \cite{Wilson:1974sk}:
\begin{equation}
S = - {N \over 2\lambda} \sum_P {\rm tr} (U_P + U_P^\dagger) \,,
\end{equation}
where the sum is over all plaquette (with only one orientation) on the lattice. In the continuum limit $a\rightarrow0$, the lattice action can reproduce the usual YM action and $\lambda$ is related to the 't Hooft coupling $\lambda = g^2 N$ in the continuum limit.

The partition function is defined by the path integral:
\begin{equation}
Z = \int {\cal D} U e^{- S} = \int \prod_{x, \mu} d U_{\mu}( x) e^{ + {N \over 2\lambda} \sum_P {\rm tr} (U_P + U_P^\dagger)} \,.
\end{equation}
The vacuum expectation value for an operator ${\cal O}$ is then given as
\begin{equation}
\langle {\cal O} \rangle = Z^{-1} \int {\cal D} U e^{- S} {\cal O} = Z^{-1} \int \prod_{ x, \mu} d U_{\mu}( x) e^{ + {N \over 2\lambda} \sum_P {\rm tr} (U_P + U_P^\dagger)} {\cal O} \,.
\end{equation}
Next we describe the bootstrap approach to compute  the expectation value of Wilson loop operators.

\subsection{Positivity bootstrap}
\label{sec:bootstrapstrategy}

The bootstrap method is an approach that constrains the expectation values of physical quantities by combining physical constraints with mathematical optimization techniques.
In quantum field theories, certain correlation functions are required to satisfy positivity conditions. 
For Wilson loops, these positivity constraints (Hermitian and reflection positivity) manifest as positive semi-definite matrices involving their expectation values.
Additionally, Wilson loops obey Schwinger-Dyson equations derived from the dynamics of the theory, imposing further algebraic constraints.

Finding bounds on Wilson loop expectations with these constraints naturally translates into a semi-definite programming (SDP) problem:\footnote{Advanced SDP techniques have been widely applied in conformal bootstrap, see \cite{Simmons-Duffin:2015qma, Poland:2018epd, Rychkov:2023wsd} for recent developments and introductions to SDP solvers.} 
positivity conditions impose convex constraints, loop equations impose further equality constraints, and SDP solvers yield rigorous upper and lower bounds on Wilson loop expectation values.
The strategy can be schematically summarized as:
%
$$
\begin{matrix}
\begin{tabular}{| c |}
\hline  Positivity conditions  \\ \hline
\end{tabular}
\\  \\
\begin{tabular}{| c |}
\hline  Loop equations  \\ \hline
\end{tabular}
 \end{matrix}
 \Bigg\} \quad
 \xLongrightarrow[\mbox{programming}]{\mbox{Semi-definite}}
 \quad
\begin{tabular}{| c |}
\hline Constraining the value \\ of Wilson loops   \\ \hline
\end{tabular}
$$
We will refer to this bootstrap approach as \emph{positivity bootstrap}, emphasizing the central role of Hermitian and reflection positivity conditions in this approach.

To illustrate the method, consider a simple case in two-dimensional SU(2) Yang-Mills theory with four Wilson loops:
\begin{equation}
\label{eq:wexample}
w_1= \WLa{midarrow}\,,\quad w_2 = \WLb{midarrow}\,,\quad w_3 = \WLc{midarrow}\,,\quad w_4 = \WLd{midarrow} \,.
\end{equation}
The positivity conditions for these loops can be expressed as:
\begin{equation}
\label{eq:positivityexample}
\left(\begin{array}{ccccc}
		1 & w_1 & w_1 & w_1 & w_1 \\
		w_1 & 1 & w_4 & w_2 & w_3 \\
		w_1 & w_4 & 1 & w_3 & w_2 \\
		w_1 & w_2 & w_3 & 1 & w_4 \\
		w_1 & w_3 & w_2 & w_4 & 1
	\end{array}\right) \succeq 0 \,,
\qquad 
\left(\begin{array}{ccc}
		1 & w_1 & w_1 \\
		w_1 & w_2 & w_3 \\
		w_1 & w_3 & w_2
	\end{array}\right) \succeq 0 \,,
\end{equation}
where the first matrix corresponds to Hermitian positivity and the second one is due to reflection positivity.
Additionally, these loops satisfy a linear relation (for SU(2) YM):
\begin{equation}
\label{eq:sdeexample}
{3\over2} \lambda \, w_1 + w_3 - w_2 + w_4 - 1 = 0 \,,
\end{equation}
given by the Schwinger-Dyson equation among these Wilson loops.
These constraints will be explained and derived in detail in Section~\ref{sec:positivity} and Section~\ref{sec:loopequation}.

\begin{figure}[t]
\centering
\subfigure[]{\includegraphics[width=0.45\linewidth]{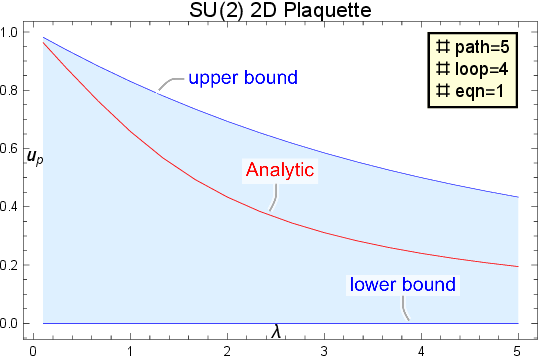}}
\qquad
\subfigure[]{\includegraphics[width=0.45\linewidth]{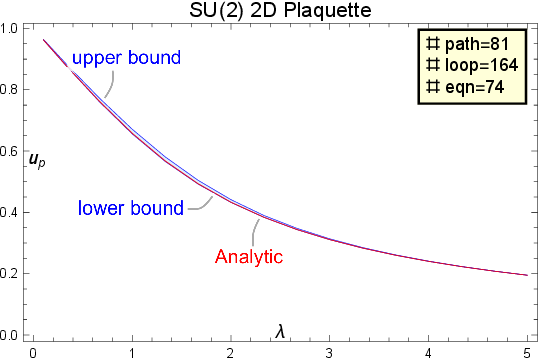}}
\caption{SDP bounds for plaquette in 2d YM.}
\label{fig:2DexampleSDPresult}
\end{figure}

Solving the constraints \eqref{eq:positivityexample} and \eqref{eq:sdeexample} yields bounds on the plaquette expectation value $w_1$, as shown in Figure~\ref{fig:2DexampleSDPresult}(a). In this simple example, the obtained bounds are relatively loose. However, by incorporating more loop variables, one can construct larger constraining systems which can give much stringent bounds, such as in Figure~\ref{fig:2DexampleSDPresult}(b).
We will present more examples in Section~\ref{sec:results2D}$-$\ref{sec:results3D4D}.

Finally, the positivity matrix can be decomposed into smaller matrices by leveraging lattice symmetries, as discussed in \cite{Kazakov:2022xuh, Kazakov:2024ool}. This reduction significantly improves computational efficiency and is reviewed in Appendix~\ref{app:symreduction}.

\section{Positivity conditions}
\label{sec:positivity}

Wilson loop observables in quantum field theories obey certain positivity conditions that play a crucial role in constraining their behavior and ensuring unitarity. Two major types of positivity conditions are Hermitian Positivity (HP) and Reflection Positivity (RP). 
In this section, we delve into these concepts and introduce a novel twist-reflection positivity.

\subsection{Hermitian positivity}

\begin{figure}[t]
\centering
\includegraphics[width=0.95\linewidth]{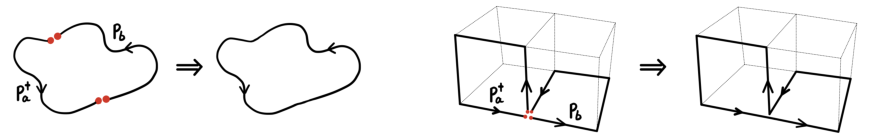}
\caption{Two paths sharing the same endpoints (red dots) can be sewn to form a Wilson loop. The figure on the right-hand side figure is a 3D lattice example in which both endpoints coincide.}
\label{fig:pathsewEx}
\end{figure}

The Hermitian Positivity (HP) is due to the positivity property of path integral that, for any functional $f(U)$, the expectation value of $\langle |f(U)|^2 \rangle$ must be non-negative:
\begin{equation}
\langle |f(U)|^2 \rangle = \int {\cal D} U \,  e^{-S}  |f(U)|^2 \geq 0\,.
\end{equation}
To form a positive condition for Wilson loops, one notes that for any matrix $U$ with components $u_{ij}, i,j =1,... N$:
\begin{equation}
{\rm tr}(U^\dagger U) = \sum_{i,j=1}^N |u_{ij}|^2 \geq 0 \,.
\end{equation}
Now, consider $U$ as a linear combination of a set of \emph{Wilson line} operators (with two open endpoints) $P_a, a=1,...,M$:
\begin{equation}
U = \sum_{a=1}^M c_a P_a \,,
\end{equation}
where $c_a$ are arbitrary real coefficients.
The Hermitian condition requires 
\begin{equation}
\langle {\rm tr}(U^\dagger U) \rangle = \sum_{a,b} c_a c_b \langle {\rm tr}(P_a^\dagger P_b) \rangle \geq 0 \,.
\end{equation}
This necessitates the matrix $A$ to be positive semi-definite, namely
\begin{equation}
A  \succcurlyeq 0 , \qquad  A_{ab}:= \langle {\rm tr}(P_a^\dagger P_b) \rangle \,.
\end{equation}
Note that in order to have the matrix elements $A_{ab}$ as gauge-invariant Wilson loop variables, all the Wilson line operators $P_a$ should share the same endpoints, so that they can be sewn to form a closed loop, as shown in Figure~\ref{fig:pathsewEx}. We will also refer to $P_a$ as \emph{path operators}.

As a concrete example, consider the following five path operators:
\begin{equation}
{\cal O} = \left\{\WPa \,,\ \WPb{midarrow}\,,\ \WPb{midreversearrow}\,,\ \WPc{midreversearrow}\,,\, \WPc{midarrow} \right\} \,,
\end{equation}
where `$\WPa$' represents the identity matrix, and both endpoints for all path operators are located at the origin of the lattice.
One can obtain its Hermitian conjugate operators by simply flipping the arrow directions
\begin{equation}
{\cal O}^\dagger = \left\{\WPa \,,\ \WPb{midreversearrow}\,,\ \WPb{midarrow}\,,\, \WPc{midarrow}\,,\ \WPc{midreversearrow} \right\} \,.
\end{equation}
Sewing the above two sets of operators, one obtains the Hermitian positivity matrix as
\begin{align}
	{\renewcommand{\arraystretch}{1.5}
	\begin{blockarray}{cccccc}
		& \WPa & \WPb{midarrow} & \WPb{midreversearrow} & \WPc{midreversearrow} & \WPc{midarrow} \\
		\begin{block}{c(ccccc)}
			\WPa & 1 & \WLa{midarrow} & \WLa{midreversearrow} & \WLa{midreversearrow} & \WLa{midarrow} \\
			\WPb{midreversearrow} & \WLa{midreversearrow} & 1 & \WLd{midreversearrow} & \WLb{midreversearrow} & \WLc{midarrow} \, \\
			\WPb{midarrow} & \WLa{midarrow} & \WLd{midarrow} & 1 & \WLc{midreversearrow} & \WLb{midarrow} \\
			\WPc{midarrow} & \WLa{midarrow} & \WLb{midarrow} & \WLc{midarrow} & 1 & \WLd{midarrow} \\
			\WPc{midreversearrow} & \WLa{midreversearrow} & \WLc{midreversearrow} & \WLb{midarrow} & \WLd{midreversearrow} & 1 \\
		\end{block}
	\end{blockarray}}
	\succeq 0 \,,
\end{align}
which contains four independent Wilson loop variables as given in \eqref{eq:wexample}. This matrix corresponds precisely to the first matrix given in \eqref{eq:positivityexample}.

\subsection{Reflection positivity}

Another more non-trivial type of positivity is the so-called Reflection Positivity (RP). Reflection positivity was first introduced by Osterwalder and Schrader in early 1970s as a key aspect to bridge the gap between an Euclidean field theory and a Minkowski quantum field theory \cite{Osterwalder:1973dx, Osterwalder:1974tc}. 
It requires that the correlation functions of a Euclidean theory remain positive under (Euclidean) time reflection.
Through analytic continuation to the Minkowski signature, reflection positivity ensures the unitarity property. 
This concept also plays an important role in statistical mechanical models \cite{Frohlich:1978px, Biskup2006}.

Below we first review the reflection positivity in lattice gauge theory \cite{Osterwalder:1977pc, Seiler:1982pw}, focusing on the construction of reflection-positivity matrices.
We will also introduce a new type of reflection positivity based on a twist reflection. 
Without loss of generality, we will consider a 3-dimensional lattice, and the three directions are along $x, y, z$-axises respectively.

\subsubsection{Standard RP}

Mathematically, reflection positivity is related to the property that the path integral may be factorized as two parts and rewritten as a modular square
\begin{equation}
\int {\cal D} U \,  e^{-S} \simeq \Big| \int {\cal D} U_+ \,  e^{-S_+}  f(U_+) \Big|^2 \geq 0\,.
\end{equation}
The key is thus to factorize the path integral in a proper way.

\begin{figure}[t]
\centering
\includegraphics[width=0.35\linewidth]{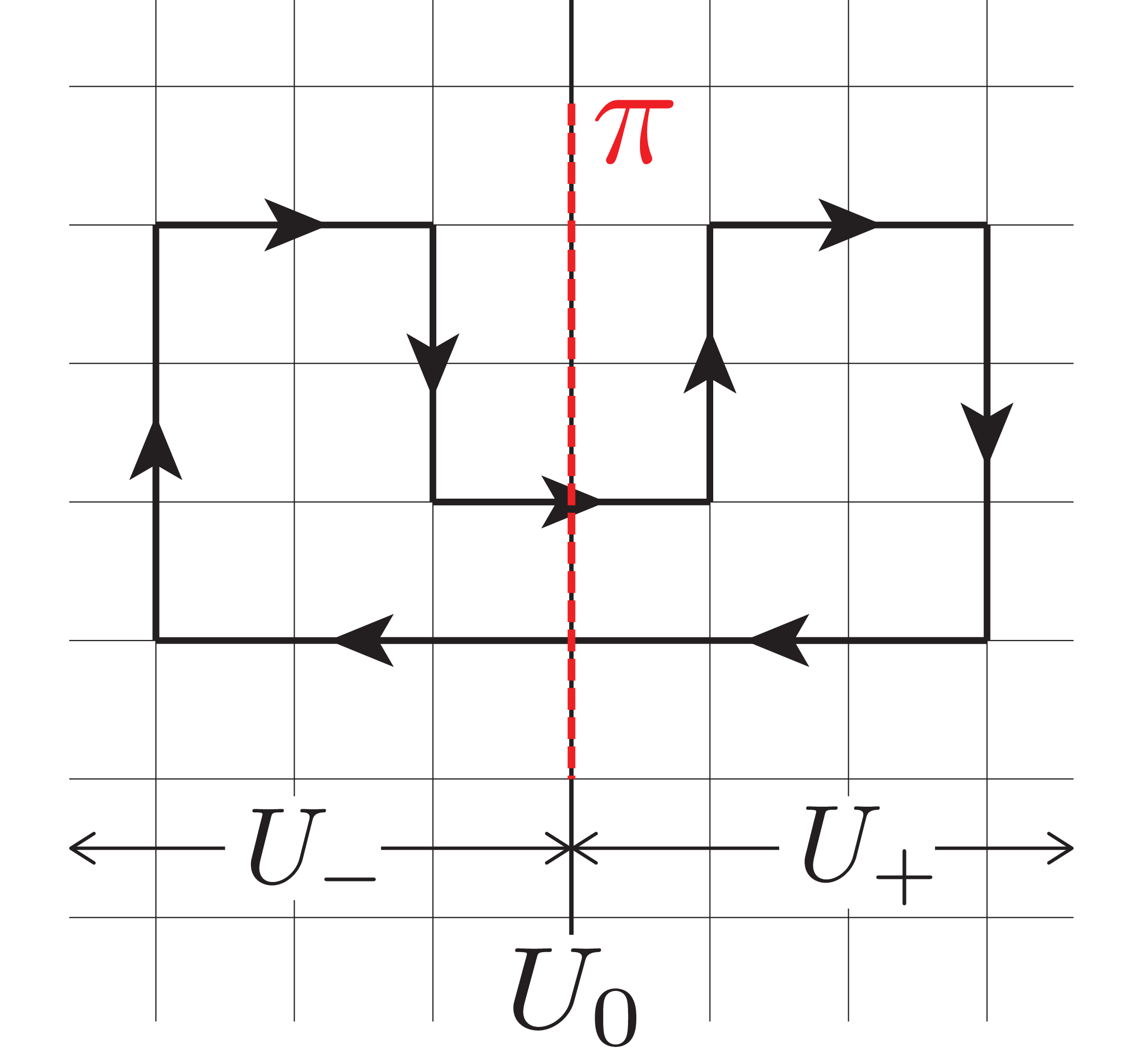}
\caption{Reflection positivity with mirror plane $\pi: x =0$.}
\label{fig:RP1}
\end{figure}

Consider the three-dimensional lattice with coordinates $(x,y,z)$. We divide the lattice into two regions by introducing a mirror plane $\pi$, which reflects points across this plane.
For simplicity, we begin by considering the mirror plane $\pi: x=0$, which divides the lattice into three parts as shown in  Figure~\ref{fig:RP1}:
\begin{itemize}
\item 
$U_+$: The region where $x>0$ (positive side of the mirror plane).
\item 
$U_-$: The region where $x<0$ (negative side of the mirror plane).
\item 
$U_0$: The region where $x=0$ (the mirror plane $\pi$).
\end{itemize}
The reflection operation $R$ maps points in $U_+$ to $U_-$ and vice versa. 
Specifically, for a point $(x,y,z)$, its reflection is given by:
\begin{equation}
R(x,y,z) = (-x,y,z) \,.
\end{equation}
The mirror plane remains invariant under this reflection.
We can decompose the path integral as 
\begin{align}
	\int {\cal D} U \,  e^{-S} = & \int {\cal D} U_+  {\cal D} U_-  {\cal D} U_0  \,  e^{-S_{U_+}-S_{U_+ \cup U_0}-S_{U_-}-S_{U_- \cup U_0} - S_{U_0}} \nonumber \\
	= & \int {\cal D} U_0  \, e^{-S_{U_0}} \Big|\int {\cal D} U_+  \, e^{-S_{U_+}-S_{U_+ \cup U_0}} \Big|^2\,.
\end{align}

To construct reflection-positive observables, we introduce the $\Theta$-operation, which combines reflection $R$ with complex conjugation. 
For a given path operator $P$, its reflected counterpart is defined as:
\begin{equation}
\Theta [ P ] = R[P]^\dagger \,,
\end{equation}
To form a Wilson loop, we consider a path $P(U_+, U_0)$ in the region $x\geq0$ with both endpoints located at the mirror plane (with $x=0$). Thus, the combined operator
$P(U_+, U_0) \Theta [P(U_+, U_0)]$ can form a Wilson loop, as shown in Figure~\ref{fig:RP1}.
Moreover, one has the following positivity condition for such a Wilson loop as
\begin{equation}
\int {\cal D} U \,  e^{-S}  \, P(U_+, U_0) \Theta [P(U_+, U_0)] = \, \int {\cal D} U_0  \, e^{-S_{U_0}} \Big|\int {\cal D} U_+  \, e^{-S_{U_+}-S_{U_+ \cup U_0}} \, P(U_+, U_0) \Big|^2 \geq 0 \,.
\end{equation}
Note that we have used the hermitian positivity for the $U_0$ integration.
Similar to the discussion in the Hermitian positivity, one can consider a linear combination of path operators $U = \sum_a c_a P_a$ and obtain a positive semi-definite matrix $A$:
\begin{equation}
\langle {\rm tr}(\Theta[U] U) \rangle \geq 0, \ \  \forall c_a \quad \Rightarrow \quad A  \succcurlyeq 0 , \ \  A_{ab}:= \langle {\rm tr}(\Theta[P_a] P_b) \rangle \,.
\end{equation}

As an example, consider the following three path operator
\begin{equation}
{\cal O}_+ = \left\{\WPa \,,\ \WPb{midarrow}\,,\ \WPb{midreversearrow} \right\} \big|_{x \geqslant 0}  \,.
\end{equation}
Their reflected counterparts under $\Theta$ are
\begin{equation}
{\cal O}_+^R = \left\{\WPa \,,\ \WPc{midarrow}\,,\ \WPc{midreversearrow} \right\} \big|_{x \leqslant 0} \,.
\end{equation}
Sewing the above two sets of operators, one can construct a reflection positive matrix as
\begin{align}
	{\renewcommand{\arraystretch}{1.5}
	\begin{blockarray}{cccc}
		& \WPa & \WPb{midarrow} & \WPb{midreversearrow} \\
		\begin{block}{c(ccc)}
			\WPa & 1 & \WLa{midarrow} & \WLa{midreversearrow} \\
			\WPc{midarrow} & \WLa{midarrow} & \WLb{midarrow} & \WLc{midarrow} \\
			\WPc{midreversearrow} & \WLa{midreversearrow} & \WLc{midreversearrow} & \WLb{midreversearrow} \\
		\end{block}
	\end{blockarray}}
	\succeq 0 \,,
\end{align}
which gives precisely the second positivity matrix given in \eqref{eq:positivityexample}.

\begin{figure}[t]
\centering
\subfigure[Mirror plane $\pi: x=a/2$.]{\includegraphics[width=0.3\linewidth]{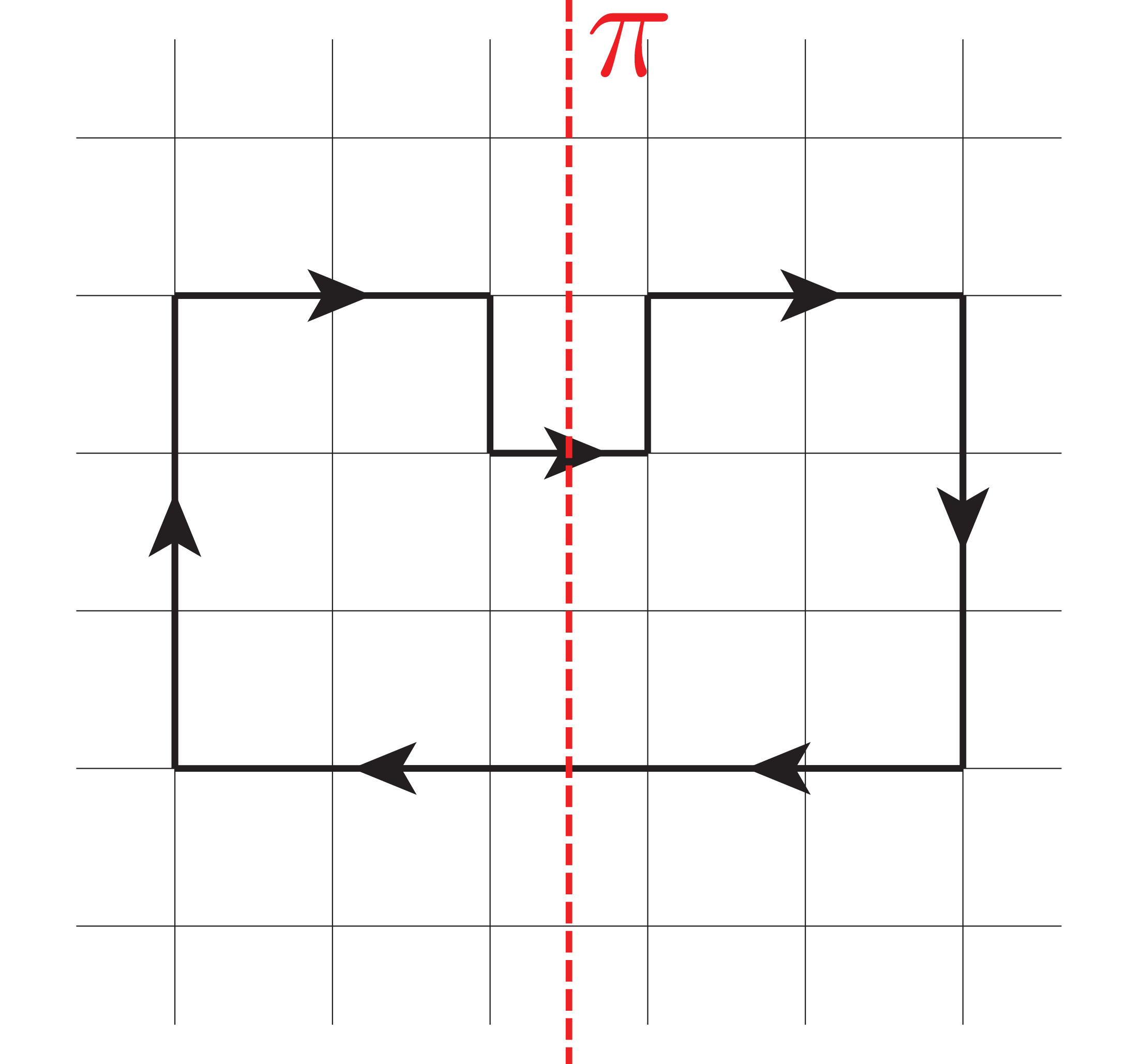}}
\qquad\qquad
\subfigure[Mirror plane $\pi: x=y$.]{\includegraphics[width=0.3\linewidth]{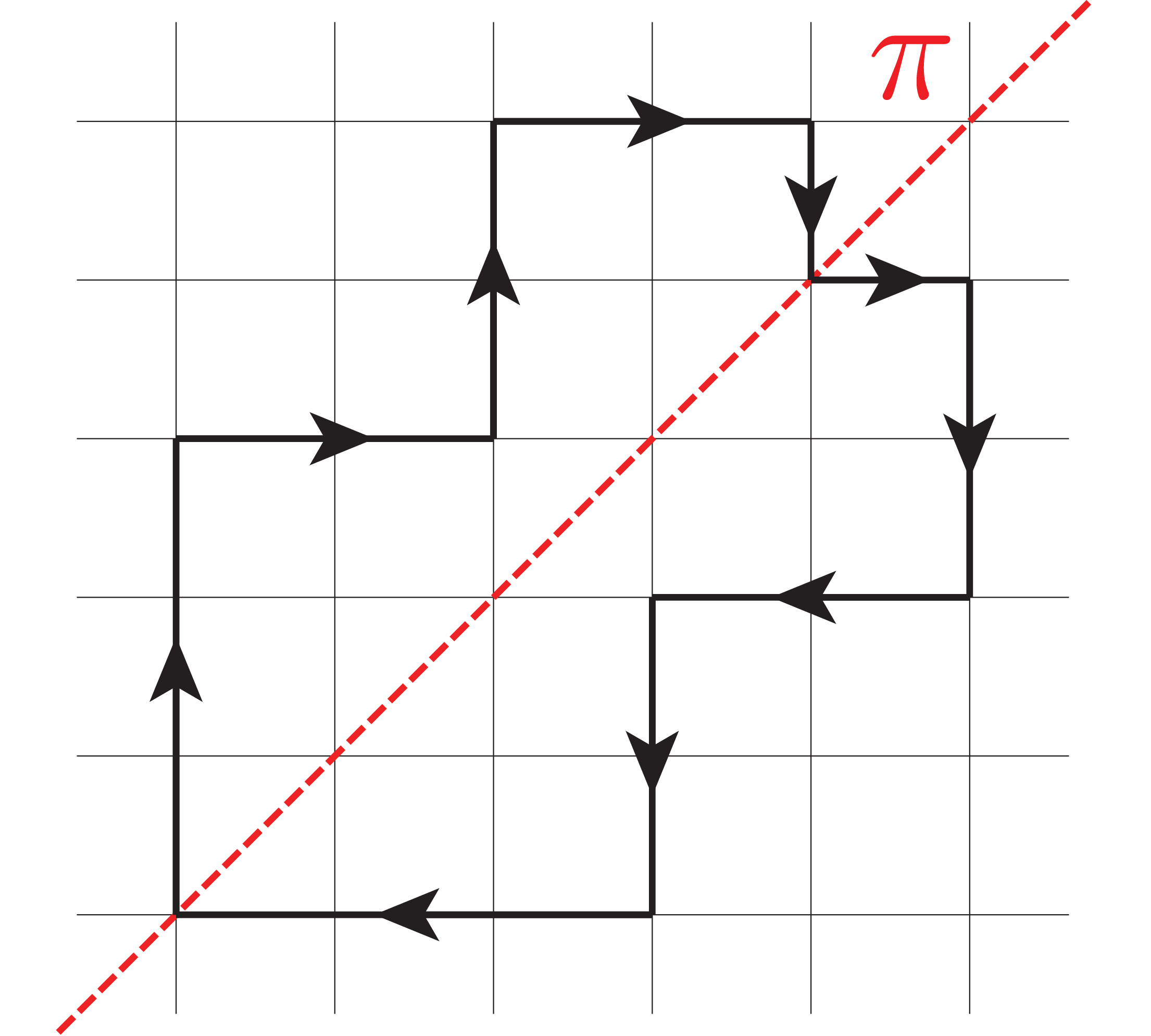}}
\caption{Other two types of reflection.}
\label{fig:RP2}
\end{figure}

Reflection positivity can be generalized in a similar way by considering different mirror planes, such as shown in Figure~\ref{fig:RP2}.

\subsubsection{Twist RP}
\label{sec:twist-reflection}

\begin{figure}[t]
\centering
\includegraphics[width=0.35\linewidth]{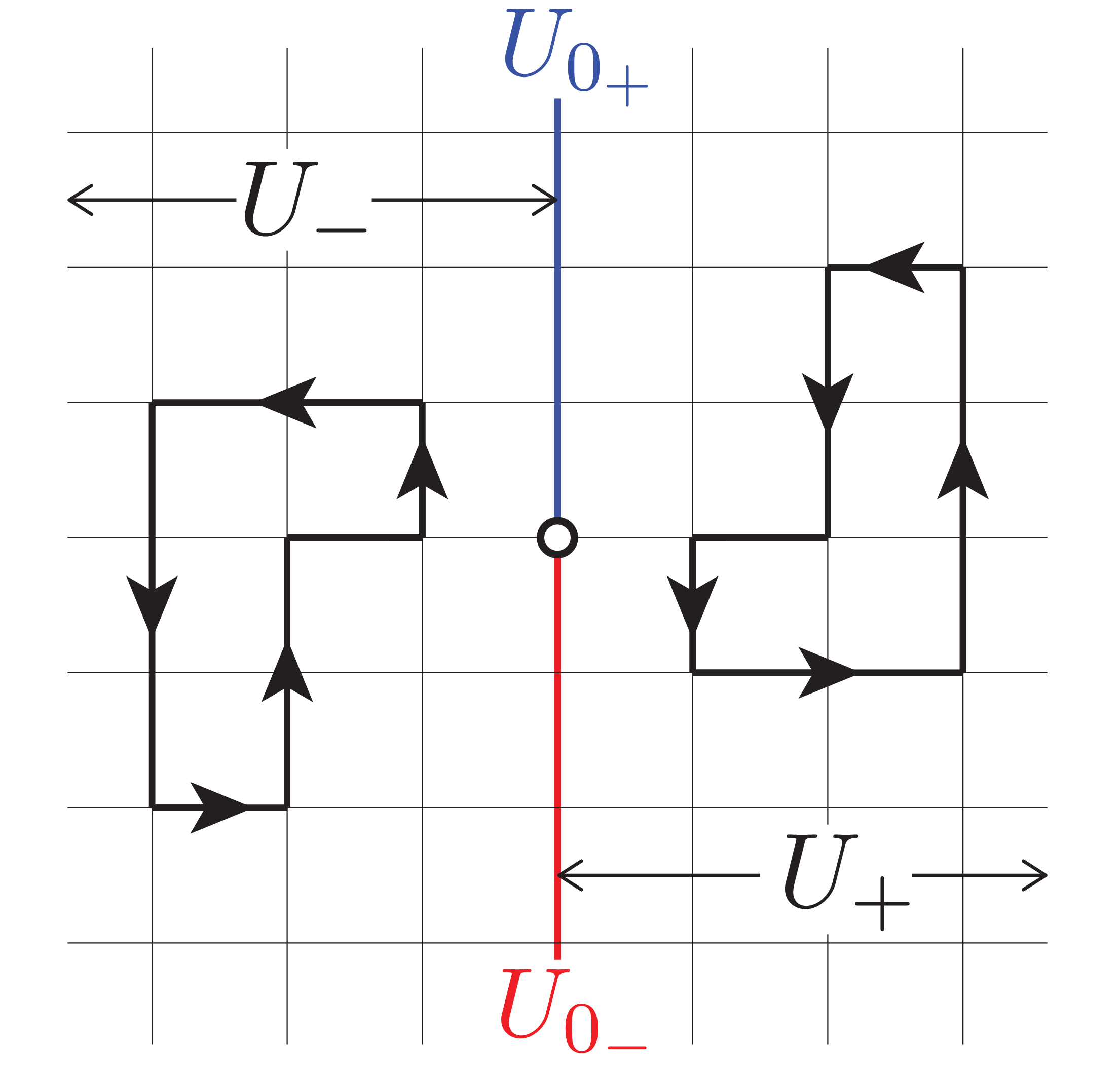}
\caption{Twist reflection.}
\label{fig:RP3}
\end{figure}

In the above reflection positivity condition, we have considered a mirror plane that maps one side of the lattice space to another side.
Here, we consider a new type of reflection operation: 
\begin{equation}
R_t(x,y,z) = (-x,-y,z) \,.
\end{equation}
We will call this a twist reflection, see Figure~\ref{fig:RP3}.\footnote{
This may be interpreted as combining the time reversal with a spatial parity transformation, which could be also referred to as PT-reflection. For simplicity and without loss of generality, we change parity only for one spacial direction $y$; one can certainly consider $\{x,y,z\}$ to $\{-x,-y,-z\}$ as well which can be analyzed in a similar way.}

Following Figure~\ref{fig:RP3}, we can divide the full lattice into four parts:
\begin{equation}
U_+: x>0 \,, \qquad U_-: x<0 \,, \qquad U_{0_+}: x=0, y>0 \,, \qquad U_{0_-}: x=0, y<0 \,.
\end{equation}
Under the reflection operation, $U_+$ maps to $U_-$, $U_{0_+}$ maps to $U_{0_-}$, and vice verse:
\begin{equation}
U_+  \leftrightarrow  U_- \,, \qquad U_{0_+} \leftrightarrow U_{0_-} \,.
\end{equation}
It is important to note that there is no reflection-invariant $U_0$ plane as in the previous case.

The path integral can be decomposed as 
\begin{equation}\label{eq:twistPTPI}
\int {\cal D} U \,  e^{-S} = \int {\cal D} U_+  {\cal D} U_-  {\cal D} U_{0_+} {\cal D} U_{0_-}  \,  e^{ - S_{U_0}-S_{U_+}-S_{U_-}-S_{U_+ \cup U_{0_+}}-S_{U_- \cup U_{0_-}} -S_{U_+ \cup U_{0_-}}-S_{U_- \cup U_{0_+}}} \,.
\end{equation}
However, unlike the previous case, due to the following mixing terms in the action  
\begin{equation}\label{eq:PTproblemterms}
-S_{U_+ \cup U_{0_+}}-S_{U_- \cup U_{0_-}} -S_{U_+ \cup U_{0_-}}-S_{U_- \cup U_{0_+}} \,,
\end{equation}
it is no longer possible to rewrite the path integral as a module square. 

Interestingly, in the two-dimensional case, we can prove that this is still possible. 
In 2D, one can choose a gauge condition such that
\begin{equation}\label{eq:2Dgaugechoice}
U_{0_+} = U_{0_-} =\mathds{1} \,.
\end{equation}
Note that $U_{0_\pm}$ form only a line in 2D, this gauge condition is easy to achieve as a sub-condition of gauge choices such as maximal-tree or temporal gauge, see \emph{e.g.}~\cite{Gattringer:2010zz}. However, in 3 (or higher) dimensions, $U_{0_+} = U_{0_-}$ will be 2 (or higher) dimensional lattice plane, where no such gauge condition exists.

Introducing the $\Theta_t$ operation
\begin{equation}
\Theta_t [ U(\{x,y,z\}) ] = U^\dagger (\{-x,-y,z\}) \,,
\end{equation}
and considering a path $P(U_+, U_0)$ where both endpoints are at the origin, 
we have positivity condition for the integrand $P(U_+, U_0) \Theta_t [P(U_+, U_0)]$ as
\begin{equation}\label{eq:2DRPequation}
\int {\cal D} U \,  e^{-S}  \, P(U_+) \Theta_t [P(U_+)] 
= \Big|\int {\cal D} U_+  \, e^{-S_{U_+}} \, P(U_+) \Big|^2 \geq 0 \,.
\end{equation}

As an example, consider the following three path operator
\begin{equation}
{\cal O}_+ = \left\{\WPa \,,\ \WPb{midarrow}\,,\ \WPb{midreversearrow} \right\}  \,.
\end{equation}
Their twist-reflected counterparts under $\Theta_t$ are
\begin{equation}
	{\cal O}_+^{R_t} = \left\{ \WPa \,,\ \WPj{midarrow} \,, \ \WPj{midreversearrow} \right\} \,.
\end{equation}
Sewing the above two sets of operators, we construct a twist-reflection positive matrix as
\begin{align}
	{\renewcommand{\arraystretch}{2}
	\begin{blockarray}{cccc}
		& \WPa & \WPb{midarrow} & \WPb{midreversearrow} \\
		\begin{block}{c(ccc)}
			\WPa & 1 & \WLa{midarrow} & \WLa{midreversearrow} \\
			\WPj{midarrow} & \WLa{midreversearrow} & \WLl{midarrow} & \WLm{midreversearrow} \\
			\WPj{midreversearrow} & \WLa{midarrow} & \WLm{midarrow} & \WLl{midreversearrow} \\
		\end{block}
	\end{blockarray}} \succeq 0 \,.
\end{align}
As we will see in Section~\ref{sec:results2D}, this condition imposes powerful new constraints in 2D YM.

\begin{figure}[t]
\centering
\includegraphics[width=0.35\linewidth]{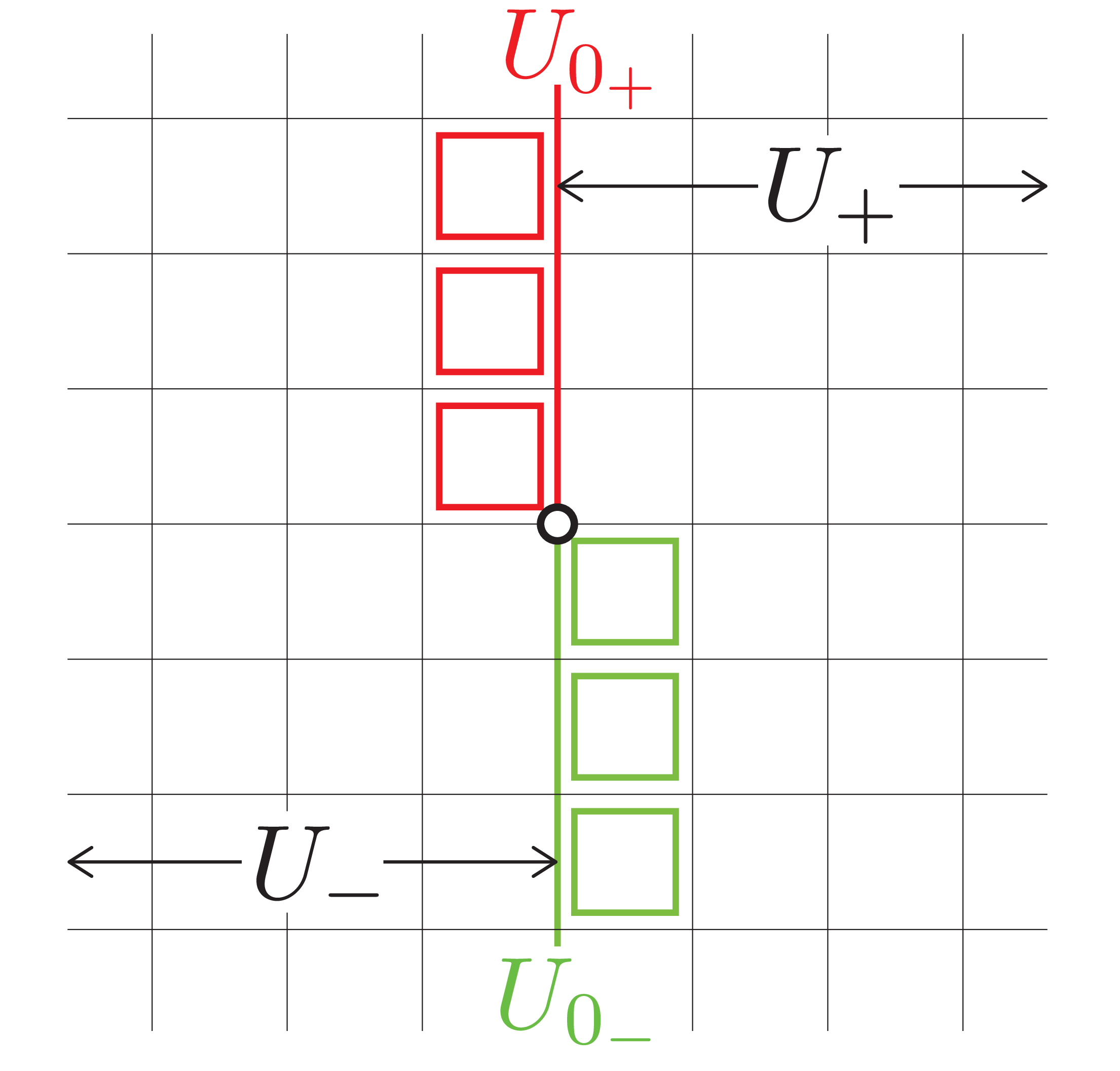}
\caption{Twist reflection. The red and green plaquettes are those causing troubles in high-dimensional cases.}
\label{fig:RP4}
\end{figure}

Unfortunately, for three or higher dimensions, the twist-reflection positivity condition is no longer true: there is no gauge choice  \eqref{eq:2Dgaugechoice}, and the path integral \eqref{eq:twistPTPI} can no longer be put in the module square form of \eqref{eq:2DRPequation}. 
Indeed, applying this condition in the bootstrap construction in 3D and 4D yields inconsistency. 

We would like to conclude with a comment on the continuous limit. One can note that the problematic terms in \eqref{eq:PTproblemterms} are associated with the plaquettes that are attached to the $x=0$ plane, namely, the single layer of plaquettes that is at $x=0$, as shown in Figure~\ref{fig:RP4}. If we take the lattice spacing $a\rightarrow0$, this layer of plaquettes may be approximately on the $x=0$ plane. Then we have
\begin{equation}
\int {\cal D} U \,  e^{-S} \simeq \int {\cal D} U_+  {\cal D} U_-  {\cal D} U_{0_+} {\cal D} U_{0_-}  \,  e^{ - S_{U_0}-S_{U_+}-S_{U_-}-S_{U_{0_+}}-S_{U_{0_-}} } \,,
\end{equation}
which is factorizable as module square. With the integrand $P(U_+, U_0) \Theta_t [P(U_+, U_0)]$, we have
\begin{align}
& \int {\cal D} U \,  e^{-S}  \, P(U_+,U_{0_+}) \Theta_t [P(U_+,U_{0_+})] \\
\simeq & \, \Big|\int {\cal D} U_+  {\cal D} U_{0_+}  \, e^{-S_{U_+}-S_{U_{0_+}}} \, P(U_+,U_{0_+}) \Big|^2 \geq 0 \,. \nonumber
\end{align}
This argument is not mathematically exact, but it seems to imply that in the continuum limit, the twist reflection could play a useful role. 
Moreover, for the lattice theory, we may expect that, if the loop operator $P(U_+,U_{0_+})$ has a scale much larger than $|a|$, or is far away from $x=0$, the positivity might be `approximately' true.

\section{Loop equations}
\label{sec:loopequation}

Wilson loops satisfy Schwinger-Dyson (SD) equations which are the quantum version of the Euler-Lagrangian equation, which is often called  Makeenko-Migdal equations \cite{Makeenko:1979pb, Migdal:1983qrz}. 
In this section, we give a detailed derivation of the equations for Wilson loop operators in lattice YM theory, following closely the notation in \cite{Anderson:2016rcw}.
We start with
\begin{equation}
\int {\cal D} U \, \delta_{\epsilon(\mu)} \Big[ e^{-S} \, W_{x}^{ab}({\Bmu}, C_\mu) \Big] = 0\,,
\end{equation}
where $W_{x}^{ab}(\boldsymbol{\mu}, C_\mu)$ is a \emph{Wilson line} with both endpoints at position ${x}$, ${\Bmu} := U_\mu(x)$ is the first link and $C_\mu$ is the remaining path. 
Here $a,b$ are fundamental color indices with $a,b =1,\ldots, N$.
The variation $\delta_\epsilon$ is with respect to the link $U_\mu$, and there are two contributions 
\begin{equation}
\label{eq:sdedelta}
- \langle W_{x}^{ab} \, \delta_{\epsilon} S \rangle + \langle \delta_{\epsilon}  W_{x}^{ab} \rangle = 0 \,.
\end{equation}

An infinitesimal variation of the link $U_\mu$ is
\begin{equation}
U'_\mu = (1+ \im \epsilon) U_\mu \,, \qquad 
U'^\dagger_\mu = U^\dagger_\mu (1-\im \epsilon) \,,
\end{equation}
where $\epsilon$ is Hermitian and traceless
\begin{equation}
\epsilon^\dagger = \epsilon \,, \qquad {\rm tr}(\epsilon)=0 \,,
\end{equation}
and one has
\begin{equation}
\delta_\epsilon U_{\mu}^{ab} = \im \epsilon^{ac} U_{\mu}^{cb} \,, \qquad 
\delta_\epsilon U^{\dagger, ab}_{\mu} = - \im U^{\dagger, ac}_{\mu} \epsilon^{cb} \,.
\end{equation}
The variation of the action is
\begin{equation}\label{eq:deltaS}
\delta_{\epsilon} S = -{N\over2\lambda} \sum_{\pm \nu \neq \mu} 
\Big[ \im \epsilon^{cd} W_{x}^{dc}(\mu \nu \bar \mu \bar \nu) - \im \epsilon^{cd} W_{x}^{dc}(\nu \mu \bar \nu \bar \mu) \Big] \,,
\end{equation}
where we abbreviate $U_\mu, U_\nu$ as $\mu,\nu$, etc, 
and for the second term, we have used the relation
\begin{equation}
(U^\dagger_{\bar \nu} U^\dagger_{\bar \mu} U^\dagger_\nu U^\dagger_{\mu}(x))^{dc} \epsilon^{cd} 
= \epsilon^{cd} (U_{\nu}(x) U_\mu U_{\bar \nu} U_{\bar\mu})^{dc} \,.
\end{equation}
The variation of $W_{x}^{ab}({\Bmu}, C_\mu)$ is
\begin{equation}\label{eq:deltaW}
\delta_{\epsilon} W_{x}^{ab}({\Bmu}, C_\mu) = \im \epsilon^{ac} W_{x}^{cb}({\Bmu}, C_\mu) 
+ \sum_{j_+=1}^{n_+} W_{{x}, {x}_+}^{ac} \im \epsilon^{cd} W_{{x}_+, {x}}^{db} 
- \sum_{j_-=1}^{n_-} W_{{x}, {x}_-}^{ac} \im \epsilon^{cd} W_{{x}_-, {x}}^{db} \,.
\end{equation}
The second and third terms are the contribution of links that overlap with $\Bmu$. There are two types of such overlapping links:
$n_+$ counts the number of links that not only overlap with $\Bmu$ but also have the same direction, with $x_+$ as the starting point, while $n_-$ counts the links that overlap with $\Bmu$ but have opposite direction, with $x_-$ as the ending point.
They can be illustrated as following figures:
\begin{equation}
\begin{gathered} {\includegraphics[height=2.5cm]{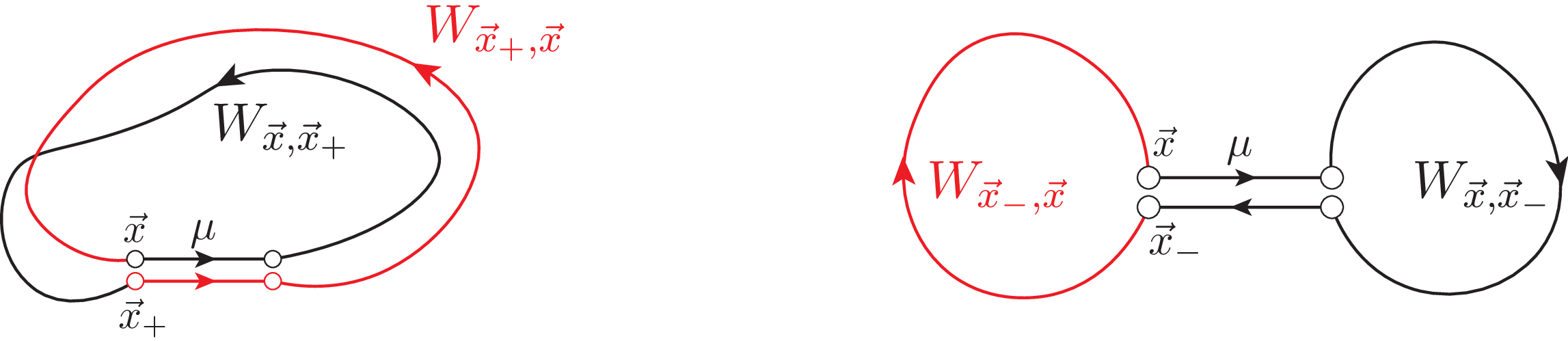} } \end{gathered} \,.
\end{equation} 
Inserting \eqref{eq:deltaS} and \eqref{eq:deltaW} into \eqref{eq:sdedelta}, we obtain the equation 
\begin{align}
& \bigg\langle {N\over2\lambda} \sum_{\pm \nu \neq \mu} 
\Big[ \im \epsilon^{cd} W_{x}^{dc}(\mu \nu \bar \mu \bar \nu) - \im \epsilon^{cd} W_{x}^{dc}(\nu \mu \bar \nu \bar \mu) \Big] W_{x}^{ab} 
\label{eq:SDE1} \\
& \quad + \im \epsilon^{ad} W_{x}^{db}(\Bmu, C_\mu) 
+ \sum_{j_+=1}^{n_+} W_{{x}, {x}_+}^{ac} \im \epsilon^{cd} W_{{x}_+, {x}}^{db} 
- \sum_{j_-=1}^{n_-} W_{{x}, {x}_-}^{ac} \im \epsilon^{cd} W_{{x}_-, {x}}^{db}  \bigg \rangle = 0  \,. \notag
\end{align}

Next, we apply the relation 
\begin{equation}
\label{eq:matrixrelation1}
\epsilon^{cd} M^{dc} = 0 \quad \Rightarrow \quad M^{dc} = \delta^{cd} {1\over N} M^{aa} \,,
\end{equation}
where the matrix $\epsilon$ is Hermitian and traceless while the matrix $M$ is a general matrix satisfying the LHS equation. 
Applying \eqref{eq:matrixrelation1} to \eqref{eq:SDE1}, we get
\begin{align}
& \bigg\langle {N\over2\lambda} \sum_{\pm \nu \neq \mu} 
\Big[  W_{x}^{dc}(\mu \nu \bar \mu \bar \nu) -W_{x}^{dc}(\nu \mu \bar \nu \bar \mu) \Big] W_{x}^{ab} (\Bmu, C_\mu)
\label{eq:SDE2} \\
& \quad + \delta^{ac} W_{x}^{db}(\Bmu, C_\mu) 
+ \sum_{j_+=1}^{n_+} W_{{x}, {x}_+}^{ac} W_{{x}_+, {x}}^{db} 
- \sum_{j_-=1}^{n_-} W_{{x}, {x}_-}^{ac} W_{{x}_-, {x}}^{db} \bigg\rangle  \notag \\
= &\ \bigg\langle \delta^{cd} \Bigg\{ {
1\over2\lambda} \sum_{\pm \nu \neq \mu} 
\Big[  W_{x}(\mu \nu \bar \mu \bar \nu) -W_{x}(\nu \mu \bar \nu \bar \mu) \Big] W_{x}^{ab} (\Bmu, C_\mu)
 \notag\\
& \qquad + {1\over N} \bigg[ W_{x}^{ab}(\Bmu, C_\mu) 
+ \sum_{j_+=1}^{n_+} W_{x}^{ab}(\Bmu, C_\mu) 
- \sum_{j_-=1}^{n_-} W_{x}^{ab}(\Bmu, C_\mu) \bigg]
\Bigg\}  \bigg\rangle \,. \notag
\end{align}
Multiplying both sides by $\delta^{ac} \delta^{bd}$ and performing contraction of color indices, we obtain the equation for Wilson loops:
\begin{align}
& \bigg\langle {N\over2\lambda} \sum_{\pm \nu \neq \mu} 
\Big[  W(\mu \nu \bar \mu \bar \nu C) -W(\nu \mu \bar \nu \bar \mu C) \Big] + N \, W(C) 
\label{eq:SDE3} \\
& \quad 
+ \sum_{j_+=1}^{n_+} W_x(C_{{x}, {x}_+}) W_x(C_{{x}_+, {x}}) 
- \sum_{j_-=1}^{n_-} W_x(C_{{x}, {x}_-}) W_x(C_{{x}_-, {x}}) \bigg\rangle  \notag \\
= &\ \bigg\langle {1\over2\lambda} \sum_{\pm \nu \neq \mu} 
\Big[  W_{x}(\mu \nu \bar \mu \bar \nu) -W_{x}(\nu \mu \bar \nu \bar \mu) \Big] W_{x}(C)+ {1\over N} (1+n_+-n_-) W(C) \bigg\rangle \,. \notag
\end{align}
Note that the product of two loops gives double-trace operators. We stress that for multiple-trace loops, it is important to keep the information of the relative position of the loops. Here, we indicate the starting position of the loop by the subscript $x$ in $W_x(C)$.

We can illustrate the various terms in \eqref{eq:SDE3} by the following figures. The variation of the action generates the first term on the LHS and the first term on the RHS:
\begin{equation}
\begin{gathered} {\includegraphics[height=3.cm]{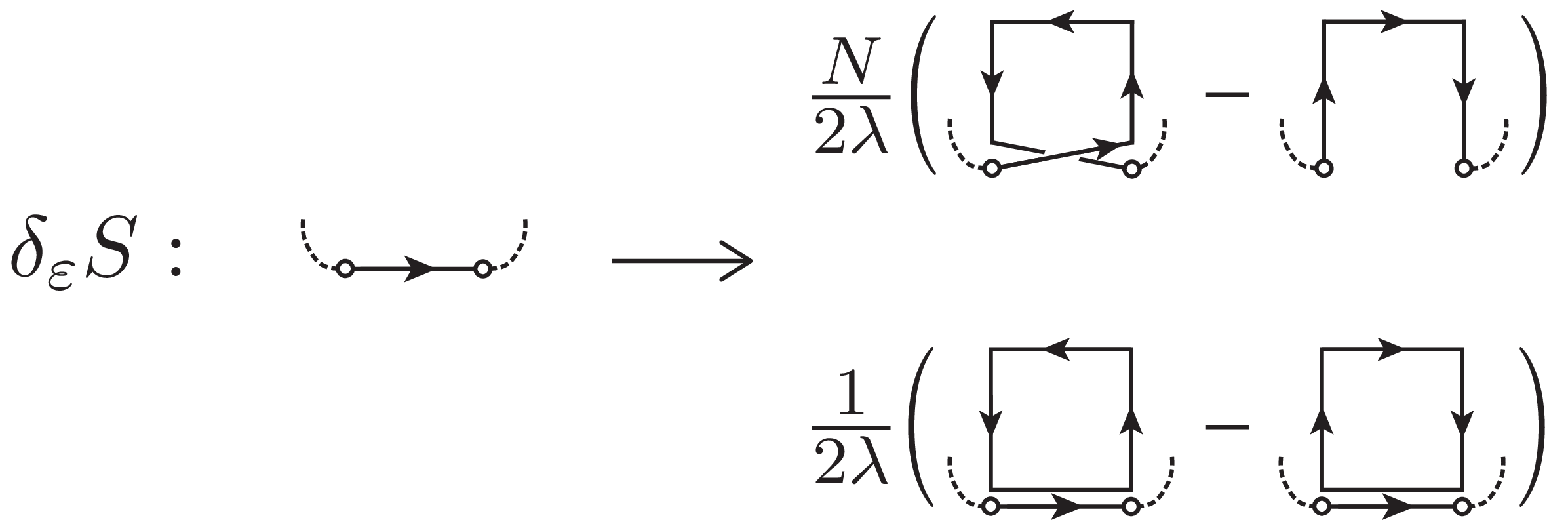} } \end{gathered} \,.
\end{equation} 
The double-trace terms in the second line of \eqref{eq:SDE3} are the contribution of overlapping links:
\begin{equation}
\begin{gathered} {\includegraphics[height=3.1cm]{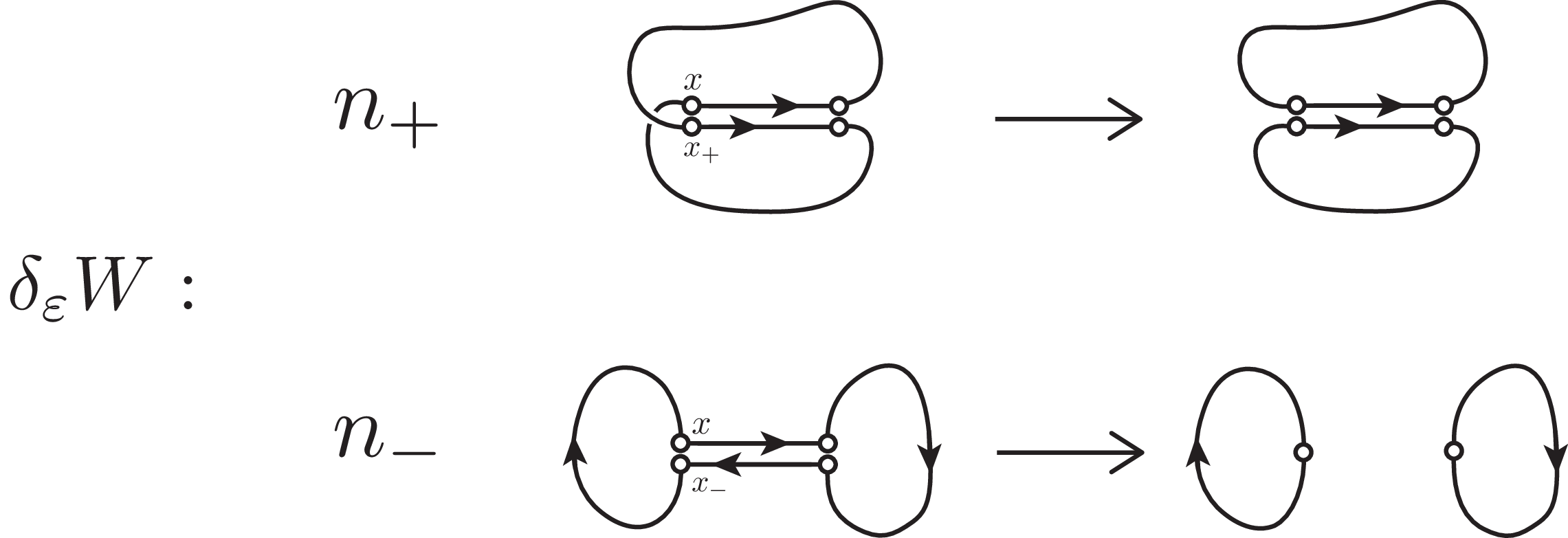} } \end{gathered} \,.
\end{equation} 

\paragraph{Normalization.}
It is convenient to work with the normalized Wilson loop $w(C)$ as
\begin{equation}\label{eq:Wnormalization}
w(C) = {1\over N} W(C) \,, \qquad w({\rm tr}(\mathds{1})) = 1 \,.
\end{equation}
In particular, the identity loop is normalized to be $1$.
The loop equation \eqref{eq:SDE3} can be rewritten as 
\begin{align}
& {N^2} \sum_{\pm \nu \neq \mu} 
\Big[  \langle w(\mu \nu \bar \mu \bar \nu C) \rangle - \langle w(\nu \mu \bar \nu \bar \mu C) \rangle \Big] \label{eq:SDE3c} \\
& + 2 \lambda \Big[ N^2 - (1+n_+-n_-) \Big] \langle w(C) \rangle \notag \\
& + 2 \lambda N^2 \bigg[ \sum_{j_+=1}^{n_+} \big\langle  w_x(C_{{x}, {x}_+}) w_x(C_{{x}_+, {x}}) \big\rangle
- \sum_{j_-=1}^{n_-} \big\langle w_x(C_{{x}, {x}_-}) w_x(C_{{x}_-, {x}}) \big\rangle \bigg]  \notag \\
& - N^2 \sum_{\pm \nu \neq \mu} 
\Big[ \big\langle w_{x}(\mu \nu \bar \mu \bar \nu)w_{x}(C) \big\rangle - \big\langle w_{x}(\nu \mu \bar \nu \bar \mu)w_{x}(C) \big\rangle \Big] =0 \,. \notag
\end{align}
In practice, we can combine the original $\Bmu$ link into the sum in $n_+$, and \eqref{eq:SDE3c} can be given as
\begin{center}
\begin{tcolorbox}[colback=white, colframe=black, width=.96\textwidth]
\begin{equation} \label{eq:SDE3d}  
\begin{aligned}
& \sum_{\pm \nu \neq \mu} 
\Big[  \langle w(\mu \nu \bar \mu \bar \nu C) \rangle - \langle w(\nu \mu \bar \nu \bar \mu C) \rangle \Big]  - {2\lambda \over N^2} ({{\check n}}_+-n_-) \langle w(C) \rangle \\
& + 2 \lambda \bigg[ \sum_{j_+=1}^{{{\check n}}_+} \big\langle  w_x(C_{{x}, {x}_+}) w_x(C_{{x}_+, {x}}) \big\rangle
- \sum_{j_-=1}^{n_-} \big\langle w_x(C_{{x}, {x}_-}) w_x(C_{{x}_-, {x}}) \big\rangle \bigg] \\
& - \sum_{\pm \nu \neq \mu} 
\Big[ \big\langle w_{x}(\mu \nu \bar \mu \bar \nu)w_{x}(C) \big\rangle - \big\langle w_{x}(\nu \mu \bar \nu \bar \mu)w_{x}(C) \big\rangle \Big] =0 \,. 
\end{aligned}
\end{equation}
\end{tcolorbox}
\end{center}
Note that ${{\check n}}_+$ counts all edges that are at the same position and have the same direction as $\mu$ (\emph{including} $\mu$ itself), while $n_-$ counts all edges that are at the same position but have opposite direction as $\mu$.

Let us apply \eqref{eq:SDE3d} for two examples in 2D YM. The simplest example of SDE is for the plaquette:
\begin{align}\label{eq:sdeexN1}
	\WLar \ \overset{\rm 2D}{\underset{}{\Longrightarrow}}  \ & \left( \, \WLc{midarrow} -\WLb{midarrow} +\WLd{midarrow} -1 \, \right) +2\lambda \left( 1 -\frac{1}{N^2}\right) \WLa{midarrow}  \\ 
	& -\left( \,\WLdaa{midarrow}{midreversearrow} -\WLdaa{midarrow}{midarrow} +\WLdab{midarrow}{midarrow} -\WLdab{midarrow}{midreversearrow} \,\right) = 0 \,. \notag
\end{align}
Another example that involves a pair of overlapping edges (${{\check n}}_+ = 2$) is
\begin{align}\label{eq:sdeexN2}
	\WLcr \ \overset{\rm 2D}{\underset{}{\Longrightarrow}} \ & \left( \, 
	\WLe{midarrow} -\WLa{midarrow} +\WLf{midarrow} -\WLa{midreversearrow} \, \right) +2\lambda \left( 1 -\frac{2}{N^2}\right) \WLc{midarrow}   \\ 
	& -\left( \, \WLdde{midarrow}{midarrow} -\WLdde{midarrow}{midreversearrow} +\WLddf{midarrow}{midarrow} -\WLddf{midarrow}{midreversearrow} \,\right) +2\lambda \left( \WLdaa{midarrow}{midreversearrow}\right) = 0 \,. \notag
\end{align}

Below we consider three special cases, the $N=\infty$, $N=2$, and $N=3$ cases.
The equations for the first two cases can be reduced such that they contain only single-trace operators. 
For the SU(3) case, one also needs to consider new equations for double-trace operators.

\subsection{Large $N$ theory}

In the large $N$ limit, we simplify the equation by ignoring all ${\cal O}(1/N)$ corrections.
Applying the large $N$ factorization property \cite{Migdal:1980au}
\begin{equation}
\langle w(C_{x x'}) w(C_{x' x}) \rangle = \langle w(C_{x x'}) \rangle \langle w(C_{x' x}) \rangle + {\cal O}(1/N^2)\,,
\end{equation}
and keep only the terms at leading $N$ order in \eqref{eq:SDE3d}, we have the large $N$ SDE
\begin{align}
& \sum_{\pm \nu \neq \mu} 
\Big[  \langle w(\mu \nu \bar \mu \bar \nu C) \rangle - \langle w(\nu \mu \bar \nu \bar \mu C) \rangle \Big]  \label{eq:SDElargeN0} \\
& + 2 \lambda \bigg[ \sum_{j_+=1}^{{{\check n}}_+} \big\langle  w(C_{{x}, {x}_+}) \rangle \langle w(C_{{x}_+, {x}}) \big\rangle
- \sum_{j_-=1}^{n_-} \big\langle w(C_{{x}, {x}_-}) \rangle \langle w(C_{{x}_-, {x}}) \big\rangle \bigg]  \notag \\
& - \sum_{\pm \nu \neq \mu} 
\Big[ \big\langle w(\mu \nu \bar \mu \bar \nu) \rangle \langle w(C) \big\rangle - \big\langle w(\nu \mu \bar \nu \bar \mu) \rangle \langle w(C) \big\rangle \Big] =0 \,. \notag
\end{align}
Note that the last line cancels so we have the final form as
\begin{center}
\begin{tcolorbox}[colback=white, colframe=black, width=.99\textwidth]
\begin{equation}\label{eq:SDElargeN}
\begin{aligned}
& \sum_{\pm \nu \neq \mu} 
\Big[  \langle w(\mu \nu \bar \mu \bar \nu C) \rangle - \langle w(\nu \mu \bar \nu \bar \mu C) \rangle \Big]   \\
& + 2 \lambda \bigg[ \sum_{j_+=1}^{{{\check n}}_+} \big\langle  w(C_{{x}, {x}_+}) \rangle \langle w(C_{{x}_+, {x}}) \big\rangle
- \sum_{j_-=1}^{n_-} \big\langle w(C_{{x}, {x}_-}) \rangle \langle w(C_{{x}_-, {x}}) \big\rangle \bigg] =0  \,. 
\end{aligned}
\end{equation}
\end{tcolorbox}
\end{center}
Recall that ${{\check n}}_+$ counts all edges that are at the same position and have the same direction as $\mu$, including $\mu$ itself.

In large $N$ theory, the previous two examples of \eqref{eq:sdeexN1} and \eqref{eq:sdeexN2} are simplified as
\begin{align}
	\WLar \ \  \overset{\rm 2D}{\underset{\textrm{large-N}}{\Longrightarrow}} \quad & 2\lambda \, \WLa{midarrow} +\WLc{midarrow} -\WLb{midarrow} +\WLd{midarrow} -1 = 0 \,, \\
	\WLcr \ \ \overset{\rm 2D}{\underset{\textrm{large-N}}{\Longrightarrow}} \quad &  
	2\, \WLe{midarrow} - 2\, \WLa{midarrow}  +2\lambda \left[ \WLc{midarrow} + \Big(\WLa{midarrow}\Big)^2 \right] = 0   \,. 
\end{align}

\subsection{SU(2) theory}

As mentioned before, for the SU(2) theory, all loops can be reduced to single-trace loops \cite{Kazakov:2024ool, Gambini:1990dt}. %
For the double-trace loop $C = C_{1,x} \otimes C_{2,x}$, we have (see Appendix~\ref{app:su3identities})
\begin{equation}\label{eq:WccSU2}
2 w(C_{1,x}) w(C_{2,x}) = w(C_{1,x}\, C_{2,x}) + w(C_{1,x}\, \bar C_{2,x}) \,.
\end{equation}
Using \eqref{eq:WccSU2}, we can reduce all double-trace loops in \eqref{eq:SDE3d} into single-trace loops. 
The final SDE for SU(2) theory is given as
\begin{center}
\begin{tcolorbox}[colback=white, colframe=black, width=.99\textwidth]
\begin{equation} \label{eq:SDESU2} 
\begin{aligned}
& \sum_{\pm\nu \neq \mu} \big[ w(\mu\nu\bar\mu\bar\nu C) - w(\nu\mu\bar\nu\bar\mu C) \big] \\
& +   \lambda \bigg[ {1 \over2} ({{\check n}}_+ - n_-) w(C) + \sum_{j_+=1}^{{{\check n}}_+} w(C_{x x_+} \bar C_{x_+ x}) - \sum_{j_-=1}^{n_-} w(C_{x x_-} \bar C_{x_- x}) \bigg] =0 . 
\end{aligned}
\end{equation}
\end{tcolorbox}
\end{center}
Note that ${{\check n}}_+$ counts all edges that are at the same position and have the same the direction as $\mu$, including $\mu$ itself.

In 2D SU(2) theory, the previous two examples of \eqref{eq:sdeexN1} and \eqref{eq:sdeexN2} are simplified as
\begin{align}
	\WLar \ \  \overset{\rm 2D}{\underset{\rm SU(2)}{\Longrightarrow}} \quad & \frac{3}{2}\lambda \, \WLa{midarrow} +\WLc{midarrow} -\WLb{midarrow} +\WLd{midarrow} -1 = 0 \,, 
	\label{eq:sdeexsu21} \\
	\WLcr \ \ \overset{\rm 2D}{\underset{\textrm{SU(2)}}{\Longrightarrow}} \quad &  
	2\, \WLe{midarrow} - 2\, \WLa{midarrow} + \lambda \left( 2\, \WLc{midarrow} + \WLb{midarrow} \right) = 0 \,. \notag
\end{align}
The first equation \eqref{eq:sdeexsu21} is precisely the one we give before in \eqref{eq:sdeexample}.

\subsection{SU(3) theory}

In the case of SU(3) theory, double-trace operators cannot be reduced to single-trace operators and should therefore be treated as independent entities.
In this context, we can set $N=3$ in \eqref{eq:SDE3d} to obtain the SU(3) equation:
\begin{center}
\begin{tcolorbox}[colback=white, colframe=black, width=.96\textwidth]
\begin{equation}\label{eq:SDEsu3} 
\begin{aligned}
& \sum_{\pm \nu \neq \mu} 
\Big[  \langle w(\mu \nu \bar \mu \bar \nu C) \rangle - \langle w(\nu \mu \bar \nu \bar \mu C) \rangle \Big]  - {2\over9} \lambda ({{\check n}}_+-n_-) \langle w(C) \rangle \\
& + 2 \lambda \bigg[ \sum_{j_+=1}^{{{\check n}}_+} \big\langle  w_x(C_{{x}, {x}_+}) w_x(C_{{x}_+, {x}}) \big\rangle
- \sum_{j_-=1}^{n_-} \big\langle w_x(C_{{x}, {x}_-}) w_x(C_{{x}_-, {x}}) \big\rangle \bigg]   \\
& - \sum_{\pm \nu \neq \mu} 
\Big[ \big\langle w_{x}(\mu \nu \bar \mu \bar \nu)w_{x}(C) \big\rangle - \big\langle w_{x}(\nu \mu \bar \nu \bar \mu)w_{x}(C) \big\rangle \Big] =0 \,. \end{aligned}
\end{equation}
\end{tcolorbox}
\end{center}
Note that ${{\check n}}_+$ counts all links that are positioned the same and have the same direction as $\mu$, including $\mu$ itself.

The above discussion applies to the SDE for single-trace operators. 
However, in SU(3), one must also consider the SDE for double-trace operators. 
Let us consider the SDE for a double-trace operator $w_x(C_1) w_y(C_2)$. 
We vary with respect to the edge $U_\mu(x)$ that is in $w_x(C_1)$.
The derivation follows the same steps as in \eqref{eq:SDE3d} but with a key new element: 
one must consider the edges in $C_2$ that overlap with $U_\mu(x)$ in $C_1$.

The resulting SDE for the double-trace operator is:
\begin{center}
\begin{tcolorbox}[colback=white, colframe=black, width=.99\textwidth]
\begin{align}
& \sum_{\pm \nu \neq \mu} 
\Big[  \langle w_x(\mu \nu \bar \mu \bar \nu C_1) w_y(C_2) \rangle - \langle w_x(\nu \mu \bar \nu \bar \mu C_1) w_y(C_2) \rangle \Big]   \notag \\
& - {2\lambda\over9}  ({{\check n}}_{1,+}-n_{1,-} + n_{2,+}-n_{2,-}) \langle w_x(C_1) w_y(C_2) \rangle \notag \\
&  + {2 \lambda \over9} \bigg[ \sum_{j_+=1}^{n_{2,+}} \big\langle w(C_{1,x} C_{2,x_+})  \big\rangle
- \sum_{j_-=1}^{n_{2,-}} \big\langle w(C_{1,x} C_{2,x_-}) \big\rangle \bigg]  \label{eq:SDE3dDT2} \\
& + 2 \lambda \bigg[ \sum_{j_+=1}^{{{\check n}}_{1,+}} \big\langle  w_x(C_{1,{x}, {x}_+}) w_x(C_{1,{x}_+, {x}}) w_y(C_2) \big\rangle
- \sum_{j_-=1}^{n_{1,-}} \big\langle w_x(C_{1,{x}, {x}_-}) w_x(C_{1,{x}_-, {x}}) w_y(C_2) \big\rangle \bigg]  \notag \\
& - \sum_{\pm \nu \neq \mu} 
\Big[ \big\langle w_x(\mu \nu \bar \mu \bar \nu)w_x(C_1) w_y(C_2) \big\rangle - \big\langle w_x(\nu \mu \bar \nu \bar \mu)w_x(C_1) w_y(C_2) \big\rangle \Big] =0 \,. \notag
\end{align}
\end{tcolorbox}
\end{center}
In this equation, ${{\check n}}_{1,+}$ counts edges in $C_1$ that match the position and direction of $\mu$ (including $\mu$ itself), while $n_{2,+}$ does the same for $C_2$. Similarly, $n_{i,-}$ counts edges in $C_i$ that share the same position but have operator direction of $\mu$.
Note that the third line in the equation involves only single-trace operators. They arise from the `sewing' of $C_1$ and $C_2$ upon the overlapping edges in $C_2$ (counted by $n_{2,\pm}$), as illustrated by the following diagrams:
\begin{equation}
\begin{gathered} {\includegraphics[width=0.85\linewidth]{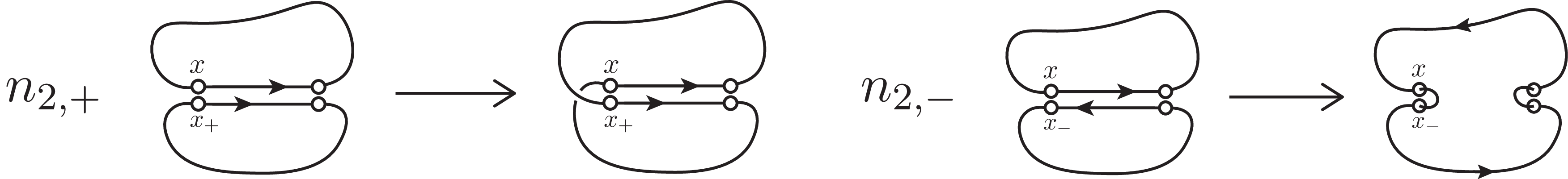} } \end{gathered} \,.
\end{equation} 

The previous two examples of \eqref{eq:sdeexN1} and \eqref{eq:sdeexN2} in SU(3) theory are given by simply setting $N=3$.
Below we give two more examples of \eqref{eq:SDE3dDT2} for double-trace loops. The first one is
\begin{align}
	\WLdaar \ \  \overset{\rm 2D}{\underset{\rm SU(3)}{\Longrightarrow}} \quad \ & \left( \, \WLdca{midarrow}{midarrow} -\WLdba{midarrow}{midarrow} +\WLdda{midarrow}{midarrow} -\WLa{midarrow} \, \right)+2\lambda\left(1-\frac{1}{9}\right) \left( \WLdaa{midarrow}{midarrow} \right) \nonumber \\
	& -\left( \, \WLta{midarrow}{midreversearrow}{midarrow} -\WLta{midarrow}{midarrow}{midarrow} +\WLtb{midarrow}{midarrow}{midarrow} -\WLtb{midreversearrow}{midarrow}{midarrow} \, \right) = 0 \,.
\end{align}
The second one involves overlapping edges corresponding to ${{\check n}}_{1,+} = n_{2,+}=1$:
\begin{align}
	\WLdabr \ \  \overset{\rm 2D}{\underset{\rm SU(3)}{\Longrightarrow}} \quad \ & \left( \, \WLddf{midarrow}{midarrow} -\WLdbb{midarrow}{midarrow}  +\WLddb{midarrow}{midarrow} -\WLa{midarrow} \, \right) +2\lambda \left( 1 -\frac{2}{9} \right) \WLdab{midarrow}{midarrow} \notag \\
	& - \left( \, \WLtc{midarrow}{midarrow}{midreversearrow} -\WLtc{midarrow}{midarrow}{midarrow} +\WLtd{midarrow}{midarrow}{midarrow} -\WLtd{midarrow}{midarrow}{midreversearrow} \, \right) +\frac{2\lambda}{9} \, \WLd{midarrow} = 0 \,.
\end{align}
We can see that the above two equations generate triple-trace loops.
In SU(3) theory, the triple-trace operators can be reduced to single- and double-trace operators using (see Appendix~\ref{app:su3identities})
\begin{align}\label{eq:SU3relations}
w(C_1) w(C_2) w(C_3) 
=& \, - {1\over9} \big[ w(C_1\, C_2 \, C_3) + w(C_3 \, C_2\, C_1) \big]  \\
& \, + {1\over3} \big[ w(C_1) w(C_2\, C_3) + w(C_2) w(C_3\, C_1) + w(C_3) w(C_1\, C_2) \big] \notag \\
& \, + {1\over3} w(\bar C_1\, C_2)w(\bar C_1\, C_3) - {1\over9} w(\bar C_1\, C_2 \,\bar C_1\, C_3) \,. \notag 
\end{align}
For example, 
\begin{align}
	\WLtb{midarrow}{midarrow}{midarrow} \, \overset{\rm SU(3)}{=} \, -\frac{2}{9} \, \WLh{midarrow} -\frac{1}{9} \, \WLi{midarrow} +\frac{1}{3}\left( \, \WLdda{midarrow}{midarrow} +2\, \WLdbc{midarrow}{midarrow} \, \right) +\frac{1}{3} \, \WLddg{midarrow}{midarrow} \,,
\end{align}
where the first two Wilson loops are the single-trace operators and the other three are double-trace operators.
For special triple- and double-trace loops consisting of identical loops, we have the following reduction formulae:
\begin{align}
	\WLtd{midarrow}{midarrow}{midarrow} \, \overset{\rm SU(3)}{=} \, \WLdab{midarrow}{midreversearrow} +\frac{1}{9} \, \WLg{midarrow} -\frac{1}{9} \,, \qquad \WLdab{midarrow}{midarrow}
	\, \overset{\rm SU(3)}{=} \, \frac{1}{3} \, \WLd{midarrow} +\frac{2}{3}\, \WLa{midarrow} \,.
\end{align}
Further discussion of SU(3) relations is given in  Appendix~\ref{app:su3identities}.

\vskip 0.3cm
\subsubsection*{Backtrack equations.}

In the above discussion of SD equations, we consider variations on edges belonging to Wilson loops. There is another generalized class of SD equations known as \emph{backtrack equations}, see \cite{Anderson:2016rcw, Kazakov:2022xuh, Kazakov:2024ool}. They can be derived by varyin the edges along a backtrack path.
A simple example is
\begin{align}
	\WLbr \ \  \overset{\rm 2D}{\underset{\rm SU(3)}{\Longrightarrow}} \quad \ & \left(\, \WLb{midarrow} -\WLc{midarrow} \,\right)-\left(\, \WLdaa{midarrow}{midarrow} -\WLdaa{midarrow}{midreversearrow} \,\right) \notag \\
	& + \left(\, \WLj{midarrow} -\WLk{midarrow}\,\right) -\left( \, \WLdac{midarrow}{midarrow} -\WLdac{midarrow}{midreversearrow} \, \right) = 0 \,. 
\label{eq:exBTloopEq}
\end{align}

Finding a complete set of backtrack equations for a large number of Wilson loops, especially in 4D SU(3) theory, is typically non-trivial. In practice, one approach is to first generate Wilson loops with various backtrack path insertions (up to a certain truncation of length), and then apply variations on the edges within the backtrack paths.
For the purposes of this study, we will generate backtrack SD equations by inserting only length-2 backtrace paths in the loops, such as in \eqref{eq:exBTloopEq}. 
We leave a systematic and efficient algorithm for deriving the full set of loop equations to future work.

\section{Path selection}
\label{sec:pathselection}

To practically solve the SDP problem, a truncation of Wilson loop variables is necessary.
In this section, before going to the explicit constructions, we discuss the strategy of choosing paths.
We first introduce a dimensional reduction method for path selection, and then we explain the choice of paths for SU(3) theory.

\subsection{Dimension reduction: plane-type paths}
\label{sec:planetypePath}

A natural way of truncation is to restrict path or loop operators to a maximum length.
Table~\ref{tab:countingLoops} lists the number of independent loop configurations of length up to 16 across different dimensions.
In practice, these loops can be represented using the letter notation introduced in Appendix~\ref{app:symreduction}, with a canonical form chosen modulo cyclic permutations, lattice symmetries, and conjugation.\footnote{
The numbers reported here are \emph{kinematically} independent, meaning loop equations are not applied. By incorporating the full set of loop equations, one can instead construct a \emph{dynamically} independent loop basis for a given length, where the constraints imposed by the loop equations are taken into account.
}
We can see that, even with a length truncation, the number of loop variables grows rapidly. A similar growth occurs for the number of independent loops after using loop equations, as shown in explicit examples in later sections.
Such large set of loop variables poses several challenges for the SDP construction. First, constructing positivity matrices and determining the complete set of loop equations is itself demanding. Second, the increased size of matrices and the larger number of independent SDP variables greatly complicate the task of solving the SDP system.

Therefore, a critical aspect of the positivity bootstrap construction is to carefully select the subset of paths used in building the positivity matrices. 
One strategy, used in \cite{Kazakov:2024ool}, is to use SD equations to generate longer paths from shorter ones, a method we also adopt. However, in 3D and 4D theories, the complexity still increases rapidly, necessitating additional reduction techniques.

\begin{table}[t]
\centering
\begin{tabular}{l | c | c | c | c | c | c | c  } 
\hline
Length of loops  	&  4 & 6 & 8 & 10 & 12 & 14 & 16  \cr \hline 
\# of 2D loops   		&  1 & 1 & 7 & 15 & 95 & 465 & 3,217   \cr \hline 
\# of 3D loops   		&  0 & 2 & 11 & 117 & 1,657 & 27,012 & 488,300  \cr \hline 
\# of 4D loops   		&  0 & 0 & 7 & 106 & 3,304 & 109,304 & {\color{black} 3,849,514}   \cr \hline 
\end{tabular} 
\caption{Number of independent loops in various dimensions. 
These loop counts are kinematically independent, meaning they are distinct geometrical shapes before any loop equations are applied.
In counting 3D loops, we exclude those that can be embedded in a 2D subspace, since such loops are already counted as 2D loops. Similarly, in 4D counting, loops that can be embedded in 2D or 3D subspaces are excluded.
\label{tab:countingLoops}
}
\end{table}

We introduce a simplification by choosing paths and loops in a one-dimensional reduced subspace.
Consider the two-dimensional case. In this case, we focus on the loops confined within a one-dimensional strip, such as
\begin{align}
	\left\{ \WLa{midarrow}\,,\ \WLb{midarrow}\,,\ \WLc{midarrow}\,,\ \WLd{midarrow} \,,\ \WLe{midarrow} \,,\ \WLf{midarrow} \,,\ \ldots \right\} \,.
\end{align}
These loops extend only along the horizontal `$x$' direction while being restricted to the vertical `$y$' range of $0\leq y \leq 1$.\footnote{This effectively `reduces' the lattice dimension by one. Note, however, there remains a finite thickness along the '$y$' direction distinguishing this setup from a genuinely lower dimensional theory.}
We call such loops \emph{ladder-type} loops.
The same choice applies to the path operators that are used to construct positivity matrices, corresponding to \emph{ladder-type paths}.
In the letter representation (see Appendix~\ref{app:symreduction}) of ladder-type paths and loops, the `b' and `-b' letters appear consecutively like
\begin{equation}
	\{ \ldots, \bp, \ldots, \bm, \ldots, \bp, \ldots, \bm, \ldots, \bp, \ldots, \bm, \ldots \} \,,
\end{equation}
where $\ldots$ consist of other letters.

\begin{table}[t]
	\centering
	\begin{tabular}{l | c | c | c | c | c | c | c | c | c } 
	\hline
	Length of loops  	&  4 & 6 & 8 & 10 & 12 & 14 & 16 & 18 & 20 \cr \hline 
	\# of general loops   	&  1 & 1 & 7 & 15 & 95 & 465 & 3,217 & 21,762 & 159,974   \cr \hline 
	\# of ladder-type loops      &  1 & 1 & 3 & 5 & 13 & 32 & 90 & 268 & 867 \cr \hline 
	\end{tabular} 
	\caption{Number of ladder-type loops is significantly smaller than general-type loops in 2D.
	\label{tab:counting2Dladder}
	}
\end{table}

Reducing loops to ladder-type ones is a significant simplification. As shown in Table~\ref{tab:counting2Dladder}, the number of ladder-type loops is considerably smaller than the general-type loops. 
Additionally, ladder-type loops have the property of forming a closed subset:
they can be generated by sewing ladder-type paths and are closed under SD equations that vary only on `b' and `-b' edges.

\begin{figure}[t]
 \centering
    \includegraphics[width=0.4\linewidth]{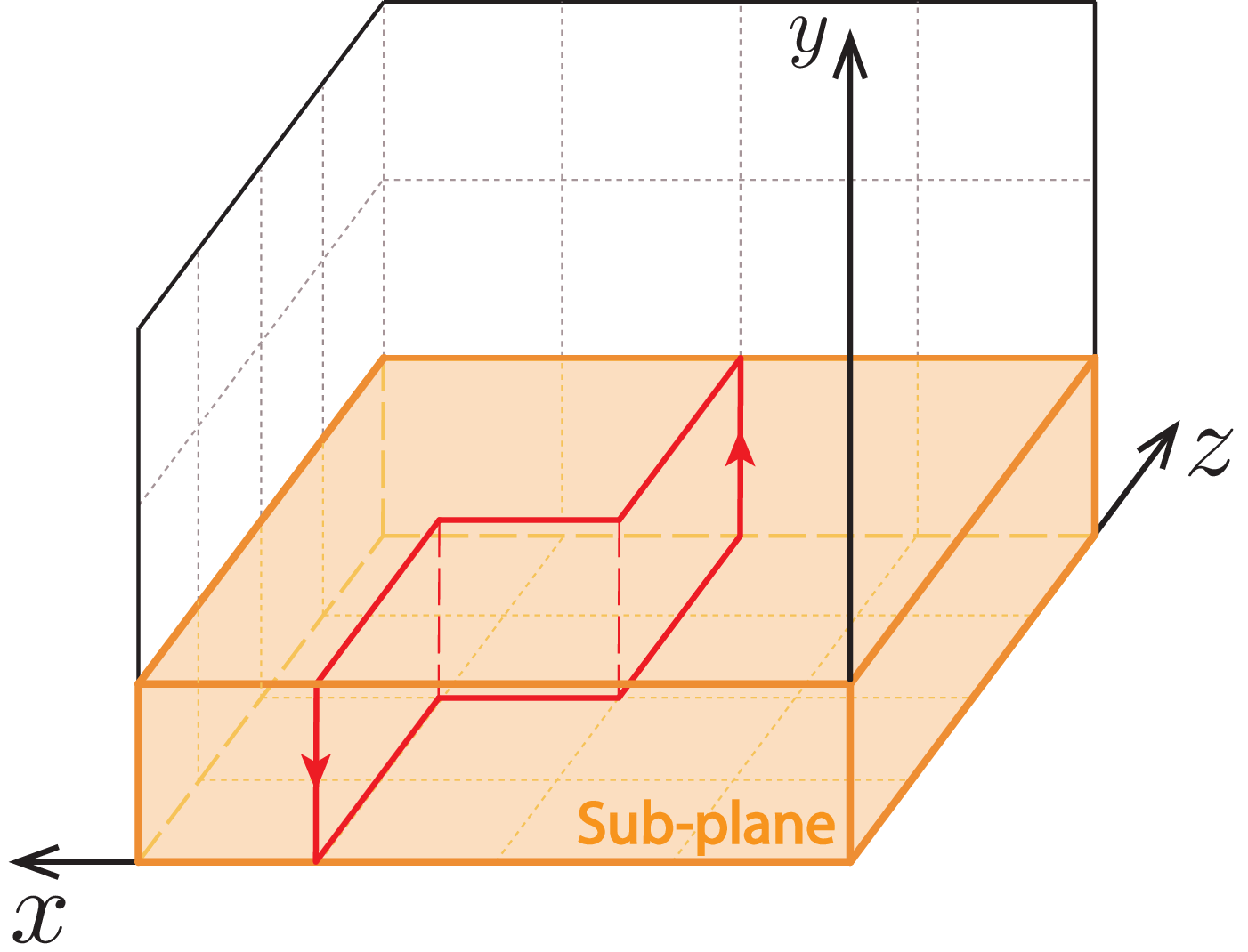}
    \caption{Sub-plane for 3d lattice.}
    \label{fig:3dplane}
\end{figure}

Similar reduction can be applied to 3D and 4D lattice theory. For example, in the 3D case, we consider dimension-reduced paths confined within a sub-plane extended along two directions, such as the $x$ and $z$ directions, as illustrated in Figure~\ref{fig:3dplane}.
We will refer to these paths in the sub-plane as \emph{plane-type paths} for both 3D and 4D theories.

Explicit examples of paths with this truncation will be given in Section~\ref{sec:results2D}$-$\ref{sec:results3D4D}.
We mention that the symmetry properties are also changed for the ladder- or plane-type paths. Since they are confined to a $(d-1)$-dimensional subplane, they have the smaller $(d-1)$-dimensional lattice symmetry rather than the full $d$-dimensional one, see Appendix~\ref{app:symreduction}.

\subsection{Choice of path in SU(3)}
\label{sec:choiceofpathsu3}

For the SU(2) and large $N$ theories, the path operators can be chosen as simple connected paths. By sewing such paths, one gets positivity matrices whose elements are single-trace Wilson loops.
On the other hand, for SU(3) gauge theory, double-trace operators are independent and must be explicitly included in the analysis. 
To achieve this, we extend our choice of paths to include combinations of a Wilson line and a Wilson loop.

For example, consider the ladder-type path (up to length 4) in 2D, we have the following 9 paths:
\begin{align}
\label{eq:pathex1}
	{\cal O} = \left\{ \WPa \,,\ \WPb{midarrow}\,,\ \WPb{midreversearrow}\,,\ \WPc{midarrow}\,,\ \WPc{midreversearrow}\,, \ \WPdaa{midarrow}\,, \ \WPdaa{midreversearrow}\,,\ \WPdab{midarrow}\,,\ \WPdab{midreversearrow} \right\} \,,
\end{align}
where `$\WPa$' represents the identity matrix.
There are two types of paths in \eqref{eq:pathex1}. The first five paths such as $\WPb{midarrow}$ are simply connected paths and will be called \emph{simple path}. On the other hand, the remaining four paths, such as $\WPdaa{midarrow}$, are not simple paths but a combination of an identity path and a plaquette loop:
\begin{equation}
\WPdaa{midarrow} =  (\WPa) \otimes (\WLa{midarrow}) \,,
\end{equation}
where the relative position of the path and the loop is indicated by the LHS.
In the next section, we will consider more general paths, such as combining a length-4 path and a plaquette as
\begin{equation}\label{eq:doublepathex2}
\left\{\WPdac{}{} \,, \ \  \WLdad{}{} \right\} \,.
\end{equation}
We will refer to these paths, which are combinations of a simple path and a Wilson loop, as \emph{composite paths}.

As an example, we give the reflection positivity matrix with respect to the paths 
\begin{equation}\label{eq:doublepathex1}
	{\cal O}_+ = \left\{ \WPa \,,\ \WPb{midarrow}\,,\ \WPb{midreversearrow}\,, \ \WPdaa{midarrow}\,, \ \WPdaa{midreversearrow} \right\} \big|_{x\geqslant 0} \,.
\end{equation}
The reflected paths (with mirror plane $x\leftrightarrow-x$) are
\begin{equation}\label{eq:doublepathex1ref}
	{\cal O}_+^R = \left\{ \WPa\,,\ \WPc{midarrow}\,,\ \WPc{midreversearrow} \,, \WPdab{midarrow}\,,\ \WPdab{midreversearrow} \right\} \big|_{x\leqslant 0} \,.
\end{equation}
Sewing the paths \eqref{eq:doublepathex1} and \eqref{eq:doublepathex1ref}, we obtain the positivity matrix
\begin{align}
	{\renewcommand{\arraystretch}{1.5}
	\begin{blockarray}{cccccc}
		& \WPa & \WPb{midarrow} & \WPb{midreversearrow} & \WPdaa{midarrow} & \WPdaa{midreversearrow} \\
		\begin{block}{c(ccccc)}
			\WPa & 1 & \WLa{midarrow} & \WLa{midreversearrow} & \WLa{midarrow} & \WLa{midreversearrow} \\
			\WPc{midarrow} & \WLa{midarrow} & \WLb{midarrow} & \WLc{midarrow} & \WLdaa{midarrow}{midarrow} & \WLdaa{midarrow}{midreversearrow} \, \\
			\WPc{midreversearrow} & \WLa{midreversearrow} & \WLc{midreversearrow} & \WLb{midreversearrow} & \WLdaa{midreversearrow}{midarrow} & \WLdaa{midreversearrow}{midreversearrow} \\
			\WPdab{midarrow} & \WLa{midarrow} & \WLdaa{midarrow}{midarrow} & \WLdaa{midarrow}{midreversearrow} & \WLdaa{midarrow}{midarrow} & \WLdaa{midarrow}{midreversearrow} \\
			\WPdab{midreversearrow} & \WLa{midreversearrow} & \WLdaa{midreversearrow}{midarrow} & \WLdaa{midreversearrow}{midreversearrow} & \WLdaa{midreversearrow}{midarrow} & \WLdaa{midreversearrow}{midreversearrow} \\
		\end{block}
	\end{blockarray}}
	\succeq 0 \,.
\end{align}
Note that we get double-trace loops as matrix elements by sewing composite paths. Sewing more general composite paths like \eqref{eq:doublepathex2} will give triple-trace loops.

It is also possible to consider other complex paths, for example, by combining pairs of simple paths. 
However, we find that such constructions impose weaker constraints. 
The most effective choice is to use the composite paths introduced above, which we will focus on in this work.
Finally, we mention that throughout the paper, we will focus on the set of paths where both endpoints coincide.

\section{Bootstrap 2D YM}
\label{sec:results2D}

We begin by applying the bootstrap method to 2D YM in this section. 
The 2D lattice gauge theory can be reduced to a matrix model characterized by one plaquette. 
The general partition functions of 2D $SU(N)$ theories can be computed as \cite{Drouffe:1983fv}\footnote{See early closed form for 2D SU(2) plaquette in \cite{Balian:1974xw}. Analytic results of large $N$ 2D YM were studied in \cite{Bars:1979xb, Gross:1980he, Wadia:2012fr}.}
\begin{align}
    Z = \sum_{n=-\infty}^{\infty} {\rm det}(I_{i-j+n}(N/\lambda)) \,, \quad 1\leqslant i,j \leqslant N \,,
\end{align}
where $I_n$ is a modified Bessel function of the first kind. 
The expectation value of the plaquette is given by
\begin{align}
\langle {\rm tr}(U_P) \rangle = \langle W_1 \rangle  = -\left( \frac{\lambda}{N} \right)^2 \partial_\lambda \log(Z)\,.
\end{align}
We will compare the analytic results with the bootstrap bounds.

Although our primary goal is the SU(3) theory, we will always begin with the simpler SU(2) case which involves only simple paths and single-trace operators. As we will see, the SU(3) case can be built upon the SU(2) construction by incorporating composite paths.

\subsection{2D SU(2)}

Let us start with the simplest 2D SU(2) case.  To clearly demonstrate the convergence of the bounds, we will consider several different choices of paths with increasing complexity, \emph{i.e.}, by including a greater number of paths.

\paragraph{Path choice 1.}
We first consider the following set of simple path configurations:
\begin{align}
	\label{eqfig:path2Da}
	\left\{ \WPa \,,\ \WPb{} \,, \ \WPd{} \,, \ \WPe{} \,, \ \WPf{}\,, \ \WPg{} \right\} \,.
\end{align}
From these initial path configurations, we apply two operations to generate a complete set of paths (here we consider only ladder-type paths as discussed in Section~\ref{sec:planetypePath}): 
(1) the symmetry operation, which corresponds to a $Z_2$ symmetry for the ladder-type paths, and (2) the conjugation operation, which involves adding an arrow in both opposite directions. 
For example, starting with the length-4 square-type path, we obtain four independent paths by symmetry and conjugation:\footnote{We refer to the path on the LHS as a \emph{path configuration} and the generated paths via symmetry and conjugation as the real paths. Since path configurations have a number significantly smaller than that of the generated paths, it provides a more efficient representation of path information.}
\begin{align}
	\WPb{} \Longrightarrow \left\{ \WPb{midarrow} \,,\ \WPb{midreversearrow} \,,\ \WPc{midarrow} \,,\ \WPc{midreversearrow} \right\} \,.
\end{align}
In total, from the configuration of \eqref{eqfig:path2Da} we generate 19 ladder-type paths.

By sewing the paths we get positivity matrices.
We will consider two positivity matrices, the Hermitian positivity and the reflection positivity with $x=0$ as the mirror line. They have dimension 19 (Hermitian) and dimension 6 (reflection), respectively. 
After the symmetry decomposition as reviewed in Appendix~\ref{app:symreduction}, we obtain a set of decomposed smaller matrices, with dimensions as
\begin{equation}
\text{dim}({\cal M}): \quad \{19\}_H \cup \{6\}_R  \ \ \overset{\rm sym.}{\underset{\rm decom.}{\Longrightarrow}} \ \   \{6, 4, 4, 5\}_H \cup \{3, 3\}_R \,,
\end{equation}
where the subscript $H$ and $R$ refer to Hermitian and Reflection, respectively. 

The matrix elements are single-trace Wilson loops, represented as sequences of letters as explained in Appendix~\ref{app:symreduction}. These loops can have multiple representations due to cyclic permutations, lattice symmetries, and conjugation (see Appendix~\ref{app:symreduction} for further details). We select a canonical representation for the Wilson loops. In the case considered here, we identify 24 independent loop variables, modulo lattice symmetry and conjugation. This number is also the number of loops that we referred to in the figures.

The loop equations, discussed in Section~\ref{sec:loopequation}, introduce additional dynamical relations among these loops.
In this case, we obtain a set of 12 loop equations. These equations impose 12 linear constraints on the 24 loop variables, reducing the number of independent variables to 12 in the positivity matrices.\footnote{In general, a set of loop equations may be not fully independent with each other, see the Path choice 3 case below and the footnote~\ref{footnote9}.}

\begin{figure}[t]
\centering
\subfigure[Path choice 1.]{\includegraphics[width=.45\linewidth]{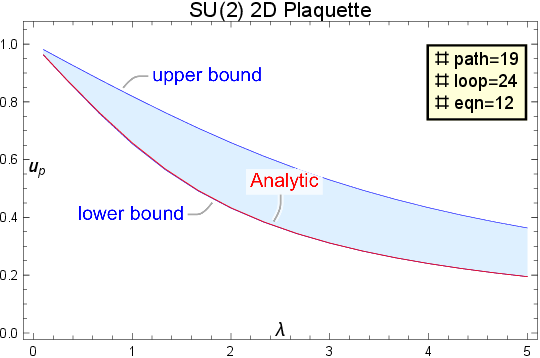}}
\qquad
\subfigure[Path choice 2.]{\includegraphics[width=.45\linewidth]{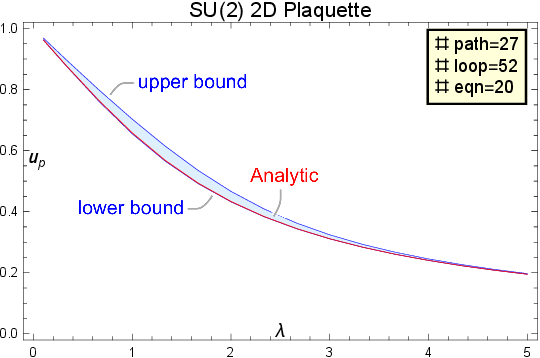}}
\qquad
\subfigure[Path choice 3.]{\includegraphics[width=.45\linewidth]{2DSU2c}}
\qquad
\subfigure[Combined all plots.]{\includegraphics[width=.45\linewidth]{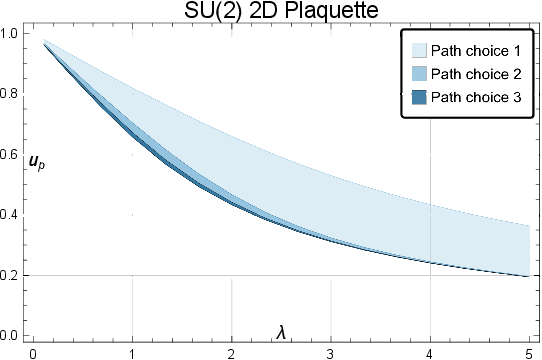}}
\caption{
Upper and lower bounds for the SU(2) plaquette expectation value in 2D YM. The three numbers displayed in each subfigure represent: (i) the number of paths, which equals the dimension of the Hermitian positive matrix (before symmetry decomposition); (ii) the number of loops, corresponding to the configuration-independent Wilson loops in the positivity matrices, accounting for lattice symmetries and conjugation; and (iii) the number of equations, referring to the set of independent loop equations. Hence, the number of independent loop variables in the positivity matrices is calculated as (\# loop - \# eqn). 
}
\label{fig:2D1}
\end{figure}

As outlined in Section~\ref{sec:bootstrapstrategy}, we are now ready to compute the bounds on the Wilson loop expectation values. 
The SDP problem for this example can be efficiently solved using {\tt SemidefiniteOptimization} function with default setting in {\tt Mathematica} \cite{mathematica}. 
The resulting bounds for the plaquette are shown in Figure~\ref{fig:2D1}(a).
Notably, even for this simple setup, the lower bound is already very close to the analytic result.

\paragraph{Path choice 2.}

To improve the upper bound, we consider adding two more path configurations into \eqref{eqfig:path2Da}:
\begin{align}
	\label{eqfig:path2Db}
	\left\{ \WPh{}\,, \ \WPi{} \right\} \,.
\end{align}
After applying symmetry and conjugation, they give 8 more ladder-type paths. 

After applying symmetry decomposition, the dimensions of reduced positivity matrices are 
\begin{equation}
\text{dim}({\cal M}): \quad \{27\}_H \cup \{10\}_R  \ \ \overset{\rm sym.}{\underset{\rm decom.}{\Longrightarrow}} \ \   \{8, 6, 6, 7\}_H \cup \{5, 5\}_R \,.
\end{equation}

The number of involved loop variables in the positivity matrices is 52, modulo lattice symmetry and conjugation.
Additionally, we obtain 20 loop equations which are independent of each other. This reduces the number of loop variables to 32 in the positivity matrices.  
Solving SDP using {\tt Mathematica}, we obtain the bounds as shown in Figure~\ref{fig:2D1}(b).
The change of the lower bound is negligible, but the upper bound receives a notable improvement.

\paragraph{Path choice 3.}

In the above two cases, we consider only the ladder-type path. 
In this choice, we extend the paths to general-type paths. 
After applying full 2D lattice symmetry (which is $D_3$ group) and conjugation for the path configurations in  \eqref{eqfig:path2Da} and \eqref{eqfig:path2Db}, we obtain 81 general-type paths. 
The dimensions of positivity matrices after symmetry decomposition are reduced to
\begin{equation}
\text{dim}({\cal M}): \quad \{81\}_H \cup \{34\}_R  \ \ \overset{\rm sym.}{\underset{\rm decom.}{\Longrightarrow}} \ \   \{8, 3, 6, 4, 10, 3, 7, 4, 6, 10\}_H \cup \{9, 8, 8, 9\}_R \,.
\end{equation}
The number of involved loop variables is 164, modulo lattice symmetry and conjugation.
In this case, we obtain 84 loop equations, providing 74 independent constraints. Namely, only 74 loop equations are independent ones.\footnote{As we show in this case, the loop equations are generally not independent of each other. In the remaining sections, we will mainly refer to the number of independent loop equations, and it is this number quoted in the SDP figures such as in Figure~\ref{fig:2D1}(c).\label{footnote9}}
Solving SDP, we obtain the bounds as shown in Figure~\ref{fig:2D1}(c).

We observe that the upper bound receives a further significant improvement, while the change of the lower bound is negligible.
This implies that to improve the upper bound, it is good to consider the path with full 2D symmetry.

\begin{figure}[t]
\centering
\subfigure[Bounds.]{\includegraphics[width=.45\linewidth]{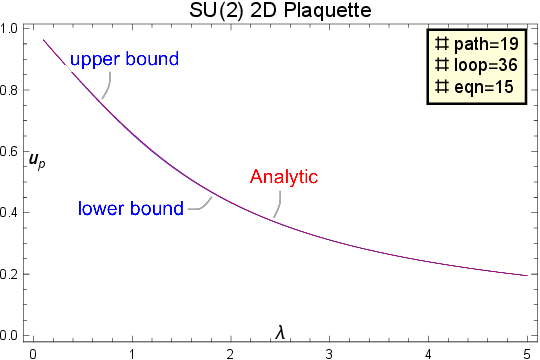}}
\qquad
\subfigure[Error of bounds: ${{\rm bound}-{\rm analytic} \over {\rm analytic}}$.]{\includegraphics[width=.48\linewidth]{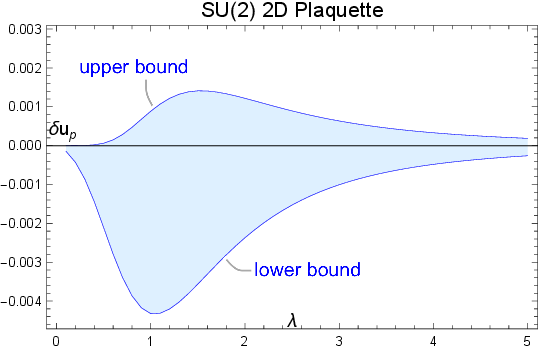}}
\caption{Positivity bounds with twist reflection condition for SU(2) YM in 2D.}
\label{fig:2D1twist}
\end{figure}

\paragraph{Twist-reflection positivity.}
We consider the twist-reflection positivity condition that was introduced in Section~\ref{sec:twist-reflection}. It turns out that with this new constraint, we can obtain a very tight upper bound. Using the simple path choice 1, the dimensions of matrices are 
\begin{equation}
\text{dim}({\cal M}): \quad \{19\}_H \cup \{6\}_R \cup \{6\}_{R_t}  \ \ \overset{\rm sym.}{\underset{\rm decom.}{\Longrightarrow}} \ \   \{6, 4, 4, 5\}_H \cup \{3, 3\}_R \cup \{3, 3\}_{R_t} \,,
\end{equation}
where $R_t$ represents the set of new matrices from the twist-reflection positivity.
The final bounds are shown in Figure~\ref{fig:2D1twist}, which are remarkably close to the analytic results.
We will find that similar bounds can be obtained for SU(3) theory.

\subsection{2D SU(3)}

In the SU(3) case, double-trace Wilson loops arise as independent operators, making it crucial to account for their effects. To address this, we will extend the path choices in the previous subsection by including composite paths, as introduced in Section~\ref{sec:choiceofpathsu3}.

\paragraph{Path choice 1.}
Apart from the simple path configurations in SU(2) case \eqref{eqfig:path2Da}, we add composite paths, as introduced in Section~\ref{sec:choiceofpathsu3}.
One set is 
\begin{align}
	\label{eqfig:pathadd2Dsu3aDT1}
	\left\{\WPdaa{}\,, \ \WPddc{} \,,\ \WPdga{} \,, \ \WPdbc{} \,, \ \WPdcc{} \right\} ,
\end{align}
which are combinations of an identity path (represented by `$\WPa$' ) and the Wilson loop configurations formed from \eqref{eqfig:path2Da}.
Another set is the length-4 path combined with a length-4 plaquette or a length-8 Wilson loop:
\begin{align}
	\label{eqfig:pathadd2Dsu3a}
	\left\{\WPdac{}{} \,, \ \WLdad{}{}\,, \ \WPdda{}{} \,,\ \WPddb{}{}\right\} \,.
\end{align}
Applying symmetry and conjugation, we obtain 50 new ladder-type composite paths. Note that one should take into account the relative directions of the path and loop in a composite path.
In total, there are 69 paths. 

\begin{figure}[t]
\centering
\subfigure[Path choice 1.]{\includegraphics[width=.45\linewidth]{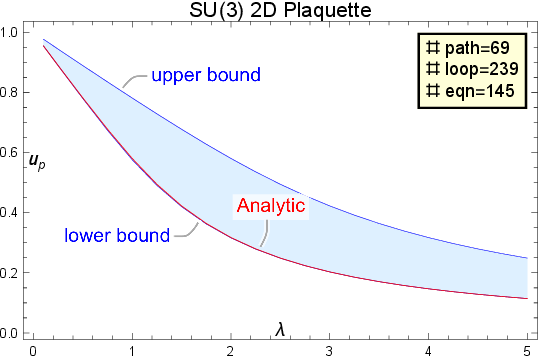}}
\qquad
\subfigure[Path choice 2.]{\includegraphics[width=.45\linewidth]{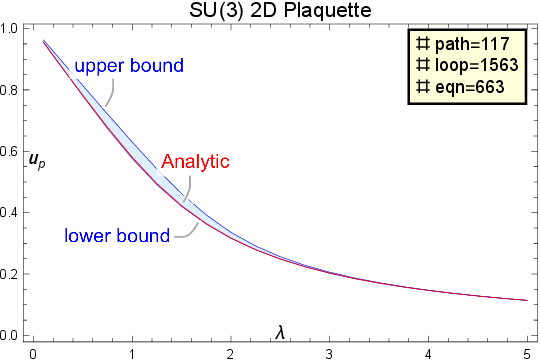}}
\qquad
\subfigure[Path choice 3.]{\includegraphics[width=.45\linewidth]{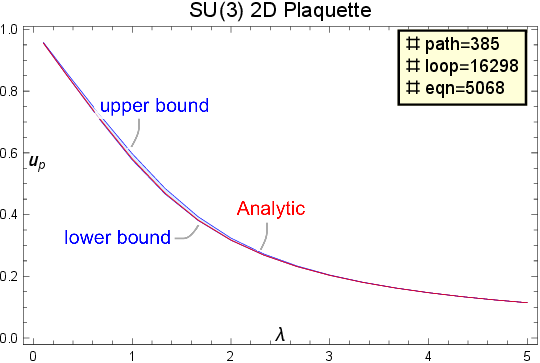}}
\qquad
\subfigure[Combined all plots.]{\includegraphics[width=.45\linewidth]{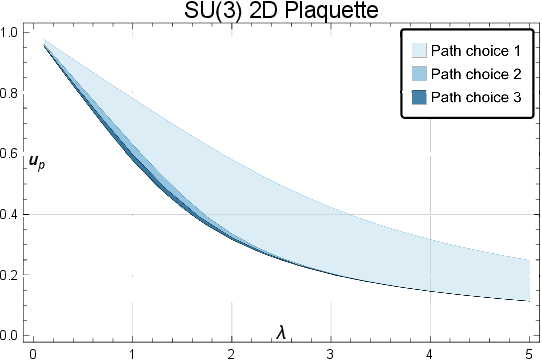}}
\caption{SU(3) in 2D.}
\label{fig:2Dsu3}
\end{figure}

We then construct the Hermitian and reflection positivity matrices from the 69 paths. 
After symmetry decomposition, the dimensions of positivity matrices are reduced to 
\begin{equation}
\text{dim}({\cal M}): \quad \{69\}_H \cup \{20\}_R  \ \ \overset{\rm sym.}{\underset{\rm decom.}{\Longrightarrow}} \ \ \{19, 16, 16, 18\}_H \cup \{10, 10\}_R \,.
\end{equation}
Note that sewing composite paths generates both double- and triple-trace Wilson loops as matrix elements.
For the triple-trace Wilson loops, we apply the reduction formula \eqref{eq:SU3relations} to express them in terms of double- and single-trace loops.
The number of independent single- and double-trace loop variables (accounting for symmetry and conjugation) is 239. We also consider the Schwinger-Dyson equations and SU(3) relations for these variables (see Section~\ref{sec:loopequation}), leading to a total of 145 loop equations.\footnote{As noted at the end of Section~\ref{sec:loopequation}, in this work we do not attempt to construct a complete set of loop equations but applying a truncation strategy for deriving backtrack equations. In practice, we find that this truncation provides a good approximation as including additional backtrack equations generally does not significantly improve the bounds.}
Solving SDP we obtain the bounds shown in Figure~\ref{fig:2Dsu3}(a).
As in the SU(2) case, the lower bound is very close to the analytic result.

\paragraph{Path choice 2.}

To improve the upper bound, together with the paths \eqref{eqfig:path2Da}, \eqref{eqfig:path2Db}, \eqref{eqfig:pathadd2Dsu3aDT1} and \eqref{eqfig:pathadd2Dsu3a}, we add four more composite path configurations:
\begin{align}
	\label{eqfig:pathadd2Dsu3b}
	\left\{ \WPdba{}{} \,,\ \WPdbb{}{} \,,\ \WPdca{}{} \,,\ \WPdcb{}{} \right\} \,,
\end{align}
After applying symmetry and conjugation, 
there are in total of 117 paths, and we construct positivity matrices from them.

The dimensions of reduced positivity matrices are 
\begin{equation}
\text{dim}({\cal M}): \quad \{117\}_H \cup \{36\}_R  \ \ \overset{\rm sym.}{\underset{\rm decom.}{\Longrightarrow}} \ \ \{31, 28, 28, 30\}_H \cup \{18, 18\}_R \,.
\end{equation}
The number of involved loop variables is 1563, subject to 663 loop equations.
Solving SDP using {\tt Mathematica}, we obtain the bounds as shown in Figure~\ref{fig:2Dsu3}(b).
The change of the lower bound is negligible, while the upper bound has a notable improvement.

\paragraph{Path choice 3.}

We consider further the general-type paths beyond ladder-type restriciton. Following the path choice 2 that combines path configurations \eqref{eqfig:path2Da}, \eqref{eqfig:path2Db}, \eqref{eqfig:pathadd2Dsu3aDT1}, \eqref{eqfig:pathadd2Dsu3a}, and \eqref{eqfig:pathadd2Dsu3b},
we apply full 2D symmetry and conjugation which give 385 general-type paths. 
The dimensions of reduced positivity matrices are 
\begin{equation}
\text{dim}({\cal M}): \ \  \{385\}_H \cup \{148\}_R  \ \ \overset{\rm sym.}{\underset{\rm decom.}{\Longrightarrow}} \ \ \{31, 18, 28, 20, 48, 18, 30, 20, 28, 48\}_H \cup \{38, 36, 36, 38\}_R \,.
\end{equation}
The number of involved loop variables is 16298, subject to 5068 loop equations.
This represents a significant increase in complexity compared to the previous ladder-type reduction, where the number of variables was smaller by an order of magnitude.

In this example, we employ MOSEK \cite{mosek} to solve the constraints due to the relatively large number of variables involved.
With a single thread in MOSEK running on an Apple M3 Max chip, each data point is computed in about 5 minutes using approximately 5 GB of memory.\footnote{
To establish a consistent benchmark, all computations we mention below were performed with one thread in MOSEK on the same hardware (MacBook Pro with Apple M3 Max), and we maintained a precision of $10^{-5}$ throughout. Increasing the number of threads can reduce computational time, while it also results in higher memory usage.}
The final bounds are shown in Figure~\ref{fig:2Dsu3}(c).
We observe that the upper bound receives a significant improvement, while the change of the lower bound is negligible. The results of choice 2 and choice 3 suggest that to improve the upper bound, one can increase the path size and also consider the full lattice symmetry.

\begin{figure}[t]
\centering
\subfigure[Bounds.]{\includegraphics[width=.45\linewidth]{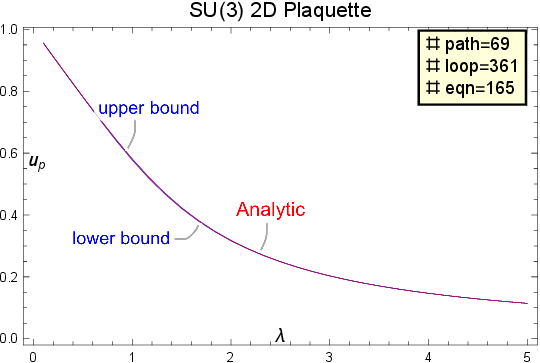}}
\qquad
\subfigure[Error of bounds: ${{\rm bound}-{\rm analytic} \over {\rm analytic}}$.]{\includegraphics[width=.48\linewidth]{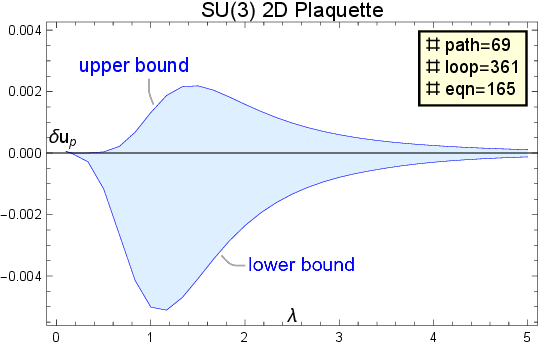}}
\caption{Positivity bounds with twisted PR condition for SU(3) YM in 2D.}
\label{fig:2Dsu3twist}
\end{figure}

\paragraph{Twist RP.}
We now impose the twist-reflection condition. With this additional constraint, we obtain a significantly tighter upper bound. Using the same path choice as in choice 1, the matrix dimensions are 
\begin{equation}
\text{dim}({\cal M}): \quad \{69\}_H \cup \{20\}_R \cup \{20\}_{R_t}  \ \ \overset{\rm sym.}{\underset{\rm decom.}{\Longrightarrow}} \ \ \{19, 16, 16, 18\}_H \cup \{10, 10\}_R \cup \{10, 10\}_{R_t} \,,
\end{equation}
where $R_t$ represents the matrices arising from twist-reflection positivity.
The resulting bounds are shown in Figure~\ref{fig:2Dsu3twist}.
We can see that, as in the SU(2) case, the bounds are remarkably close to the analytic results, deviating by only a few thousandths.

Our study of  2D SU(2) and SU(3) theories reveals useful patterns that can guide the construction of 3D and 4D theories.
Notably, the behavior of the upper and lower bounds depends on the choice of paths.
The results suggest a strategy for computing the lower and upper bound separately:
\begin{itemize}
\item 
For the lower bound, one can focus on the ladder-type paths, increase the loop length without extending the scale.

\item
For the upper bound, one can consider the paths with full lattice symmetry and include large-scale paths.

\end{itemize}
However, since the twist-reflection positivity does not hold in higher-dimensional theories, we will not consider this condition further.

\section{Bootstrap 3D and 4D YM}
\label{sec:results3D4D}

Next, we turn to the more challenging 3D and 4D YM theory. Our construction will largely follow that of the 2D case, so we will be brief, focusing on presenting the main data and results. 

As the spacetime dimension increases, the number of path or loop variables grows rapidly. Therefore, it is crucial to keep a relatively small size of operator system. In this work, our primary goal is to give a proof of principle that the bootstrap strategy works in practice. We will demonstrate this by employing a `relatively' small set of paths, without attempting to obtain the optimal bounds.

Since the high dimensional theories are not solvable in closed form, we compare the obtained bounds with the results of the strong and weak coupling expansions as well as Monte Carlo simulation results, and we find complete consistency between the results. The strong and weak coupling expansions are reviewed in Appendix~\ref{app:expansion}.

As in the 2D case, we first consider the SU(2) theory, followed by the SU(3) construction. We will consider three positivity matrices: Hermitian positivity and two reflection positivities, with $x=0$ (denoted $R_1$)  and $x=z$ (denoted $R_2$) as the mirror planes.

\subsection{3D SU(2)}

\paragraph{Path choice 1.}
We consider the following set of plane-type path configurations as 
\begin{equation}
\adjustbox{valign=c}{\includegraphics[width=.95\linewidth]{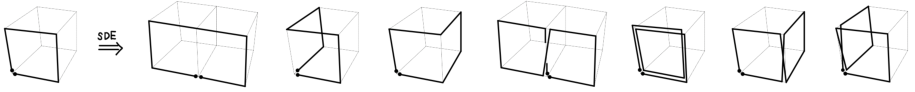} }
\label{eqfig:3Dpathsde1}
\end{equation} 
which are generated by applying SD equation on a plaquette path.
Here, the vertical direction is the $y$ direction, or equivalently, the edge `b' direction; and we apply SDE only on the `b' edges to generate the plane-type paths.
After applying symmetry and conjugation for the 8 configurations in \eqref{eqfig:3Dpathsde1}, and also including the identity matrix as a path, we have in total of 85 plane-type paths.

\begin{figure}[t]
\centering
\subfigure[Path choice 1.]{\includegraphics[width=.45\linewidth]{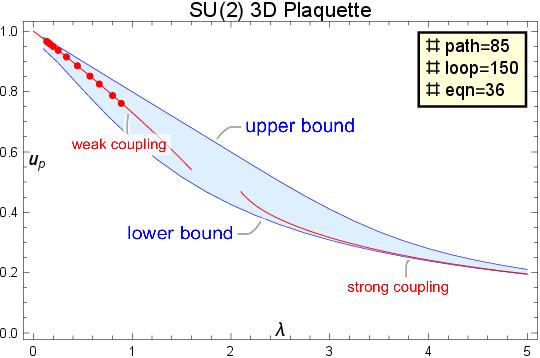}}
\qquad
\subfigure[Path choice 2.]{\includegraphics[width=.45\linewidth]{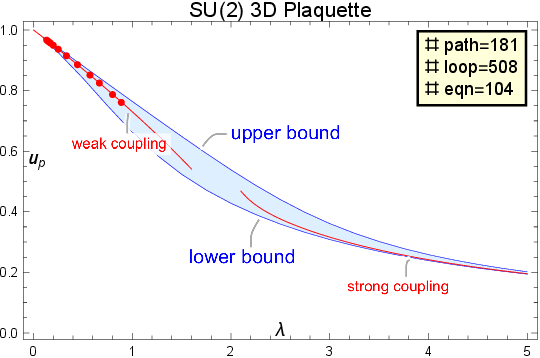}}
\qquad
\subfigure[Path choice 3 and 4.]{\includegraphics[width=.45\linewidth]{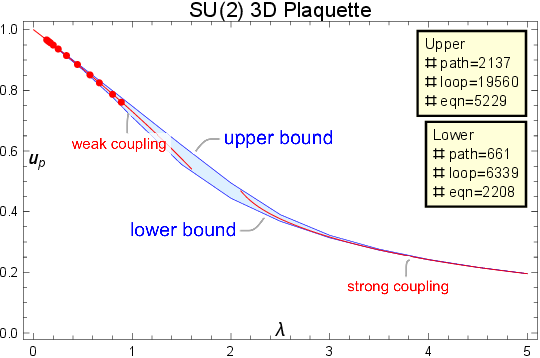}}
\qquad
\subfigure[Combined plots.]{\includegraphics[width=.45\linewidth]{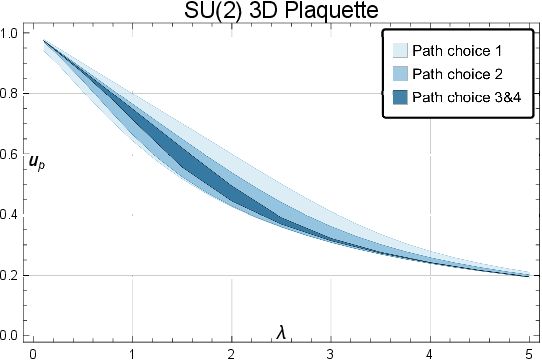}}
\caption{SDP bounds for SU(2) plaquette in 3D. Red curves are weak and strong coupling expansion (see Appendix~\ref{app:expansion}), and the red dots are the MC data \cite{Athenodorou:2016ebg}.}
\label{fig:3Dsu2plot}
\end{figure}

By sewing the paths we construct the three positivity matrices with dimensions given by
\begin{equation}
\text{dim}({\cal M}): \ \{85\}_H \cup \{46\}_{R_1} \cup \{30\}_{R_2} \,,
\end{equation}
where $R_1$ and $R_2$ denote the reflection positivities with $x=0$ and $x=z$ mirror planes respectively.
After symmetry decomposition, the reduced positivity matrices have dimensions:
\begin{equation}
\text{dim}({\cal M}): \ \{9, 3, 4, 7, 10, 6, 4, 3, 7, 11\}_H \cup \{13, 10, 12, 11\}_{R_1} \cup \{8, 7, 7, 8\}_{R_2} \,.
\end{equation}
The number of involved loop variables is 150, subject to 36 loop equations.
Solving SDP using {\tt Mathematica}, we obtain the bounds as shown in Figure~\ref{fig:3Dsu2plot}(a).

\paragraph{Path choice 2.}

The bounds can have a notable improvement by simply adding following length-6 path configurations:
\begin{equation}
\begin{gathered} {\includegraphics[width=.9\linewidth]{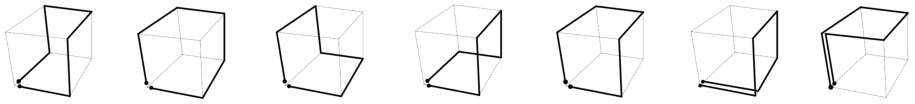} } \end{gathered} .
\label{eqfig:3DpathaddL6}
\end{equation} 
After applying symmetry and conjugation, they give 32 more ladder-type paths. 
There are in total of 181 paths.

By sewing the paths we construct the three positivity matrices with dimensions given by
\begin{equation}
\text{dim}({\cal M}): \ \{181\}_H \cup \{94\}_{R_1} \cup \{58\}_{R_2} \,.
\end{equation}
After symmetry decomposition, the dimensions of reduced matrices are 
\begin{equation}
\text{dim}({\cal M}): \ \{16, 8, 11, 12, 22, 11, 11, 8, 14, 23\}_H \cup \{25, 22, 24, 23\}_{R_1} \cup \{16, 13, 13, 16\}_{R_2} \,.
\end{equation}
The number of involved loop variables is 508, subject to 101 loop equations.
Solving SDP using {\tt Mathematica}, we obtain the bounds as shown in Figure~\ref{fig:3Dsu2plot}(b).
We can see a good improvement for the upper bound and also for the lower bound at weak coupling.

\paragraph{Path choice 3.}
Our next goal is to improve the lower bound. As implied in the 2D case, we can consider paths with longer lengths, without increasing the scale.
Together with the previous choice \eqref{eqfig:3Dpathsde1} and \eqref{eqfig:3DpathaddL6}, we add the following path configurations:
\begin{itemize}
\item 
Acting SDE on the length-6 paths in \eqref{eqfig:3Dpathsde1}, we get four more length-8 path configurations.
\begin{equation}
\begin{gathered} {\includegraphics[height=1.7cm]{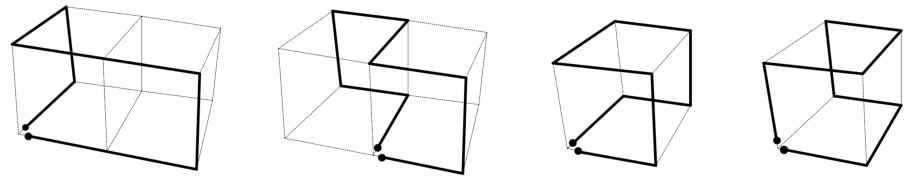} } \end{gathered} \ .
\label{eqfig:3DpathmoreL8}
\end{equation}
\item 
To get length-10 and length-12 type paths, we choose to act SDE on the four types of length-8 paths in \eqref{eqfig:3Dpathsde1}. Moreover, we act only on the `b' edges that are located at the origin. In this way, we get 15 types of length-10 and 14 types of length-12 path configurations.\footnote{It is straightforward exercise to produce these configuration by using SD equations and we will not present them here.}

\end{itemize}
After applying symmetry and conjugation, there are in total of 661 paths.

The three positivity matrices have dimensions:
\begin{equation}
\text{dim}({\cal M}): \ \{661\}_H \cup \{276\}_{R_1} \cup \{124\}_{R_2} \,,
\end{equation}
and the reduced positivity matrices have dimensions: 
\begin{equation}
\text{dim}({\cal M}): \ \{50, 34, 38, 45, 82, 44, 38, 34, 48, 83\}_H \cup \{71, 67, 70, 68\}_{R_1} \cup \{33, 29, 29, 33\}_{R_2} \,.
\end{equation}
The number of involved loop variables is 6339, subject to 2189 loop equations.

Using MOSEK, each data point is computed in about 40 seconds using 1 GB of memory.
We obtain the lower bound as shown in Figure~\ref{fig:3Dsu2plot}(c).
We see that the lower bound has a good match with both the weak and strong coupling expansion.

\paragraph{Path choice 4.}
We now present an improved upper bound. 
As implied in the 2D case, we extend the Choice 3 to general-type paths, by combining \eqref{eqfig:3Dpathsde1}, \eqref{eqfig:3DpathaddL6} and \eqref{eqfig:3DpathmoreL8}, along with the length-10 type paths mentioned below \eqref{eqfig:3DpathmoreL8}. 
After applying full 3D symmetry and conjugation, we obtain a total of 2137 paths.

The three positivity matrices have dimensions:
\begin{equation}
\text{dim}({\cal M}): \ \{2137\}_H \cup \{916\}_{R_1} \cup \{404\}_{R_2} \,,
\end{equation}
and the dimensions of reduced positivity matrices  are 
\begin{align}
\text{dim}({\cal M}): \ & 
\{29, 23, 21, 18, 51, 39, 71, 66, 68, 61, 22, 27, 18, 21, 49, 39, 68, 62, 71, 67\}_H  \\
& \cup \{64, 51, 56, 59, 114, 57, 57, 52, 62, 115\}_{R_1} \cup \{54, 51, 49, 48, 48, 49, 53, 52\}_{R_2} \,. \nonumber
\end{align}
The number of involved loop variables is 19560, constrained by 5229 loop equations.

Using MOSEK, each data point is computed in approximately 1000 seconds and requires about 10 GB of memory.
The resulting upper bound is presented in Figure~\ref{fig:3Dsu2plot}(c), which has a notable improvement compared to Choice 3 with plane-type truncation.
However, the lower bound in Choice 4 is less restrictive than that in Choice 3, even though Choice 4 involves a larger SDP system and is more time-consuming to solve.
This difference supports our strategy of optimizing lower and upper bounds by selecting different paths.\footnote{
The bounds for 3D SU(2) plaquette were also investigated in \cite{Kazakov:2024ool} using a length-truncation strategy. In that case, they achieved a tighter bound with 93561 independent SDP variables and a set of matrices of size about 200, which are a substantially larger system than ours. Since our truncation is a subset of the choice in \cite{Kazakov:2024ool}, our bounds is weaker but reasonably close to their results. As our primary focus is to extend the analysis to SU(3) and four dimensions, we do not pursue to further improve the bounds here.
}

\subsection{3D SU(3)}

To study the SU(3) theory, we generalize the choice of simple paths in SU(2) by including composite paths.

\paragraph{Path choice 1.}
The simple paths are the same as obtained from \eqref{eqfig:3Dpathsde1}. To unify the setup, these simple paths may be taken as a special class of composite paths that are combined with a trivial `identity' loop, and we denote them as
\begin{equation}
(\text{path}_{L\leq8}) \otimes (\text{loop}_{L=0}) \,,
\end{equation}
which contain 85 paths after applying symmetry and conjugation.

To introduce new composite paths, we consider following two classes. The first class is to combine length-4 paths with length-4 loops, which is denoted as
\begin{equation}
(\text{path}_{L=4}) \otimes (\text{loop}_{L=4}) \,. 
\end{equation}
They can be obtained by applying symmetry and conjugation to the following composite path configurations: 
\begin{equation}
\begin{gathered} {\includegraphics[height=1.7cm]{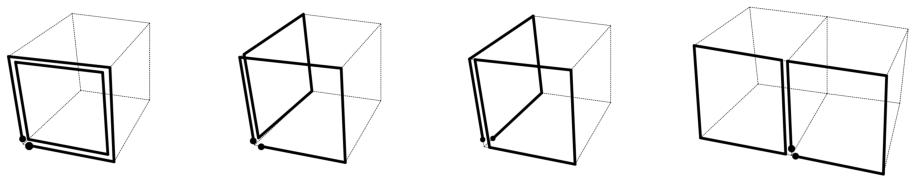} } \end{gathered} .
\label{eqfig:doublepathL4L4}
\end{equation}
This gives in total of $64=8\times8$ composite paths.

The second class is the combination of the length-0 path (the identity matrix) and Wilson loops given by taking the trace of the simple paths from \eqref{eqfig:3Dpathsde1}:
\begin{equation}
(\text{path}_{L=0}) \otimes (\text{loop}_{L\leq8}) \,,
\end{equation}
which give $84$ paths.
In total there are $233=85+64+84$ paths. 

\begin{figure}[t]
\centering
\subfigure[Path choice 1.]{\includegraphics[width=.45\linewidth]{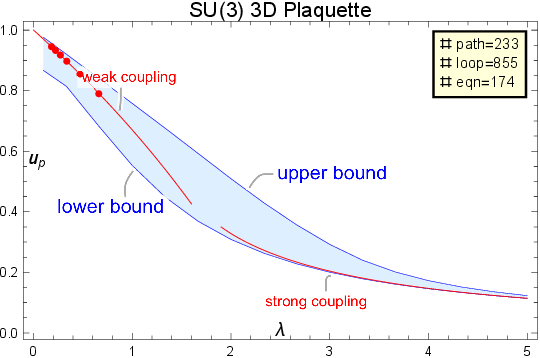}}
\qquad
\subfigure[Path choice 2.]{\includegraphics[width=.45\linewidth]{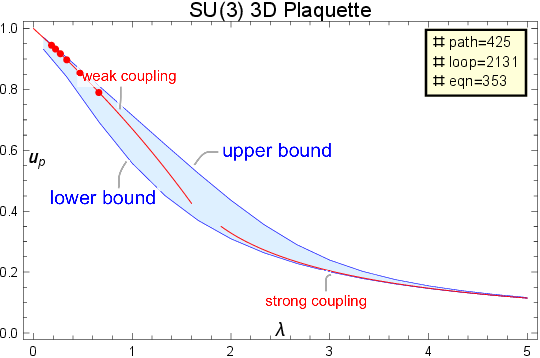}}
\qquad
\subfigure[Path choice 3 and 4.]{\includegraphics[width=.45\linewidth]{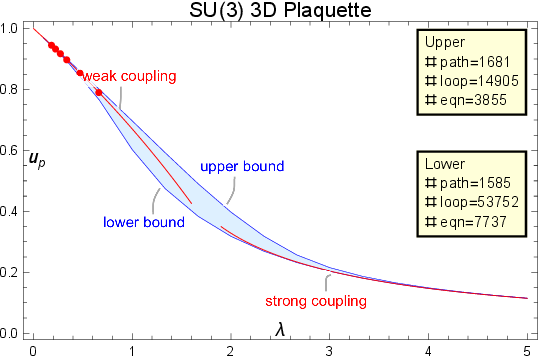}}
\qquad
\subfigure[Combined all plots.]{\includegraphics[width=.45\linewidth]{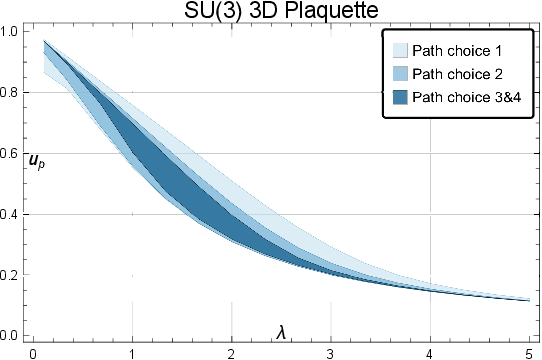}}
\caption{SDP bounds for SU(3) plaquette in 3D. Red curves are weak and strong coupling expansion (see Appendix~\ref{app:expansion}), and the red dots are the MC data \cite{Athenodorou:2016ebg}.}
\label{fig:3Dsu3plot}
\end{figure}

As in the SU(2) case, we construct three positivity matrices with dimensions given by
\begin{equation}
\text{dim}({\cal M}): \ \{233\}_H \cup \{128\}_{R_1} \cup \{76\}_{R_2} \,.
\end{equation}
The dimensions of matrices after symmetry decomposition are 
\begin{equation}
\text{dim}({\cal M}): \ 
\{23, 8, 10, 20, 28, 18, 10, 8, 20, 30\}_H  
\cup \{36, 28, 34, 30\}_{R_1} \cup \{20, 18, 18, 20\}_{R_2} \,. 
\end{equation}
The number of involved loop variables is 855, subject to 174 loop equations.\footnote{In the counting of loops variables, we reduce all triple-trace loops into independent double and single-trace loops by using SU(3) reduction formulae.}
Solving SDP using {\tt Mathematica}, we obtain the bounds as given in Figure~\ref{fig:3Dsu3plot}(a).

\paragraph{Path choice 2.}

The simple paths are the same as SU(2) case by combining \eqref{eqfig:3Dpathsde1} and \eqref{eqfig:3DpathaddL6},
Similarly, we can denote these simple paths as
\begin{equation}
(\text{path}_{L\leq8}) \otimes (\text{loop}_{L=0}) \,.
\end{equation}
The composite paths contain length-4 paths times length-4 loops which is the same as in the choice 1, together with the length-0 path combined with length-8 loops:
\begin{equation}
(\text{path}_{L=4}) \otimes (\text{loop}_{L=4}) \,, \qquad (\text{path}_{L=0}) \otimes (\text{loop}_{L\leq8}) \,.
\end{equation}
In total, there are $425=181+64+180$ paths. 

The three positivity matrices have dimensions:
\begin{equation}
\text{dim}({\cal M}): \ \{425\}_H \cup \{224\}_{R_1} \cup \{132\}_{R_2} \,,
\end{equation}
and the reduced positivity matrices have dimensions as 
\begin{equation}
\text{dim}({\cal M}): \ 
\{37, 18, 24, 30, 52, 28, 24, 18, 34, 54\}_H  
\cup \{60, 52, 58, 54\}_{R_1} \cup \{36, 30, 30, 36\}_{R_2} \,. 
\end{equation}
The number of involved loop variables is 2131, subject to 353 loop equations.

Solving SDP using {\tt Mathematica}, we obtain the bounds as shown in Figure~\ref{fig:3Dsu3plot}(b).
We see a notable improvement for the upper bound and also the lower bound in the weak coupling.

\paragraph{Path choice 3.}

The simple paths are the same as in previous choice 3 in SU(2), which includes up to length-12 paths.
We take the simple paths as
\begin{equation}
(\text{path}_{L\leq12}) \otimes (\text{loop}_{L=0}) \,.
\end{equation}
We add new composite paths. One class is the length-8 paths in \eqref{eqfig:3Dpathsde1} combined with length-4 plaquette loops:
\begin{equation}
(\text{path}_{L\leq8}) \otimes (\text{loop}_{L=4}) \,, 
\end{equation}
which gives $672=84\times8$ paths.
Another class of composite paths is the length-0 path combined with loops up to length-8 (combining \eqref{eqfig:3Dpathsde1} and \eqref{eqfig:3DpathmoreL8}):
\begin{equation}
(\text{path}_{L=0}) \otimes (\text{loop}_{L\leq8}) \,,
\end{equation}
which give $252$ paths.
In total, $1585=661+672+252$ paths. 

The three positivity matrices have dimensions:
\begin{equation}
\text{dim}({\cal M}): \ \{1585\}_H \cup \{674\}_{R_1} \cup \{316\}_{R_2} \,,
\end{equation}
and the dimensions of symmetry-reduced positivity matrices are 
\begin{align}
\text{dim}({\cal M}): \ 
\{119, 81, 89, 110, 197, 108, 89, 81, 116, 199\}_H  
\cup \{174, 163, 172, 165\}_{R_1} \cup \{83, 75, 75, 83\}_{R_2} \,. 
\end{align}
The number of involved loop variables is 53752, subject to 7737 loop equations.

Applying MOSEK with a single thread takes about 2 hours and 50 GB of memory for each data point.
We obtain the lower bound as shown in Figure~\ref{fig:3Dsu3plot}(c), which have a notable improvement in particular at weak coupling.

\paragraph{Path choice 4.}

The simple paths are the same as in previous choice 4 in SU(2).
We denote the simple paths as
\begin{equation}
(\text{path}_{L\leq8}) \otimes (\text{loop}_{L=0}) \,.
\end{equation}
The new composite paths are length-4 paths combined with length-4 loops, and the length-0 path combined with length-8 loops:
\begin{equation}
(\text{path}_{L\leq4}) \otimes (\text{loop}_{L\leq4}) \,, \qquad (\text{path}_{L=0}) \otimes (\text{loop}_{L\leq8}) \,,
\end{equation}
which gives $576=24\times24$ and $552$ paths, respectively.
In total, $1681=553+576+552$ paths. 

The three positivity matrices have dimensions:
\begin{equation}
\text{dim}({\cal M}): \ \{1681\}_H \cup \{824\}_{R_1} \cup \{428\}_{R_2} \,,
\end{equation}
and the symmetry-reduced positivity matrices have dimensions:
\begin{align}
\text{dim}({\cal M}): \ & 
\{11, 5, 6, 3, 15, 9, 20, 18, 17, 13, 4, 9, 3, 6, 13, 9, 17, 14, 20, 19\}_H  \\
& \cup \{23, 13, 18, 18, 35, 16, 19, 14, 21, 36\}_{R_1} \cup \{24, 21, 19, 18, 18, 19, 23, 22\}_{R_2} \,. \nonumber
\end{align}
The number of involved loop variables is 14905, subject to 3855  loop equations.

Using MOSEK, each data point takes approximately 11 minutes and requires 5 GB of memory.
The resulting upper bound is shown in Figure~\ref{fig:3Dsu3plot}(c), which shows a notable improvement over Choice 2. 

Our results of the 3D SU(3) YM theory demonstrate a clear convergence of the bounds as the number of paths increases in forming a large constraining system.
The resulting bounds are fully consistent with the perturbation expansion results at both weak and strong coupling. In particular, we observe that the upper bound is tighter at weak coupling, while the lower bound is more restrictive at strong coupling. This feature will be more obvious in the 4D case.

\subsection{4D SU(2)}

The 4D theory is much more challenging since the number of paths and loop variables increases significantly.
Therefore, it is crucial to keep a reasonably small set of paths.

\paragraph{Path choice 1.}
As in the 3D case, we first consider a simple set of path configurations up to length 8, which are generated by applying SDE only on the `b' edges of the plaquette. They are the same as in  \eqref{eqfig:3Dpathsde1}.
After applying 4D plane-type symmetry and conjugation, these give a total of 211 plane-type paths. 
Note that the identity matrix is always included as a path.

\begin{figure}[t]
\centering
\subfigure[Path choice 1.]{\includegraphics[width=.45\linewidth]{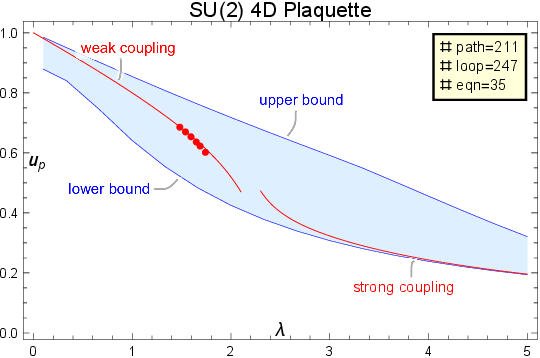}}
\qquad
\subfigure[Path choice 2.]{\includegraphics[width=.45\linewidth]{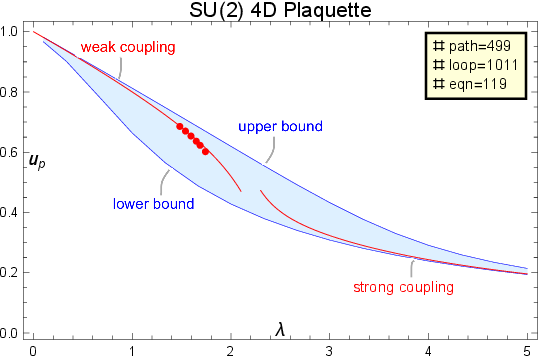}}
\qquad
\subfigure[Path choice 3 and 4.]{\includegraphics[width=.45\linewidth]{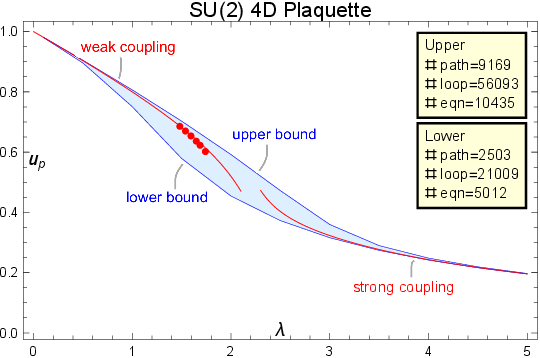}}
\qquad
\subfigure[Combined all plots.]{\includegraphics[width=.45\linewidth]{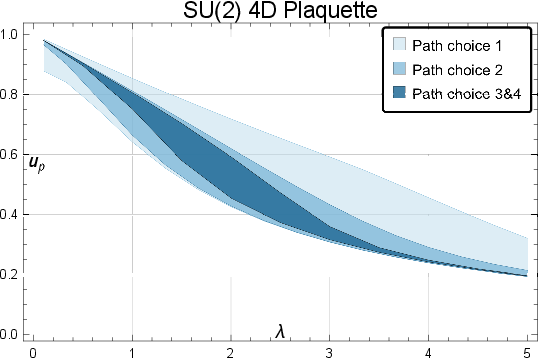}}
\caption{SDP bounds for SU(2) plaquette in 4D. Red curves are weak and strong coupling expansion (see Appendix~\ref{app:expansion}), and the red dots are the MC data \cite{Athenodorou:2021qvs}.}
\label{fig:4Dsu2plot}
\end{figure}

We construct three positivity matrices with dimensions given by
\begin{equation}
\text{dim}({\cal M}): \ \{211\}_H \cup \{144\}_{R_1} \cup \{100\}_{R_2} \,.
\end{equation}
The dimensions of symmetry-reduced positivity matrices are 
\begin{align}
\text{dim}({\cal M}): \ &\{9, 3, 11, 10, 4, 7, 3, 6, 4, 10, 11, 3, 7, 4\}_H \\
& \cup \{17, 3, 4, 14, 17, 15, 4, 3, 14, 18\}_{R_1} \cup \{19, 10, 14, 7, 17, 11, 15, 7\}_{R_2} \,. \nonumber
\end{align}
The number of involved loop variables is 247, subject to 35 loop equations.
Solving SDP using {\tt Mathematica}, we obtain the bounds as shown in Figure~\ref{fig:4Dsu2plot}(a).

\paragraph{Path choice 2.}

We consider adding more length-6 path configurations by combining \eqref{eqfig:3Dpathsde1} and \eqref{eqfig:3DpathaddL6}.
After applying symmetry and conjugation, there are in total of 499 paths.
The three positivity matrices with dimensions given by
\begin{equation}
\text{dim}({\cal M}): \ \{499\}_H \cup \{336\}_{R_1} \cup \{224\}_{R_2} \,,
\end{equation}
and the symmetry-reduced matrices have dimensions:
\begin{align}
\text{dim}({\cal M}): \ &\{16, 8, 23, 22, 11, 19, 8, 11, 11, 22, 23, 8, 19, 11\}_H \\
& \cup \{36, 8, 11, 31, 41, 32, 11, 8, 33, 42\}_{R_1} \cup \{39, 22, 32, 19, 35, 23, 35, 19\}_{R_2} \,. \nonumber
\end{align}
The number of involved loop variables is 1011, subject to 119 loop equations.
Solving SDP using {\tt Mathematica}, we obtain the bounds as shown in Figure~\ref{fig:4Dsu2plot}(b).
A notable improvement can be observed for the upper bound, as well as the lower bound at small coupling.

\paragraph{Path choice 3.}
Together with the previous choices, we add the following path configurations:
\begin{itemize}
\item 
Acting SDE to the length-6 path configurations in \eqref{eqfig:3Dpathsde1}, we obtain 6 new length-8 path configurations. Four of these coincide with the 3D paths in \eqref{eqfig:3DpathmoreL8}, while the remaining two are unique to the 4D case.
\item 
To generate length-10 and length-12 type paths, we act SDE on the length-8 path configurations in \eqref{eqfig:3Dpathsde1}, restricting the operation to the `b' edges located at the origin. This procedure yields 18 length-10 and 16 length-12 type plane-type path configurations.

\end{itemize}
After incorporating symmetry and conjugation, there are in total of 2503 paths.

The three positivity matrices with dimensions given by
\begin{equation}
\text{dim}({\cal M}): \ \{2503\}_H \cup \{1442\}_{R_1} \cup \{828\}_{R_2} \,,
\end{equation}
and the dimensions of the symmetry-reduced matrices are
\begin{align}
\text{dim}({\cal M}): \ &\{56, 41, 7, 7, 96, 14, 102, 58, 92, 55, 51, 44, 7, 7, 95, 14, 103, 55, 92, 58\}_H \\
 \cup & \{128, 55, 58, 122, 179, 124, 58, 55, 124, 180\}_{R_1} \cup \{126, 93, 112, 83, 122, 94, 115, 83\}_{R_2} \,. \nonumber
\end{align}
The number of involved loop variables is 21009, subject to 5012 loop equations.

Using MOSEK, each data point is computed in about 30 minutes using 8.5 GB of memory.
Solving SDP we obtain the lower bound as shown in Figure~\ref{fig:4Dsu2plot}(c).
A notable improvement can be seen compared to Figure~\ref{fig:4Dsu2plot}(b).

\paragraph{Path choice 4.}
Now we consider general-type paths to improve the upper bound. We combine the previous choice \eqref{eqfig:3Dpathsde1} and \eqref{eqfig:3DpathaddL6}, and also the 18 length-10 path configurations considered in Choice 3. 
After applying full 4D symmetry and conjugation, we obtain a total of 9169 paths.
We see that the number of paths increases significantly comparing to the planar-type truncation.

The three positivity matrices have dimensions:
\begin{equation}
\text{dim}({\cal M}): \ \{9169\}_H \cup \{5070\}_{R_1} \cup \{2732\}_{R_2} \,,
\end{equation}
and the dimensions of reduced positivity matrices  are 
\begin{align}
\text{dim}({\cal M}): \ & 
\{29, 3, 3, 18, 6, 46, 9, 9, 62, 72, 30, 60, 27, 74, 66, 75, 69, 70, 134, 57, 22, 3, 3, 21, 6, 43,   \nonumber\\
& \ \  9, 9,66, 67, 27, 63, 30, 71, 69, 71, 66, 76, 134, 57\}_H  \\
\cup & \{87, 72, 30, 27, 159, 57, 187, 136, 178, 128, 77, 79, 27, 30, 156, 57, 185, 129, 181, 137\}_{R_1} \nonumber\\
\cup & \{117, 63, 68, 110, 177, 105, 59, 62, 102, 163, 105, 69, 64, 110, 175, 106, 62, 59, 107, 167\}_{R_2} . \nonumber
\end{align}
There are 56093 loop variables, constrained by 10435 loop equations.
We find that, the Hermitian positivity matrices after the symmetry decomposition are very complicated, partly due to the complexity of the representations for the $B_4$ group. In practice, we only use the reflection positivity conditions to obtain the bounds.

Using MOSEK, each data point is computed in about 7 hours using 64 GB of memory.
Solving SDP we obtain the upper bound as shown in Figure~\ref{fig:4Dsu2plot}(c).
We see that there is a good improvement compared to the choice 3 with ladder-type paths.
On the other hand, the lower bound in Choice 4 is less constraining than in Choice 3, despite its larger system size.

We find that the upper bound matches nicely with the weak coupling expansion, while the lower bound provides a very good approximation at strong coupling.\footnote{As in the 3D SU(2) case, data points of tighter bounds (with significantly larger positivity matrices) for 4D SU(2) plaquette were presented in \cite{Kazakov:2024ool}.}

\subsection{4D SU(3)}

Finally, we consider the most challenging 4D SU(3) case.

\paragraph{Path choice 1.}
The simple paths are the same as in the SU(2) case.
The composite paths are length-4 paths combined with length-4 loops, and the length-0 path combined with length-8 loops:
\begin{equation}
(\text{path}_{L\leq4}) \otimes (\text{loop}_{L\leq4}) \,, \qquad (\text{path}_{L=0}) \otimes (\text{loop}_{L\leq8}) \,,
\end{equation}
which gives $144=12\times12$ and $210$ paths, respectively.
In total, $565=211+144+210$ paths. 

As in the SU(2) case, we construct three positivity matrices with dimensions given by
\begin{equation}
\text{dim}({\cal M}): \ \{565\}_H \cup \{388\}_{R_1} \cup \{264\}_{R_2} \,.
\end{equation}
The dimensions of symmetry-reduced matrices are 
\begin{align}
\text{dim}({\cal M}): \ &\{23, 8, 30, 28, 10, 18, 8, 18, 10, 28, 30, 8, 18, 10\}_H \\
& \cup \{46, 8, 10, 38, 46, 42, 10, 8, 38, 48\}_{R_1} \cup \{50, 28, 36, 18, 46, 30, 38, 18\}_{R_2} \,. \nonumber
\end{align}
The number of involved loop variables is 1385, subject to 195 loop equations.

\begin{figure}[t]
\centering
\subfigure[Path choice 1.]{\includegraphics[width=.45\linewidth]{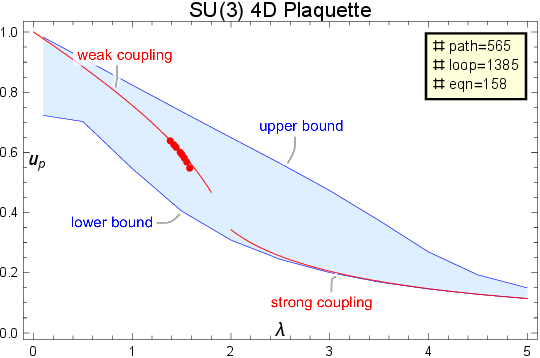}}
\qquad
\subfigure[Path choice 2.]{\includegraphics[width=.45\linewidth]{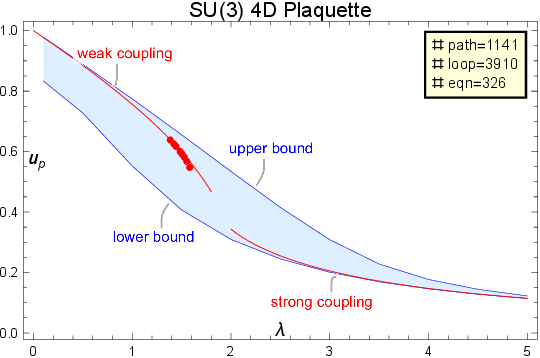}}
\qquad
\subfigure[Path choice 3 and 4.]{\includegraphics[width=.45\linewidth]{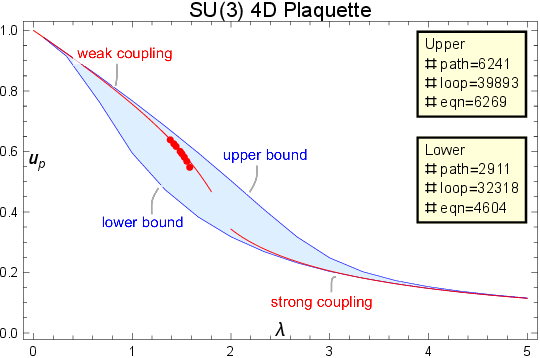}}
\qquad
\subfigure[Combined all plots.]{\includegraphics[width=.45\linewidth]{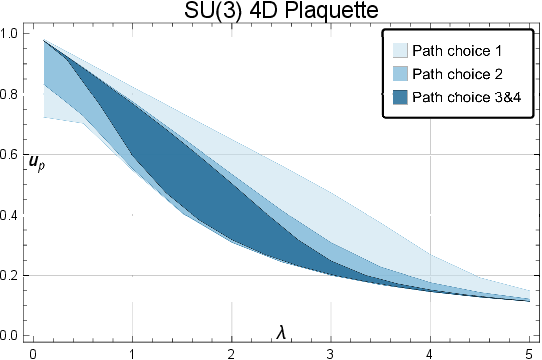}}
\caption{SDP bounds for SU(3) plaquette in 4D. Red curves are weak and strong coupling expansion (see Appendix~\ref{app:expansion}), and the red dots are the MC data \cite{Athenodorou:2021qvs}.}
\label{fig:4Dsu3plot}
\end{figure}

Using MOSEK, it takes about 4 seconds for a data point.
Solving SDP we obtain the bounds as given in Figure~\ref{fig:4Dsu3plot}(a).

\paragraph{Path choice 2.}

The simple paths are the same as in the SU(2) case.
The new composite paths are length-4 paths times length-4 loops, and the length-0 path times length-8 loops:
\begin{equation}
(\text{path}_{L\leq4}) \otimes (\text{loop}_{L\leq4}) \,, \qquad (\text{path}_{L=0}) \otimes (\text{loop}_{L\leq8}) \,,
\end{equation}
in which the number of length-8 loops is 498.
In total, $1141=499+144+498$ paths. 

The three positivity matrices with dimensions given by
\begin{equation}
\text{dim}({\cal M}): \ \{1141\}_H \cup \{772\}_{R_1} \cup \{512\}_{R_2} \,,
\end{equation}
and the symmetry-reduced matrices have dimensions as 
\begin{align}
\text{dim}({\cal M}): \ &\{37, 18, 54, 52, 24, 42, 18, 28, 24, 52, 54, 18, 42, 24\}_H \\
& \cup \{84, 18, 24, 72, 94, 76, 24, 18, 76, 96\}_{R_1} \cup \{90, 52, 72, 42, 82, 54, 78, 42\}_{R_2} \,. \nonumber
\end{align}
The number of involved loop variables is 3910, subject to 516 loop equations.

Using MOSEK, it takes about 30 seconds for a data point.
Solving SDP we obtain the bounds as shown in Figure~\ref{fig:4Dsu3plot}(b).

\paragraph{Path choice 3.}

To avoid an explosion of the number of paths, we make a truncation for the length-10 and length-12 paths compared to the SU(2) case. 
Consider first the simple paths.
We take the plaquette path as a seed, generating first the length-6 and length-8 paths via SDE, then from the length-8 paths, we further generate length-10 and length-12 paths. The strategy can be illustrated as 
\begin{equation}
	\adjustbox{valign=c}{\includegraphics[width=0.7\linewidth]{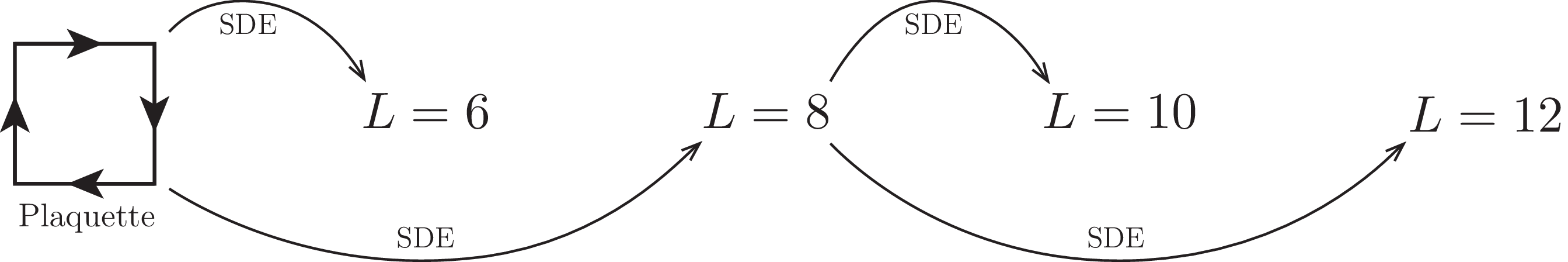} } .
	\label{eq:4DSU3choice3}
\end{equation}
Here we restrict the paths to the plane type, and the SDE is also only applied to `b' edges located at the origin.
We also include all other length-6 paths that appeared in choice 2.
This gives in total of 1543 simple paths.

For the composite paths, we consider the length-8 paths combined with length-4 loops, 
\begin{equation}
(\text{path}_{L\leq8}) \otimes (\text{loop}_{L\leq4}) \,, 
\end{equation}
where the length-8 paths contain only those generated by SDE in \eqref{eq:4DSU3choice3}. This gives $1368 = 144\times 12$ paths.
In total, we consider $2911=1543+1368$ paths.

The three positivity matrices with dimensions given by
\begin{equation}
\text{dim}({\cal M}): \ \{2911\}_H \cup \{1714\}_{R_1} \cup \{912\}_{R_2} \,,
\end{equation}
and the dimensions of matrices after symmetry decomposition is 
\begin{align}
\text{dim}({\cal M}): \ &\{76, 47, 6, 6, 122, 12, 127, 62, 103, 59, 71, 50, 6, 6, 121, 12, 128, 59, 103, 62\}_H \\
\cup \{161, & 59, 62, 151, 212, 157, 62, 59, 153, 213\}_{R_1} \cup \{145, 109, 117, 85, 141, 110, 120, 85\}_{R_2} \,. \nonumber
\end{align}
The number of involved loop variables is 32318, subject to 4604 loop equations.

Using MOSEK, each data point is computed in about 40 minutes using 18 GB of memory.
Solving SDP we obtain the lower bound as shown in Figure~\ref{fig:4Dsu3plot}(c).
We can see an obvious improvement for the lower bound in particular at the weak coupling.

\paragraph{Path choice 4.}

In this final choice, we consider general-type paths to improve the upper bound.
The simple paths are the same as in the SU(2) case.
The new composite paths are length-4 paths times length-4 loops, and  length-0 paths times length-8 loops:
\begin{equation}
(\text{path}_{L\leq4}) \otimes (\text{loop}_{L\leq4}) \,, \qquad (\text{path}_{L=0}) \otimes (\text{loop}_{L\leq8}) \,,
\end{equation}
which gives $2304=48\times 48$ and $1968$ paths, respectively.
In total, there are $6241=1969+2304+1968$ paths. 

We obtain the three positivity matrices with dimensions given by
\begin{equation}
\text{dim}({\cal M}): \ \{6241\}_H \cup \{3780\}_{R_1} \cup \{2268\}_{R_2} \,.
\end{equation}
After full 4D symmetry decomposition, the dimensions of matrices are 
\begin{align}
\text{dim}({\cal M}): \ & 
\{32, 1, 1, 11, 2, 42, 1, 1, 43, 63, 18, 38, 12, 62, 40, 58, 44, 42, 100, 30, 14, 16, \\
& \quad 30, 2, 2, 56, 54, 12, 44, 18, 56, 44, 44, 40, 60, 100, 30\}_H  \nonumber\\
& \cup \{84, 52, 18, 12, 136, 30, 150, 100, 134, 80, 60, 70, 12, 18, 130, 30, 146, 82, 140, 102\}_{R_1} \nonumber\\
& \cup \{120, 40, 56, 98, 152, 92, 38, 42, 88, 128, 88, 58, 42, 104, 148, 98, 42, 38, 96, 136\}_{R_2} \,. \nonumber
\end{align}
The number of involved loop variables is 39893, subject to 6269 loop equations.
We observe a significant growth in the number of loop variables compared to the plane-type truncation. 
Furthermore, the Hermitian positivity matrices are particularly complicated, partly due to the complexity of the representations for the $B_4$ group. 
In practice, to solve the SDP efficiently, we restrict ourselves to the reflection positivity conditions.

Using MOSEK, each data point requires approximately 2 hours of computation and 33 GB of memory
The resulting upper bound is shown in Figure~\ref{fig:4Dsu3plot}(c).

As in the previous cases, the above results for our most complicated 4D SU(3) case also demonstrate a clear convergences of the bounds. In particular, from Figure~\ref{fig:4Dsu3plot}(c) (and similarly in Figure~\ref{fig:4Dsu3plot}(b)), we observe that the upper bound is accurate for $\lambda <1$, while the lower bound holds well for $\lambda>2$. We leave further comment to the discussion section below.

\section{Discussion and outlook}
\label{sec:outlook}

In this study, we investigate the positivity bootstrap method in SU(3) lattice YM theory.
Unlike previously studied large N and SU(2) YM theories, which only involve single-trace operators, 
the inclusion of double-trace contributions significantly increases the number of path and loop variables, as well as the size of the positivity matrices. 
Despite these challenges, we observe clear convergence of the bounds. 
Notably, using a similar set of path configurations, the convergence behavior in SU(3) mirrors that of the SU(2) case, suggesting that the method is applicable to more general gauge theories.  

To simplify the construction, we introduce a dimension reduction strategy by considering the plane-type paths and loops. 
This approach significantly reduces the system's complexity while maintaining good convergence properties.
Moreover, our findings suggest that separating lower and upper bounds is beneficial, with plane-type paths being particularly effective for improving the lower bound.
We provide detailed descriptions of our construction, including the specific choices of paths, the dimensions of the positivity matrices and loop variables, as well as the resulting bounds. 
Derivations of loop equations in SU(3) theory, including those for double-trace loops, are also presented. 

\begin{figure}[t]
\centering
\subfigure[]{\includegraphics[width=.32\linewidth]{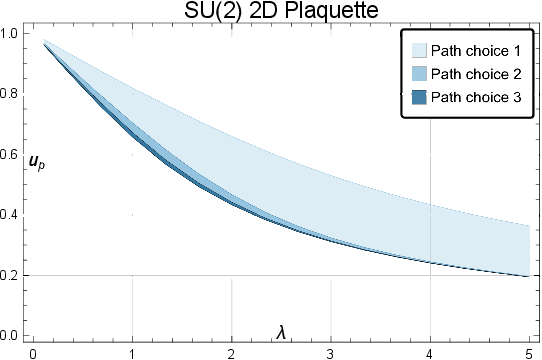}}
\ 
\subfigure[]{\includegraphics[width=.32\linewidth]{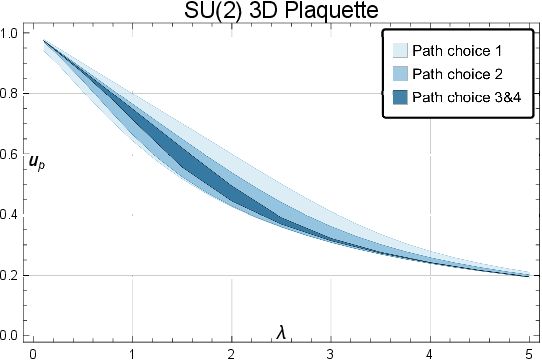}}
\ 
\subfigure[]{\includegraphics[width=.32\linewidth]{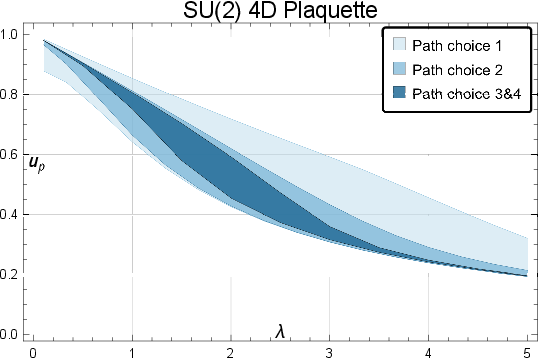}}

\subfigure[]{\includegraphics[width=.32\linewidth]{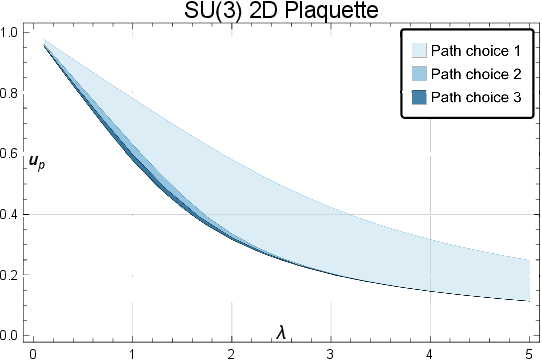}}
\ 
\subfigure[]{\includegraphics[width=.32\linewidth]{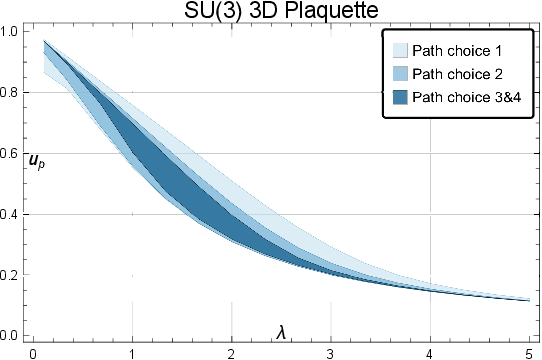}}
\ 
\subfigure[]{\includegraphics[width=.32\linewidth]{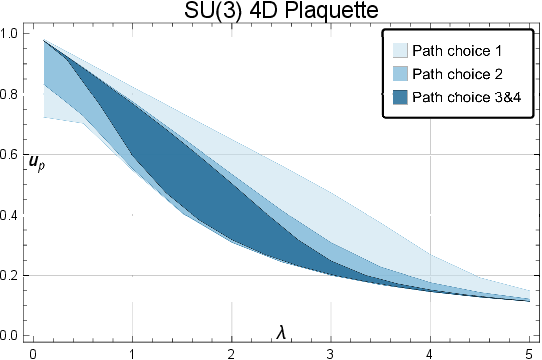}}
\caption{A collection of plots for the bounds of plaquette expectation value in SU(2) and SU(3) YM from 2D to 4D.}
\label{fig:allplots}
\end{figure}

Our studies span from the exactly solvable two-dimensional YM theory to the more complex four-dimensional SU(3) case. A collection of plots for the bounds of the plaquette expectation value is summarized in Figure~\ref{fig:allplots}. In all cases, we see a clear trend of convergence as the dimension of the positivity matrices increases. 

In the 2D case, we also introduce and prove a novel twist-reflection positivity condition, which imposes remarkably strong constraints. Although this condition is not exactly satisfied in higher dimensional theories, it may serve as a useful approximation, meriting further exploration.

For 3D and 4D YM theories, our primary goal is to demonstrate a proof of principle rather than achieve the optimal bounds. 
Enhancing the positivity bounds is a natural next step, by incorporating more general paths and including a more comprehensive set of loop equations. There are three key challenges: (1) how to select the most effective paths, (2) how to derive the full set of loop equations efficiently, and (3) how to solve large-scale SDP systems.  
These challenges are critical for advancing the bootstrap approach and should guide future investigations.

A fundamental question is how the convergence properties evolve as more paths and loops are included. 
Naturally, convergence properties vary significantly across dimensions, as revealed in our study.
In lower-dimensional theories (2D and 3D), relatively flat expectation value curves indicate the potential for achieving tight bounds. 
In contrast, in 4D SU(3) theory, the expectation value of the Wilson loop exhibits a notable transition when the coupling parameter $\lambda$ increases from $1$ to $2$.
This behavior aligns with the expected first-order phase transition for high $N$, as observed in lattice simulations \cite{Creutz:1981qr, Lucini:2005vg}, which poses challenges for maintaining tight bounds in this region. 
We find that upper bounds are more reliable before the transition, while lower bounds are more effective beyond the transition, a pattern observed in our 4D SU(3) results.

Besides the questions discussed above, there are many other directions to explore. 
Exploring more general types of Wilson loop operators could reveal deeper physical properties, such as verifying the area law or extracting the glueball spectrum. Additionally, investigating gauge theories with dynamical matter fields is an important next step, particularly in the context of QCD. We hope to address these questions in future studies.

\acknowledgments

It is a pleasure to thank Zhiming Cai, Ying Chen, Bo Feng, Feng-Kun Guo, Mingxing Luo, Jian-Ping Ma, Wei Sun, Yi-Bo Yang, and Qiang Zhao for discussions. 
This work is supported by the Chinese Academy of Sciences (Grant No. YSBR-101) and by the National Natural Science Foundation of China (Grants No.~12425504, 12175291, 12047503). 
We also acknowledge the support of the HPC Cluster of ITP-CAS.

\appendix

\section{Letter representation and symmetry decomposition}
\label{app:symreduction}

A loop can be represented by a list of ordered letters, each letter representing an edge in the lattice. This will be referred to as `letter representation'.
In the most complicated 4-dimensional lattice, we have eight types of edges 
\begin{equation}
\textrm{edges of 4-dimensional lattice}: \{\ap, \am, \bp, \bm, \cp, \cm, \dpp, \dmm \} \,,
\end{equation}
where `a,b,c,d' represent four dimensions, and `-' implies the flip of the direction.
The 2- and 3-dimensional lattices are subspace of the 4-dimensional lattice. For example, 
\begin{equation}
\textrm{edges of 2-dimensional lattice}: \{\ap, \am, \bp, \bm \} \,.
\end{equation}

As some examples of the loops:
\begin{align}
& \WL[\ap,\bp,\am,\bm] : \  \WLa{midarrow} \,, \quad
\WL[\ap, \ap, \bp, \bp, \ap, \bm, \am, \am, \am, \bm] : \ \WLn \,, 
\label{eq:2Dloopexamples} \\
& \WL[\ap, \ap, \bp, \ap, \bp, \cp, \am, \bm, \cm, \am, \cp, \am, \bm, \cm] : \  \begin{gathered} {\includegraphics[height=2.cm]{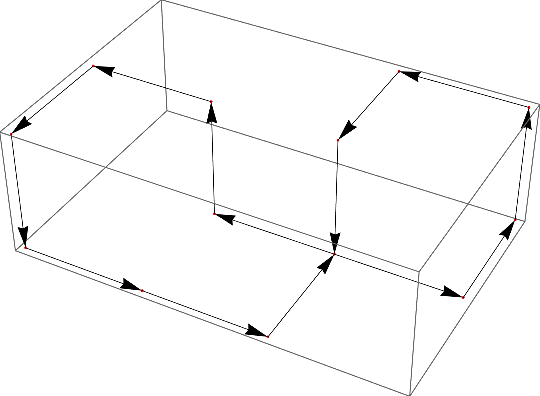} } \end{gathered} \,,
\end{align}
where the first two are 2-dimensional loops and the last one is a 3-dimensional one.
Similar representation applies to Wilson lines, or equivalently, the path operators.
It's important to note that Wilson loops are invariant under cyclic permutations of the letters, whereas the order of letters in a path operator is unique.

\paragraph{Symmetry and conjugation in the letter representation.}

The symmetry operations of a loop on the square lattice include rotation and reflection operators.
For example, the 2-dimensional lattice has following the 8 symmetry operations:
\begin{align}
	\{&\{\ap\to \ap, \bp\to \bp\},\{\ap\to \bp, \bp\to \ap\},\{\ap\to \ap, \bp\to \bm\},\{\ap\to \am, \bp\to \bp\}, \\
	&\{\ap\to \bp, \bp\to \am\},\{\ap\to \bm, \bp\to \ap\},\{\ap\to \am, \bp\to
   \bm\},\{\ap\to \bm, \bp\to \am\}\} \,, \nonumber
\end{align}
which form the dihedral group $D_4$ of 2-dimensional square lattice. 
It is straightforward to generalize to 3- and 4-dimensional lattices, and one has
\begin{align}
	& \textrm{2D lattice:~  8 operations} \,, \nonumber \\
	& \textrm{3D lattice:~  48 operations} \,, \nonumber \\
	& \textrm{4D lattice:~  384 operations} \,, \nonumber
\end{align}
where the number of symmetry operators increases significantly.

We comment that the expectation value of the Wilson loops operators is Hermitian conjugation invariant, i.e.,
\begin{equation}\label{eq:hermitianWL}
	\langle \WL \rangle = \langle \WL^\dagger \rangle \,,
\end{equation}
where hermitian conjugation of a given loop corresponds to reversing the order of the edges and flipping the direction of all edges in the letter representation, for example:
\begin{align}
	\WL[\ap,\bp, \am, \bm]^\dagger = \WL[\bp, \ap, \bm, \am] \,.
\end{align}

\paragraph{Symmetry decomposition.}
The symmetry decomposition of Wilson loop operators will significantly reduce the size of the semi-positive definite matrix when the set of paths possess symmetries. We briefly introduce this strategy, see also \cite{Kazakov:2022xuh, Kazakov:2024ool}.

More concretely, if the set of paths $\{ P_i\}$ is invariant under the actions of a certain group $G$, which is equivalent to
\begin{equation}
	\langle (g \circ O)|(g \circ O) \rangle = 
	\langle O|O \rangle \,, \qquad \forall g \in G \,,
\end{equation}
with $O$ a general linear combination of $P_i$ as $O = \sum_{i} c_i P_i$, the corresponding semi-positive definite matrix ${\cal M}$ can be decomposed by the irreducible representation of group $G$ as
\begin{equation}
	{\cal M} \to \bigoplus_k (\textrm{Rep}_k)^{\oplus m_k} \,,
\end{equation}
where with $m_k$ being the multiplicity of the $k$-th irreducible representation $\textrm{Rep}_k$. The dimension of the full matrix is
\begin{equation}
\textrm{dim}({\cal M}) = \sum_{k} d_k m_k \,,
\end{equation}
where $d_k$ is the dimension of the $k$-th irreducible representation.

To reach this decomposition of ${\cal M}$, one can introduce projection operators acting on \emph{paths} as
\begin{equation}
	p_{\alpha \beta,k} = \frac{d_k }{\textrm{dim}(G)} \sum_{g\in G} r^{(k)}_{\alpha \beta} (g^{-1}) \, g \,, \qquad \alpha, \beta = 1, \ldots , d_k\,,
\end{equation}
where $\textrm{dim}(G)$ counts the number of group elements and $r^{(k)}_{\alpha \beta}(g)$ are the matrix elements of the $k$-th irreducible representation for $g$. It will project the set of paths into the basis of different irreducible representations. GAP \cite{GAP4} is a convenient tool to obtain the irreducible representation matrix. For convenience, the representation matrix given by GPA should be transformed to a \emph{real} matrix through a matrix similarity transformation.

Importantly, the dimension of the semi-positive definite matrix can be reduced to $\sum_{k} m_k$ instead of $\sum_{k} d_k m_k$, due to the property of Invariant Semidefinite Programming \cite{Bachoc2012}. More precisely, it is enough to recover the constraints of the original matrix by selecting only one projection operator for each irreducible representation, \emph{e.g.}, $p_{1 1,k}$. See \cite{Kazakov:2024ool} for further details.

Finally, we comment on the symmetry group of paths. The symmetry group of any set of paths $\{P_i\}$ is always a subgroup of $B_n \times Z_2$, where $B_n$ is the hyperoctahedral group as the symmetry of $n$-dimensional lattice and $Z_2$ corresponds to the conjugation invariance of loop operators.
Specifically, for the plane-type paths introduced in Section~\ref{sec:planetypePath}, the maximal symmetry group is the subgroup $B_{n-1} \times Z_2$.
We summarize the symmetry groups involved in this paper in Table~\ref{tab:symmetrygroup}.

\begin{table}[t]
	\begin{center}
		\begin{tabular}{|c|c|c|} 
		\hline
			{Dimension} & {General path} & {Plane-type Path}\\
			\hline
			2 & $B_2$ $\times$ $Z_2$ & $Z_2$ $\times$ $Z_2$\\
			\hline
			3 & $B_3$ $\times$ $Z_2$ & $B_2$ $\times$ $Z_2$\\
			\hline
			4 & $B_4$ $\times$ $Z_2$ &
			$B_3$ $\times$ $Z_2$\\
			\hline
		\end{tabular}
	\end{center}
\caption{Symmetry groups: $G_{\rm lattice} \times Z_2$, where the $Z_2$ is due to the conjugation invariance. }
\label{tab:symmetrygroup}
	\end{table}

\section{SU(3) identities}
\label{app:su3identities}

In this appendix we provide various useful SU(3) identities based on the observation in \cite{Kazakov:2024ool}.
The SU(3) matrix satisfies
\begin{equation}
1 = \det(X) = {1\over3!} \epsilon^{a_1 a_2 a_3} \epsilon_{b_1 b_2 b_3} X_{a_1}^{~b_1} X_{a_2}^{~b_2} X_{a_3}^{~b_3}  \,, \qquad \forall \, X \in SU(3) \,.
\end{equation}
Expanding the RHS we have
\begin{align}
{\rm tr}(X) {\rm tr}(X) {\rm tr}(X) - 3 {\rm tr}(X) {\rm tr}(XX) + 2 {\rm tr}(XXX) = 6   \,, \qquad \forall \, X \in SU(3) \,. \label{eq:su3triplereduce3X}
\end{align}

For triple-trace products, using
\begin{equation}
\delta^{a_1 a_2 a_3}_{b_1 b_2 b_3} \, X_{a_1}^{~b_1} Y_{a_2}^{~b_2} Z_{a_3}^{~b_3} = \delta^{a_1 a_2 a_3}_{b_1 b_2 b_3} \, X_{a_1}^{~b_1} Y_{a_2}^{~b_2} Z_{a_3}^{~b_3} \det(X^\dagger)  \,, \qquad \forall \, X, Y, Z \in SU(3) \,,
\end{equation}
one can prove the following identity \cite{Kazakov:2024ool} 
\begin{align}
{\rm tr}(X) {\rm tr}(Y) {\rm tr}(Z) = & \, -{\rm tr}(XYZ) -{\rm tr}(ZYX) + {\rm tr}(X) {\rm tr}(YZ) + {\rm tr}(Y) {\rm tr}(ZX) + {\rm tr}(Z) {\rm tr}(XY)  \nonumber\\
& \, + {\rm tr}(X^\dagger Y) {\rm tr}(X^\dagger Z) - {\rm tr}(X^\dagger Y X^\dagger Z) \,, \qquad \forall \, X, Y, Z \in SU(3) \,, \label{eq:su3triplereduceXYZ}
\end{align}
which can be used to reduce any triple-trace operator to a linear combination of double- and single-trace operators.

A few special cases follow. Taking $Y$ to be $X$ or $X^\dagger$ in  \eqref{eq:su3triplereduceXYZ}, we have
\begin{align}
{\rm tr}(X) {\rm tr}(X) {\rm tr}(Z) & =  -2{\rm tr}(XXZ) + 2{\rm tr}(X) {\rm tr}(XZ) + {\rm tr}(Z) {\rm tr}(XX)  + 2{\rm tr}(X^\dagger Z) \,, \label{eq:su3triplereduceXXZ} \\
{\rm tr}(X) {\rm tr}(X^\dagger) {\rm tr}(Z) & = {\rm tr}(Z) + {\rm tr}(X) {\rm tr}(X^\dagger Z) + {\rm tr}(X^\dagger) {\rm tr}(XZ)  \label{eq:su3triplereduceXXdagZ} \\
& \quad + {\rm tr}(X^\dagger X^\dagger){\rm tr}(X^\dagger Z) - {\rm tr}(X^\dagger X^\dagger X^\dagger Z) \,, \qquad \forall \, X, Z \in SU(3) \,.  \notag
\end{align}
Further taking $Z$ to be $X$ in \eqref{eq:su3triplereduceXXZ} and \eqref{eq:su3triplereduceXXdagZ}, we have
\begin{align}
{\rm tr}(X) {\rm tr}(X) {\rm tr}(X) & =  3 {\rm tr}(XX){\rm tr}(X) -2{\rm tr}(XXX) + 6 \,, \label{eq:su3triplereduceXXX} \\
{\rm tr}(X) {\rm tr}(X) {\rm tr}(X^\dagger)  & =  {\rm tr}(XX){\rm tr}(X^\dagger)  + 2{\rm tr}(X^\dagger X^\dagger) + 4 {\rm tr}(X)   \,, \qquad \forall \, X \in SU(3) \,, \label{eq:su3triplereduceXXXdag} 
\end{align}
and taking $Z$ to be the identity matrix in \eqref{eq:su3triplereduceXXZ} and \eqref{eq:su3triplereduceXXdagZ}, we have
\begin{align}
{\rm tr}(X) {\rm tr}(X) & =  {\rm tr}(X X) + 2 {\rm tr}(X^\dagger) \,, \label{eq:su3triplereduceXX1} \\
{\rm tr}(X) {\rm tr}(X^\dagger)  & =  {\rm tr}(X^\dagger X^\dagger){\rm tr}(X^\dagger) - {\rm tr}(X^\dagger X^\dagger X^\dagger) +3 \notag \\
 & = {\rm tr}(X X){\rm tr}(X) - {\rm tr}(XXX) +3  \,, \qquad \forall \, X \in SU(3) \,. \label{eq:su3triplereduceXXdag1} 
\end{align}
Using \eqref{eq:su3triplereduceXXdag1}, we can rewrite \eqref{eq:su3triplereduceXXX} as
\begin{align}
{\rm tr}(X) {\rm tr}(X) {\rm tr}(X) & =  3{\rm tr}(X) {\rm tr}(X^\dagger)+  {\rm tr}(XXX)  -3  \,. \label{eq:su3triplereduceXXXv2} 
\end{align}

The permutations for  the last line of \eqref{eq:su3triplereduceXYZ} implies that 
\begin{align}
& \quad {\rm tr}(X^\dagger Y) {\rm tr}(X^\dagger Z) - {\rm tr}(X^\dagger Y X^\dagger Z) \\
& = {\rm tr}(Y^\dagger X) {\rm tr}(Y^\dagger Z) - {\rm tr}(Y^\dagger X Y^\dagger Z) \notag \\
& = {\rm tr}(Z^\dagger X) {\rm tr}(Z^\dagger Y) - {\rm tr}(Z^\dagger X Z^\dagger Y) \,, \qquad \forall \, X, Y, Z \in SU(3) \,.\nonumber
\end{align}
We can further take $Z$ to be identity matrix, which gives
\begin{align}
& \quad {\rm tr}(X Y^\dagger) {\rm tr}(Y^\dagger) - {\rm tr}(X Y^\dagger Y^\dagger)  \\
& = {\rm tr}(Y X^\dagger) {\rm tr}(X^\dagger) - {\rm tr}(Y X^\dagger X^\dagger) \notag\\
& = {\rm tr}(X) {\rm tr}(Y) - {\rm tr}(X Y) \,, \qquad \forall \, X, Y \in SU(3) \,.\nonumber
\end{align}

\subsubsection*{Reduction formula for SU(3) Wilson loops}

The above relations can be directly applied to Wilson loops since link variables are group elements.
Below we summarize the relation for the Wilson loops that can be used directly in the main text. Note that we have used the normalization condition \eqref{eq:Wnormalization}.

The most general formula is
\begin{align}
w(C_1) w(C_2) w(C_3) 
=& \, - {1\over9} \big[ w(C_1\, C_2 \, C_3) + w(C_3 \, C_2\, C_1) \big]  \\
& \, + {1\over3} \big[ w(C_1) w(C_2\, C_3) + w(C_2) w(C_3\, C_1) + w(C_3) w(C_1\, C_2) \big] \notag \\
& \, + {1\over3} w(\bar C_1\, C_2)w(\bar C_1\, C_3) - {1\over9} w(\bar C_1\, C_2 \,\bar C_1\, C_3)  \,,
\end{align}
where $\bar C_i$ is the reverse of the $C_i$.
We emphasize that in this formula all loops $C_i$ should be understood as starting from the same position $x$ as $C_{i, x}$. If $C_i$ are disconnected, one should add certain backtrack paths so that they have the same contact point.

From the above relation, one can derive several other formulae as
\begin{align}
w(\bar C_1 C_2) w(\bar C_1 C_3) - {1\over3}w(\bar C_1 C_2 \bar C_1 C_3) 
& = w(\bar C_2 C_1) w(\bar C_2 C_3) - {1\over3}w(\bar C_2 C_1 \bar C_2 C_3)  \\
& = w(\bar C_3 C_1) w(\bar C_3 C_2) - {1\over3}w(\bar C_3 C_1 \bar C_3 C_2)  \,.\nonumber \\
w(C_1) w(C_2) - {1\over3} w(C_1 C_2) & = w(C_1 \bar C_2) w(\bar C_2) - {1\over3}w(C_1 \bar C_2 \bar C_2)  \\
& = w(C_2 \bar C_1) w(\bar C_1) - {1\over3}w(C_2 \bar C_1 \bar C_1) \,.\nonumber
\end{align}

\begin{align}\label{eq:SU3relations2}
w(C) w(C) w(C)  & = w(C) w(\bar C)+ {1\over9} w(CCC)  - {1\over9}  \,,  \\
w(C) w(C) & = {1\over3} w(CC) + {2\over3} w(\bar C) \,, \\
w(CC)w(C) & = w(C) w(\bar C) + {1\over3} w(CCC) - {1\over3}  \,.
\end{align}

\paragraph{SU(2) relations.} For completeness, we also give the SU(2) relations. 
Using the SU(2) matrix property, it is easy to find 
\begin{equation}
{\rm tr}(X) {\rm tr}(Y) = {\rm tr}(XY) + {\rm tr}(X Y^\dagger) = {\rm tr}(X Y) + {\rm tr}(X^\dagger Y) \,, \qquad \forall  \, X, Y \in SU(2) \,.
\end{equation}
Note that in SU(2) one has ${\rm tr}(X) = {\rm tr}(X^\dagger)$.
With the normalization condition \eqref{eq:Wnormalization}, we have
\begin{align}\label{eq:WccSU2app}
w(C_1) w(C_2) & = {1\over2} w(C_1\, C_2) + {1\over2} w(C_1\, \bar C_2) \,, \\
w(C) w(C) & = {1\over2} w(C C) + {1\over2} \,,
\end{align}
where both loops $C_i$ should be understood as starting from a same position $x$ as $C_{i, x}$.

\section{Strong and weak coupling expansion}
\label{app:expansion}

In this section, we review the strong and weak coupling expansion of the single plaquette Wilson loop. They provide a good comparison of our bootstrap results. For convenience, we collect them here and use the convention $\beta = N^2/\lambda$.

\paragraph{Strong coupling expansion.}
The strong coupling expansion of free energy $F$ for the $SU(2)$ and $SU(3)$ theory was obtained in \cite{Balian:1974xw}, see also the review in \cite{Drouffe:1983fv}. The expression of the single plaquette Wilson loop can then be extracted from the free energy.

The core idea is evaluating the character expansion of the integrand kernel in the partition functions as
\begin{align}
	\exp(-S) \equiv \exp\left(\beta \sum_{p} \chi_{\rm f}(W_p)\right) = \prod_p \sum_{r} \tilde{\beta}_r \chi_r(W_p) \,,
\end{align}
where $\chi_r$ is the character of the irreducible representation $r$ of the $SU(N)$ group and $\tilde{\beta}_r$ are the new expansion parameters (which will be defined below). For instance, the fundamental representation $r={\rm f}$ takes
\begin{align}
	\chi_{\rm f}(W_p) = d_{\rm f}^{-1} \mathrm{tr}(D^{\rm f}(W_p)) \sim \frac{\mathrm{tr}(W_p)}{N} \,,
\end{align}
where $D^{\rm f}(W_p)$ is a unitary matrix as the fundamental representation matrix of the group element $W_p$, and $d_{\rm f} = N$ is the representation dimension.
The partial function $Z$ for general $d \geqslant 2$ dimensional cases is given as
\begin{align}
	\frac{\log(Z)}{d(d-1)/2} = \ln(\tilde{\beta}_0) +(d-2) \left[ \frac{1}{3} S_6 +(2d-5)S_{10} + {\cal O}(\beta^{11}) \right] \,,
\end{align}
where $S_n$ is defined by
\begin{align}
	S_n = \sum_{r \neq 0} \left( \tilde{\beta}_r/\tilde{\beta}_0 \right)^n \,.
\end{align}

For the gauge group SU(2), it is well-known that the irreducible representation is characterized by the half-integer $j=0,1/2,1,\cdots$, and the expansion coefficients $\tilde{\beta}_r$ is given by $\tilde{\beta}_{j} = I_{2j+1}(\beta)/I_1(\beta)$. The gauge group $SU(3)$ is charactered by the two integer $(\lambda, \mu)$ and $\tilde\beta_r$ is given by
\begin{align}
	\tilde{\beta}_{(\lambda,\mu)} = \sum_{n=-\infty}^{\infty} \det(I_{l_j+i-j+n}(\beta/3)) \,,
\end{align}
where $1 \leqslant i,j \leqslant 3$ and $l_1 =\lambda+\mu$, $l_2 = \mu$, $l_3 = 0$.
For convenience, we give the concrete results as follows
\begin{align}
	\frac{\log(Z_{\text{SU(2)}})}{d(d-1)/2} = & \frac{\beta ^2}{8}-\frac{\beta ^4}{384}+\frac{\beta ^6 (3 d-5)}{9216} +\mathcal{O}(\beta^8) \,, \\
	\frac{\log(Z_{\text{SU(3)}})}{d(d-1)/2} = & \frac{\beta ^2}{36}+\frac{\beta ^3}{648}-\frac{\beta ^5}{93312}+\frac{\beta ^6 (16 d-113)}{90699264} +\mathcal{O}(\beta^7) \,. \notag
\end{align}

The expectation value of the plaquette Wilson loop can be extracted from the free energy as
\begin{align}
	\langle {\rm tr}(U_P) \rangle = \langle W_1 \rangle = \frac{1}{d(d-1)/2}\frac{\partial \log Z}{\partial \beta} \,.
\end{align}

\paragraph{Weak coupling expansion.}
Below we list the perturbation expansion results for the plaquette at weak coupling that we used in the plot.
In 3D YM theory, based on the traditional Feynman diagram method, the results (up to four loops) were obtained in \cite{Hietanen:2004ew, Panagopoulos:2006ky} as
\begin{align}
	\langle W_1 \rangle_{\text{weak}}^{3D\text{ SU(2)}} = & 1- \beta^{-1} -0.2326265(2)\beta^{-2} -0.34214(1)\beta^{-3} +\mathcal{O}(\beta^{-4}) \,, \\
	\langle W_1 \rangle_{\text{weak}}^{3D\text{ SU(3)}} = & 1- \frac{8}{3}\beta^{-1} -1.951315(1)\beta^{-2} -6.8612(2)\beta^{-3} +\mathcal{O}(\beta^{-4}) \,,
\end{align}
where we truncate the known results to three loops.

In the 4D YM theory, the analytical results up to three-loop result were calculated in \cite{Alles:1993dn, Alles:1998is}. The higher series expansions were obtained by numerical methods. In the SU(2) theory, results up to of order $\mathcal{O}(\beta^{-16})$ were obtained based on the series expansion of density of states \cite{Denbleyker:2008ss}. In SU(3), the results up to of order $\mathcal{O}(\beta^{-35})$ were obtained in \cite{Bali:2014fea} by using the numerical stochastic perturbation theory. We give the concrete expressions, truncated as we used in the plot, as follows:
\begin{align}
	\langle W_1 \rangle_{\text{weak}}^{4D\text{ SU(2)}} = & 1 -0.7498 \beta^{-1} -0.1511 \beta^{-2} -0.1427 \beta^{-3} -0.1747\beta^{-4} -0.2435 \beta^{-5} \\
	& -0.368 \beta^{-6} -0.5884 \beta^{-7} -0.98 \beta^{-8} -1.6839 \beta^{-9} -2.9652 \beta^{-10} -5.326 \beta^{-11} \notag \\
	& -9.7234 \beta^{-12} -17.995 \beta^{-13} -33.690 \beta^{-14} -63.702 \beta^{-15} +\mathcal{O}(\beta^{-16}) \,, \notag \\
	\langle W_1 \rangle_{\text{weak}}^{4D\text{ SU(3)}} = & 1 -2\beta^{-1} -1.2208 \beta^{-2} -2.9621 \beta^{-3} -9.417 \beta^{-4} -34.39 \beta^{-5} \\
	& -136.8 \beta^{-6} -577.4 \beta^{-7} -2545 \beta^{-8} -11590 \beta^{-9} -54160 \beta^{-10} +\mathcal{O}(\beta^{-11}) \,. \notag
\end{align}

\providecommand{\href}[2]{#2}\begingroup\raggedright\endgroup


\begin{thebibliography}{10}

\bibitem{JaffeWitten}
A.~Jaffe and E.~Witten, {\it {QUANTUM YANG–MILLS THEORY}}, . Description of
  the Millennium Prize Problems of CMI.

\bibitem{Seiberg:1994rs}
N.~Seiberg and E.~Witten, {\it {Electric - magnetic duality, monopole
  condensation, and confinement in N=2 supersymmetric Yang-Mills theory}},
  {\em Nucl. Phys. B} {\bf 426} (1994) 19--52,
  [\href{http://arxiv.org/abs/hep-th/9407087}{{\tt hep-th/9407087}}]. [Erratum:
  Nucl.Phys.B 430, 485--486 (1994)].

\bibitem{Beisert:2010jr}
N.~Beisert et~al., {\it {Review of AdS/CFT Integrability: An Overview}},  {\em
  Lett. Math. Phys.} {\bf 99} (2012) 3--32,
  [\href{http://arxiv.org/abs/1012.3982}{{\tt arXiv:1012.3982}}].

\bibitem{Rattazzi:2008pe}
R.~Rattazzi, V.~S. Rychkov, E.~Tonni, and A.~Vichi, {\it {Bounding scalar
  operator dimensions in 4D CFT}},  {\em JHEP} {\bf 12} (2008) 031,
  [\href{http://arxiv.org/abs/0807.0004}{{\tt arXiv:0807.0004}}].

\bibitem{Anderson:2016rcw}
P.~D. Anderson and M.~Kruczenski, {\it {Loop Equations and bootstrap methods in
  the lattice}},  {\em Nucl. Phys. B} {\bf 921} (2017) 702--726,
  [\href{http://arxiv.org/abs/1612.08140}{{\tt arXiv:1612.08140}}].

\bibitem{Kazakov:2022xuh}
V.~Kazakov and Z.~Zheng, {\it {Bootstrap for lattice Yang-Mills theory}},  {\em
  Phys. Rev. D} {\bf 107} (2023), no.~5 L051501,
  [\href{http://arxiv.org/abs/2203.11360}{{\tt arXiv:2203.11360}}].

\bibitem{Kazakov:2024ool}
V.~Kazakov and Z.~Zheng, {\it {Bootstrap for Finite N Lattice Yang-Mills
  Theory}},  \href{http://arxiv.org/abs/2404.16925}{{\tt arXiv:2404.16925}}.

\bibitem{Li:2024wrd}
Z.~Li and S.~Zhou, {\it {Bootstrapping the Abelian lattice gauge theories}},
  {\em JHEP} {\bf 08} (2024) 154, [\href{http://arxiv.org/abs/2404.17071}{{\tt
  arXiv:2404.17071}}].

\bibitem{Lin:2020mme}
H.~W. Lin, {\it {Bootstraps to strings: solving random matrix models with
  positivity}},  {\em JHEP} {\bf 06} (2020) 090,
  [\href{http://arxiv.org/abs/2002.08387}{{\tt arXiv:2002.08387}}].

\bibitem{Kazakov:2021lel}
V.~Kazakov and Z.~Zheng, {\it {Analytic and numerical bootstrap for one-matrix
  model and \textquotedblleft{}unsolvable\textquotedblright{} two-matrix
  model}},  {\em JHEP} {\bf 06} (2022) 030,
  [\href{http://arxiv.org/abs/2108.04830}{{\tt arXiv:2108.04830}}].

\bibitem{Cho:2022lcj}
M.~Cho, B.~Gabai, Y.-H. Lin, V.~A. Rodriguez, J.~Sandor, and X.~Yin, {\it
  {Bootstrapping the Ising Model on the Lattice}},
  \href{http://arxiv.org/abs/2206.12538}{{\tt arXiv:2206.12538}}.

\bibitem{Cho:2023ulr}
M.~Cho and X.~Sun, {\it {Bootstrap, Markov Chain Monte Carlo, and LP/SDP
  hierarchy for the lattice Ising model}},  {\em JHEP} {\bf 11} (2023) 047,
  [\href{http://arxiv.org/abs/2309.01016}{{\tt arXiv:2309.01016}}].

\bibitem{Han:2020bkb}
X.~Han, S.~A. Hartnoll, and J.~Kruthoff, {\it {Bootstrapping Matrix Quantum
  Mechanics}},  {\em Phys. Rev. Lett.} {\bf 125} (2020), no.~4 041601,
  [\href{http://arxiv.org/abs/2004.10212}{{\tt arXiv:2004.10212}}].

\bibitem{Berenstein:2021dyf}
D.~Berenstein and G.~Hulsey, {\it {Bootstrapping Simple QM Systems}},
  \href{http://arxiv.org/abs/2108.08757}{{\tt arXiv:2108.08757}}.

\bibitem{Berenstein:2022unr}
D.~Berenstein and G.~Hulsey, {\it {Semidefinite programming algorithm for the
  quantum mechanical bootstrap}},  {\em Phys. Rev. E} {\bf 107} (2023), no.~5
  L053301, [\href{http://arxiv.org/abs/2209.14332}{{\tt arXiv:2209.14332}}].

\bibitem{Nancarrow:2022wdr}
C.~O. Nancarrow and Y.~Xin, {\it {Bootstrapping the gap in quantum spin
  systems}},  {\em JHEP} {\bf 08} (2023) 052,
  [\href{http://arxiv.org/abs/2211.03819}{{\tt arXiv:2211.03819}}].

\bibitem{Lin:2023owt}
H.~W. Lin, {\it {Bootstrap bounds on D0-brane quantum mechanics}},  {\em JHEP}
  {\bf 06} (2023) 038, [\href{http://arxiv.org/abs/2302.04416}{{\tt
  arXiv:2302.04416}}].

\bibitem{Lin:2024vvg}
H.~W. Lin and Z.~Zheng, {\it {Bootstrapping ground state correlators in matrix
  theory. Part I}},  {\em JHEP} {\bf 01} (2025) 190,
  [\href{http://arxiv.org/abs/2410.14647}{{\tt arXiv:2410.14647}}].

\bibitem{Guo:2023gfi}
Y.~Guo and W.~Li, {\it {Solving anharmonic oscillator with null states:
  Hamiltonian bootstrap and Dyson-Schwinger equations}},  {\em Phys. Rev. D}
  {\bf 108} (2023), no.~12 125002, [\href{http://arxiv.org/abs/2305.15992}{{\tt
  arXiv:2305.15992}}].

\bibitem{Mathaba:2023non}
K.~Mathaba, M.~Mulokwe, and J.~a.~P. Rodrigues, {\it {Large N master field
  optimization: the quantum mechanics of two Yang-Mills coupled matrices}},
  {\em JHEP} {\bf 02} (2024) 054, [\href{http://arxiv.org/abs/2306.00935}{{\tt
  arXiv:2306.00935}}].

\bibitem{Hessam:2021byc}
H.~Hessam, M.~Khalkhali, and N.~Pagliaroli, {\it {Bootstrapping Dirac
  ensembles}},  {\em J. Phys. A} {\bf 55} (2022), no.~33 335204,
  [\href{http://arxiv.org/abs/2107.10333}{{\tt arXiv:2107.10333}}].

\bibitem{Berenstein:2024ebf}
D.~Berenstein, G.~Hulsey, and P.~N.~T. Lloyd, {\it {Numerical exploration of
  the bootstrap in spin chain models}},
  \href{http://arxiv.org/abs/2406.17844}{{\tt arXiv:2406.17844}}.

\bibitem{Khalkhali:2024kiv}
M.~Khalkhali, N.~Pagliaroli, A.~Parfeni, and B.~Smith, {\it {Bootstrapping the
  critical behavior of multi-matrix models}},
  \href{http://arxiv.org/abs/2409.07565}{{\tt arXiv:2409.07565}}.

\bibitem{Cho:2024kxn}
M.~Cho, B.~Gabai, J.~Sandor, and X.~Yin, {\it {Thermal Bootstrap of Matrix
  Quantum Mechanics}},  \href{http://arxiv.org/abs/2410.04262}{{\tt
  arXiv:2410.04262}}.

\bibitem{Scheer:2024eyu}
M.~G. Scheer, {\it {Hamiltonian Bootstrap}},
  \href{http://arxiv.org/abs/2410.00810}{{\tt arXiv:2410.00810}}.

\bibitem{Cho:2024owx}
M.~Cho, C.~O. Nancarrow, P.~Tadi\'c, Y.~Xin, and Z.~Zheng, {\it {Coarse-grained
  Bootstrap of Quantum Many-body Systems}},
  \href{http://arxiv.org/abs/2412.07837}{{\tt arXiv:2412.07837}}.

\bibitem{Lawrence:2024mnj}
S.~Lawrence, B.~McPeak, and D.~Neill, {\it {Bootstrapping time-evolution in
  quantum mechanics}},  \href{http://arxiv.org/abs/2412.08721}{{\tt
  arXiv:2412.08721}}.

\bibitem{Zeng:2023jek}
M.~Zeng, {\it {Feynman integrals from positivity constraints}},  {\em JHEP}
  {\bf 09} (2023) 042, [\href{http://arxiv.org/abs/2303.15624}{{\tt
  arXiv:2303.15624}}].

\bibitem{MazziottiPRA2001}
D.~A. Mazziotti and R.~M. Erdahl, {\it Uncertainty relations and reduced
  density matrices: Mapping many-body quantum mechanics onto four particles},
  {\em Phys. Rev. A} {\bf 63} (Mar, 2001) 042113.

\bibitem{NakataEtal2001}
M.~Nakata, H.~Nakatsuji, M.~Ehara, M.~Fukuda, K.~Nakata, and K.~Fujisawa, {\it
  Variational calculations of fermion second-order reduced density matrices by
  semidefinite programming algorithm},  {\em The Journal of Chemical Physics}
  {\bf 114} (05, 2001) 8282--8292.

\bibitem{MazziottiPRL2004}
D.~A. Mazziotti, {\it Realization of quantum chemistry without wave functions
  through first-order semidefinite programming},  {\em Phys. Rev. Lett.} {\bf
  93} (Nov, 2004) 213001.

\bibitem{Barthel:2012Solving}
T.~Barthel and R.~Hübener, {\it Solving condensed-matter ground-state problems
  by semidefinite relaxations},  {\em Phys. Rev. Lett.} {\bf 108} (2012),
  no.~20 200404, [\href{http://arxiv.org/abs/1106.4966}{{\tt
  arXiv:1106.4966}}].

\bibitem{Baumgratz:2012Lower}
T.~Baumgratz and M.~B. Plenio, {\it Lower bounds for ground states of condensed
  matter systems},  {\em New J. Phys.} {\bf 14} (2012), no.~2 023027,
  [\href{http://arxiv.org/abs/1106.5275}{{\tt arXiv:1106.5275}}].

\bibitem{Lin:2020Variational}
L.~Lin and M.~Lindsey, {\it Variational embedding for quantum many-body
  problems},  \href{http://arxiv.org/abs/1910.00560}{{\tt arXiv:1910.00560}}.

\bibitem{Haim:2020VariationalCorrelations}
A.~Haim, R.~Kueng, and G.~Refael, {\it Variational-correlations approach to
  quantum many-body problems},  \href{http://arxiv.org/abs/2001.06510}{{\tt
  arXiv:2001.06510}}.

\bibitem{Wang:2023hss}
J.~Wang, J.~Surace, I.~Fr\'erot, B.~Legat, M.-O. Renou, V.~Magron, and
  A.~Ac\'\i{}n, {\it {Certifying Ground-State Properties of Many-Body
  Systems}},  {\em Phys. Rev. X} {\bf 14} (2024), no.~3 031006,
  [\href{http://arxiv.org/abs/2310.05844}{{\tt arXiv:2310.05844}}].

\bibitem{Gao:2024etm}
Q.~Gao, R.~A. Lanzetta, P.~Ledwith, J.~Wang, and E.~Khalaf, {\it {Bootstrapping
  the Quantum Hall problem}},  \href{http://arxiv.org/abs/2409.10619}{{\tt
  arXiv:2409.10619}}.

\bibitem{Wilson:1974sk}
K.~G. Wilson, {\it {Confinement of Quarks}},  {\em Phys. Rev. D} {\bf 10}
  (1974) 2445--2459.

\bibitem{Simmons-Duffin:2015qma}
D.~Simmons-Duffin, {\it {A Semidefinite Program Solver for the Conformal
  Bootstrap}},  {\em JHEP} {\bf 06} (2015) 174,
  [\href{http://arxiv.org/abs/1502.02033}{{\tt arXiv:1502.02033}}].

\bibitem{Poland:2018epd}
D.~Poland, S.~Rychkov, and A.~Vichi, {\it {The Conformal Bootstrap: Theory,
  Numerical Techniques, and Applications}},  {\em Rev. Mod. Phys.} {\bf 91}
  (2019) 015002, [\href{http://arxiv.org/abs/1805.04405}{{\tt
  arXiv:1805.04405}}].

\bibitem{Rychkov:2023wsd}
S.~Rychkov and N.~Su, {\it {New developments in the numerical conformal
  bootstrap}},  {\em Rev. Mod. Phys.} {\bf 96} (2024), no.~4 045004,
  [\href{http://arxiv.org/abs/2311.15844}{{\tt arXiv:2311.15844}}].

\bibitem{Osterwalder:1973dx}
K.~Osterwalder and R.~Schrader, {\it {AXIOMS FOR EUCLIDEAN GREEN'S FUNCTIONS}},
   {\em Commun. Math. Phys.} {\bf 31} (1973) 83--112.

\bibitem{Osterwalder:1974tc}
K.~Osterwalder and R.~Schrader, {\it {Axioms for Euclidean Green's Functions.
  2.}},  {\em Commun. Math. Phys.} {\bf 42} (1975) 281.

\bibitem{Frohlich:1978px}
J.~Frohlich, R.~Israel, E.~H. Lieb, and B.~Simon, {\it {Phase Transitions and
  Reflection Positivity. 1. General Theory and Long Range Lattice Models}},
  {\em Commun. Math. Phys.} {\bf 62} (1978) 1--34.

\bibitem{Biskup2006}
M.~{Biskup}, {\it {Reflection positivity and phase transitions in lattice spin
  models}},  {\em Methods of Contemporary Mathematical Statistical Physics,
  Lecture Notes in Mathematics} (2009) math--ph/0610025,
  [\href{http://arxiv.org/abs/math-ph/0610025}{{\tt math-ph/0610025}}].

\bibitem{Osterwalder:1977pc}
K.~Osterwalder and E.~Seiler, {\it {Gauge Field Theories on the Lattice}},
  {\em Annals Phys.} {\bf 110} (1978) 440.

\bibitem{Seiler:1982pw}
E.~Seiler, {\em {Gauge Theories as a Problem of Constructive Quantum Field
  Theory and Statistical Mechanics}}, vol.~159.
\newblock Lecture Notes in Physics, Vol. 159, Springer-Verlag., 1982.

\bibitem{Gattringer:2010zz}
C.~Gattringer and C.~B. Lang, {\em {Quantum chromodynamics on the lattice}},
  vol.~788.
\newblock Springer, Berlin, 2010.

\bibitem{Makeenko:1979pb}
Y.~M. Makeenko and A.~A. Migdal, {\it {Exact Equation for the Loop Average in
  Multicolor QCD}},  {\em Phys. Lett. B} {\bf 88} (1979) 135. [Erratum:
  Phys.Lett.B 89, 437 (1980)].

\bibitem{Migdal:1983qrz}
A.~A. Migdal, {\it {Loop Equations and 1/N Expansion}},  {\em Phys. Rept.} {\bf
  102} (1983) 199--290.

\bibitem{Migdal:1980au}
A.~A. Migdal, {\it {PROPERTIES OF THE LOOP AVERAGE IN QCD}},  {\em Annals
  Phys.} {\bf 126} (1980) 279--290.

\bibitem{Gambini:1990dt}
R.~Gambini and J.~Griego, {\it {A Geometric approach to the Makeenko-Migdal
  equations}},  {\em Phys. Lett. B} {\bf 256} (1991) 437--441.

\bibitem{Drouffe:1983fv}
J.-M. Drouffe and J.-B. Zuber, {\it {Strong Coupling and Mean Field Methods in
  Lattice Gauge Theories}},  {\em Phys. Rept.} {\bf 102} (1983) 1.

\bibitem{Balian:1974xw}
R.~Balian, J.~M. Drouffe, and C.~Itzykson, {\it {Gauge Fields on a Lattice. 3.
  Strong Coupling Expansions and Transition Points}},  {\em Phys. Rev. D} {\bf
  11} (1975) 2104. [Erratum: Phys.Rev.D 19, 2514 (1979)].

\bibitem{Bars:1979xb}
I.~Bars and F.~Green, {\it {Complete Integration of U ($N$) Lattice Gauge
  Theory in a Large $N$ Limit}},  {\em Phys. Rev. D} {\bf 20} (1979) 3311.

\bibitem{Gross:1980he}
D.~J. Gross and E.~Witten, {\it {Possible Third Order Phase Transition in the
  Large N Lattice Gauge Theory}},  {\em Phys. Rev. D} {\bf 21} (1980) 446--453.

\bibitem{Wadia:2012fr}
S.~R. Wadia, {\it {A Study of U(N) Lattice Gauge Theory in 2-dimensions}},
  \href{http://arxiv.org/abs/1212.2906}{{\tt arXiv:1212.2906}}.

\bibitem{mathematica}
W.~R. Inc., ``Mathematica, {V}ersion 14.1.''
\newblock Champaign, IL, 2024.

\bibitem{mosek}
M.~ApS, {\em MOSEK Optimizer API for Python. Version 10.2.}, 2024.

\bibitem{Athenodorou:2016ebg}
A.~Athenodorou and M.~Teper, {\it {SU(N) gauge theories in 2+1 dimensions:
  glueball spectra and k-string tensions}},  {\em JHEP} {\bf 02} (2017) 015,
  [\href{http://arxiv.org/abs/1609.03873}{{\tt arXiv:1609.03873}}].

\bibitem{Athenodorou:2021qvs}
A.~Athenodorou and M.~Teper, {\it {SU(N) gauge theories in 3+1 dimensions:
  glueball spectrum, string tensions and topology}},  {\em JHEP} {\bf 12}
  (2021) 082, [\href{http://arxiv.org/abs/2106.00364}{{\tt arXiv:2106.00364}}].

\bibitem{Creutz:1981qr}
M.~Creutz and K.~J.~M. Moriarty, {\it {Phase Transition in SU(6) Lattice Gauge
  Theory}},  {\em Phys. Rev. D} {\bf 25} (1982) 1724--1726.

\bibitem{Lucini:2005vg}
B.~Lucini, M.~Teper, and U.~Wenger, {\it {Properties of the deconfining phase
  transition in SU(N) gauge theories}},  {\em JHEP} {\bf 02} (2005) 033,
  [\href{http://arxiv.org/abs/hep-lat/0502003}{{\tt hep-lat/0502003}}].

\bibitem{GAP4}
The GAP~Group, {\em {GAP -- Groups, Algorithms, and Programming, Version
  4.13.1}}, 2024.

\bibitem{Bachoc2012}
C.~Bachoc, D.~C. Gijswijt, A.~Schrijver, and F.~Vallentin, {\em Invariant
  Semidefinite Programs}, pp.~219--269.
\newblock Springer US, New York, NY, 2012.

\bibitem{Hietanen:2004ew}
A.~Hietanen, K.~Kajantie, M.~Laine, K.~Rummukainen, and Y.~Schroder, {\it
  {Plaquette expectation value and gluon condensate in three dimensions}},
  {\em JHEP} {\bf 01} (2005) 013,
  [\href{http://arxiv.org/abs/hep-lat/0412008}{{\tt hep-lat/0412008}}].

\bibitem{Panagopoulos:2006ky}
H.~Panagopoulos, A.~Skouroupathis, and A.~Tsapalis, {\it {Free energy and
  plaquette expectation value for gluons on the lattice, in three dimensions}},
   {\em Phys. Rev. D} {\bf 73} (2006) 054511,
  [\href{http://arxiv.org/abs/hep-lat/0601009}{{\tt hep-lat/0601009}}].

\bibitem{Alles:1993dn}
B.~Alles, M.~Campostrini, A.~Feo, and H.~Panagopoulos, {\it {The Three loop
  lattice free energy}},  {\em Phys. Lett. B} {\bf 324} (1994) 433--436,
  [\href{http://arxiv.org/abs/hep-lat/9306001}{{\tt hep-lat/9306001}}].

\bibitem{Alles:1998is}
B.~Alles, A.~Feo, and H.~Panagopoulos, {\it {Asymptotic scaling corrections in
  QCD with Wilson fermions from the three loop average plaquette}},  {\em Phys.
  Lett. B} {\bf 426} (1998) 361--366,
  [\href{http://arxiv.org/abs/hep-lat/9801003}{{\tt hep-lat/9801003}}].
  [Erratum: Phys.Lett.B 553, 337--338 (2003)].

\bibitem{Denbleyker:2008ss}
A.~Denbleyker, D.~Du, Y.~Liu, Y.~Meurice, and A.~Velytsky, {\it {Series
  expansions of the density of states in SU(2) lattice gauge theory}},  {\em
  Phys. Rev. D} {\bf 78} (2008) 054503,
  [\href{http://arxiv.org/abs/0807.0185}{{\tt arXiv:0807.0185}}].

\bibitem{Bali:2014fea}
G.~S. Bali, C.~Bauer, and A.~Pineda, {\it {Perturbative expansion of the
  plaquette to ${\cal O}(\alpha^{35})$ in four-dimensional SU(3) gauge
  theory}},  {\em Phys. Rev. D} {\bf 89} (2014) 054505,
  [\href{http://arxiv.org/abs/1401.7999}{{\tt arXiv:1401.7999}}].

\end{thebibliography}
\end{document}